\documentclass[12pt,notitlepage,nofootinbib]{book}

\usepackage{graphicx}
\usepackage{amsmath}
\usepackage{amssymb}
\usepackage{dcolumn}
\usepackage{bm}
\usepackage{color}
\usepackage{dsfont}

\newcommand \beq{\begin{eqnarray}}
\newcommand \eeq{\end{eqnarray}}

\newcommand \be{\begin{equation}}
\newcommand \ee{\end{equation}}

\title{\Large \bf ASPECTS OF CONFINEMENT WITHIN NON-ABELIAN GAUGE THEORIES}

\author{Urko Reinosa}

\date{\it Centre de Physique Th\'eorique, CNRS, Ecole polytechnique,\\ IP Paris, F-91128 Palaiseau, France.}

\begin{document}

\maketitle
\begin{figure}[h]
\begin{center}
\includegraphics[height=0.57\textheight]{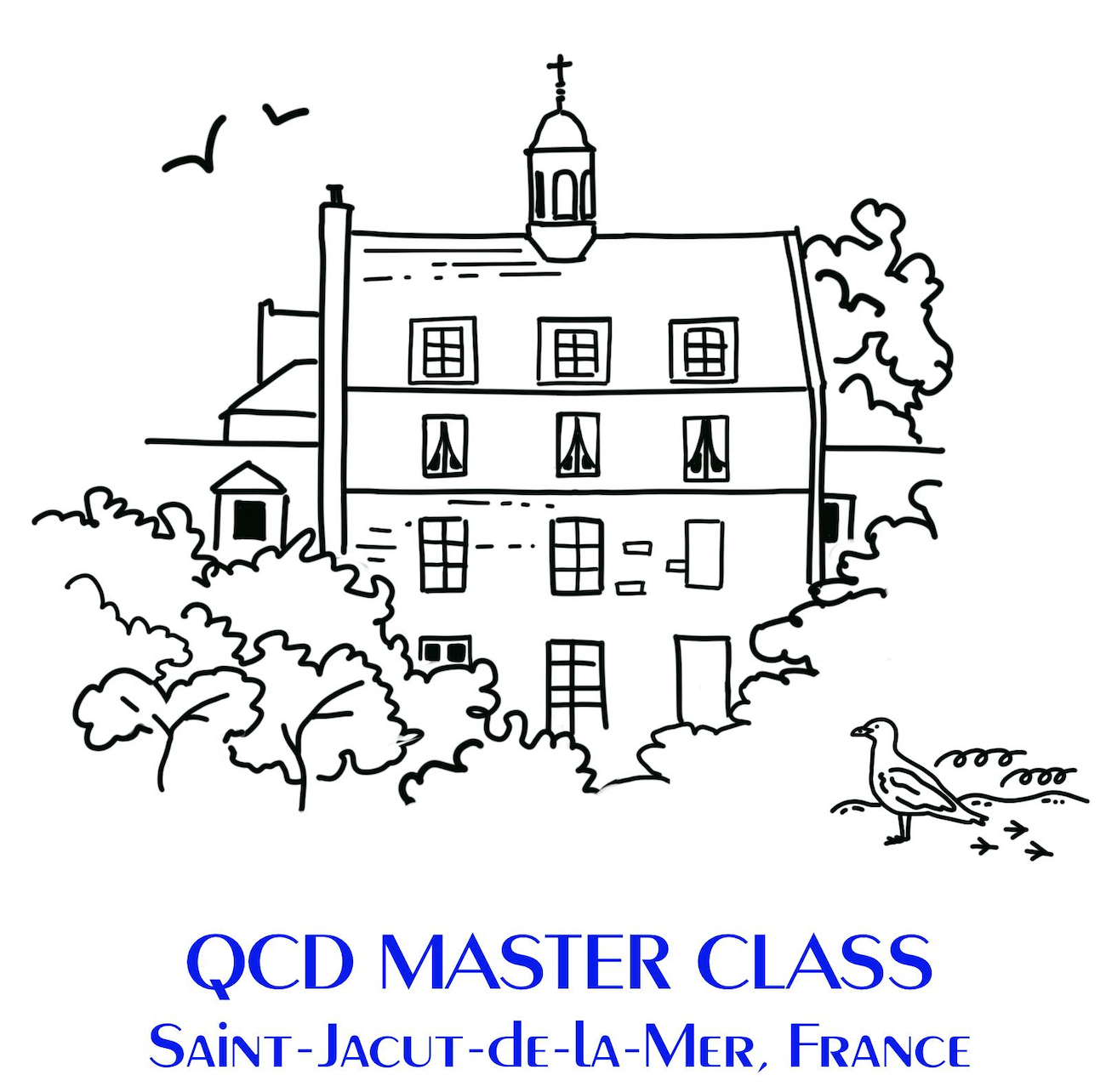}
\end{center}
\end{figure}

\tableofcontents

\chapter*{Acknowledgements}

My warmest thanks to the very active students that attended the 2023 iteration of the QCD Master Class at Saint-Jacut-de-la-Mer, as well as to the other lecturers that shared their QCD knowledge.

Special thanks also to the two excellent organizers, Fran\c{c}ois Arleo (who dreams of driving a Ferrari) and St\'ephane Munier (who is not so sure about Fran\c{c}ois driving skills), and also to the fabulous support teams at Subatech and the Saint-Jacut Abbaye.

\chapter*{Introduction}

The phenomenon of {\it confinement} is a vast topic whose definition and/or manifestation can take different forms. It is not the purpose of these lectures to cover all these aspects and we refer for instance to Ref.~\cite{Greensite:2011zz} for a more detailed presentation. In these lectures, instead, our goal is to analyze a few aspects from the perspective of the phase properties of a system of quarks and gluons in thermal equilibrium. We shall even restrict to a thermal bath of gluons, see below.

Quarks and gluons are the building blocks that allow for the formulation of the fundamental theory of the Strong interaction a.k.a. {\it Quantum Chromodynamics} or QCD. Yet, at low energies, these building blocks are not directly observable but rather {\it confined} within hadronic bound states. Consequently, the low temperature phase is referred to as the {\it confined phase.} On the other hand, {\it asymptotic freedom} in QCD implies that the strength of the interaction diminishes with increasing energy scale. One deduces that, in one way or another, at large enough temperatures, the quarks and the gluons should be liberated into what is commonly referred to as a {\it deconfined phase.} 

\begin{figure}[t]
\begin{center}
\includegraphics[height=0.4\textheight]{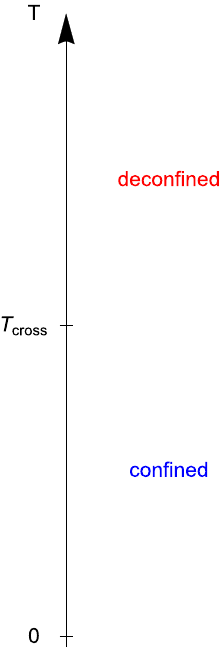}
\caption{Schematic phase diagram of QCD as a function of $T$.}\label{fig:phase}
\end{center}
\end{figure}

These considerations lead to the schematic {\it phase diagram} represented in Fig.~\ref{fig:phase} as a function of the temperature $T$. Interestingly enough, this schematic picture has been put into solid grounds thanks to {\it lattice QCD.} The latter finds indeed that the system undergoes a transition as the temperature is increased. This transition is commonly referred to as the {\it confinement/deconfinement transition.} It is however not a sharp transition but rather a smooth {\it crossover} characterized by a rapid but continuous variation of thermodynamical observables.

\begin{figure}[t]
\begin{center}
\includegraphics[height=0.4\textheight]{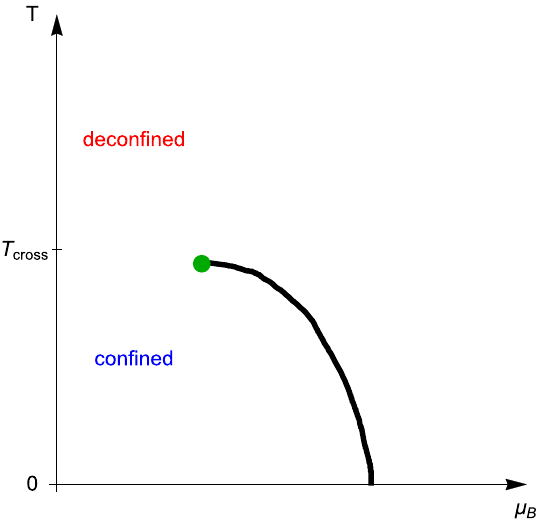}
\caption{One possible scenario for the QCD phase diagram as a function of $T$ and $\mu_B$}\label{fig:phase2}
\end{center}
\end{figure}

Given these premises, it is very tempting to add new directions to the QCD phase diagram in the form of chemical potentials associated to conserved charges. One particularly relevant case is that of the phase diagram in the presence of a {\it baryonic chemical potential} $\mu_B$, which gives access to situations with a non-vanishing baryonic charge density, see Fig.~\ref{fig:phase2}. Among the hotly debated questions in this context, one can cite the hypothetical existence of a {\it critical end-point} within the QCD phase diagram that would separate the crossover region from a first-order transition line. Unfortunately, lattice QCD does not provide a definite answer to this question since it looses efficiency and eventually fails within the finite baryonic chemical potential region due to the {\it infamous sign problem.} 

This calls for the development of new computational techniques. These can be based on well tested {\it phenomenological models}  (such as the Nambu-Jona Lasinio model, various matrix models or the Curci-Ferrari model) or on more first principle approaches known as {\it non-perturbative continuum methods,} of which the Dyson-Schwinger equations or the Functional Renormalization Group are the main representatives.\\

\noindent{These lectures aim at two things:}
\begin{itemize}
\item[$-$] first, irrespectively of the particular approach followed to tackle the QCD phase diagram, we shall introduce some tools that help discussing the confinement/deconfinement transition;

\item[$-$] second, within one particular approach, based on the Curci-Ferrari model, we shall illustrate the use of these various notions in order the describe to the confinement/deconfinement transition.
\end{itemize}

\noindent{In these lectures, for the sake of simplicity, we shall limit ourselves to the case of zero chemical potential where the various concepts and approaches can be tested against lattice results. Also, we shall not consider QCD but the related pure Yang-Mills theory for which a confinement/deconfinement transition can be defined in a sense to be discussed below. The transition in this case is an actual, sharp transition, which can be studied thanks to order parameters and which is interpreted in terms of the breaking of center symmetry. This limit also allows one to decouple the confinement/deconfinement transition from the other important transition in QCD associated to the spontaneous breaking of chiral symmetry.}

\chapter{Poyakov Loop and Center Symmetry}

{\bf \underline{Goal:}} Introduce basic, first principle notions that are useful when addressing the question of the deconfinement transition in Yang-Mills theories.\\

 This includes the Euclidean Yang-Mills action, the order parameter for the transition, a.k.a. the Polyakov loop, and the associated center symmetry.

\section{Thermal Yang-Mills Theory}
In these lectures, our focus is on thermodynamical aspects of YM theory. In particular, we want to discuss what are the possible phases of a system of gluons in equilibrium, as a function of the temperature $T$. This question can in principle be addressed once one knows the {\it partition function} $Z$. The latter can be written as a functional integral over the gluon field, involving an Euclidean version of the YM action.

\subsection{Quantum Mechanical Example}
Let us first recall how the notion of {\it Euclidean action} arises. To this purpose, we use a simpler example borrowed from non-relativistic quantum mechanics, see for instance \cite{Bellac:2011kqa} for more details. Consider a particle moving in a potential $V(q)$. The Lagrangian of such system writes
\beq
L(q,\dot{q})=\frac{1}{2}m\dot{q}^2-V(q)\,,
\eeq 
to which corresponds the Hamiltonian $\smash{H(q,p)=\frac{p^2}{2m}+V(q)}$. To the latter, one can formally associate a {\it partition function}
\beq
Z\equiv {\rm Tr}\,e^{-\beta H}\,,\label{eq:Z}
\eeq
where $\smash{\beta\equiv 1/T}$ is the inverse temperature and ${\rm Tr}$ denotes the trace over a complete set of states. 

One possibility to evaluate the trace in Eq.~(\ref{eq:Z}) is to use the eigenstates $|q\rangle$ of the position operator. One obtains
\beq
Z=\int dq\,\langle q|e^{-\beta H}|q\rangle\,.\label{eq:pos}
\eeq
Now, the matrix element in this formula looks pretty much like a {\it transition amplitude} $\langle q'|e^{-iHt}|q\rangle$ between two positions $q$ and $q'$, which, in the context of Feynman quantization, admits the {\it path integral representation}
\beq
\langle q'|e^{-iHt}|q\rangle\propto \int_{q(0)=q}^{q(t)=q'} \!\!\!{\cal D}q\,\exp\left\{i\int_0^t ds\,L(q(s),\dot{q}(s))\right\},\label{eq:Feynman}
\eeq
where the paths are constrained to obey the boundary conditions $\smash{q(0)=q}$ and $\smash{q(t)=q'}$. In the present case, however, two important differences are worth noticing with respect to the matrix element in Eq.~(\ref{eq:Feynman}). First, in Eq.~(\ref{eq:pos}), we have $\smash{q'=q}$, so what is actually needed is the amplitude to revisit position $q$ after the system as evolved for some time. Second, this evolution takes place along a fictitious, {\it imaginary time,} from an initial time $0$ to a final time $-i\beta$. Correspondingly, the paths to be considered depend upon an imaginary time $s$  that varies along the compact interval $[0,-i\beta]$. 

All in all, this means that the partition function (\ref{eq:pos}) can be formally rewritten as
\beq
Z=\int dq\,\int_{q(0)=q}^{q(-i\beta)=q} {\cal D}q\,\exp\left\{i\int_{[0,-i\beta]} \!\!\!ds\,L(q(s),\dot{q}(s))\right\}.
\eeq
Note that the two integrals in this formula can be combined into a single integral 
\beq
\int dq\,\int_{q(0)=q}^{q(-i\beta)=q} {\cal D}q=\int_{q(0)=q(-i\beta)} {\cal D}q\,,
\eeq 
which sums over all paths obeying $\smash{q(0)=q(-i\beta)}$, known as {\it periodic boundary conditions.} Moreover upon writing the imaginary time $\smash{s=-i\tau}$ in terms of a real {\it Euclidean time $\tau$,} and thus $\smash{ds=-id\tau}$, we find
\beq
& & i\int_{[0,-i\beta]} ds\,L(q(s),\dot{q}(s))\nonumber\\
& & \hspace{1.5cm}=\,i\int_{[0,-i\beta]} ds\,\left[\frac{1}{2}m\dot{q}^2(s)-V(q(s))\right]\nonumber\\
& & \hspace{1.5cm}=\,\int_0^{\beta} d\tau\,\left[\frac{1}{2}m\dot{q}^2(-i\tau)-V(q(-i\tau))\right].
\eeq
Finally, upon re-parameterising the paths as $\smash{q_E(\tau)\equiv q(-i\tau)}$, and thus $\smash{\dot{q}_E(\tau)=-i\dot{q}(-i\tau)}$, this rewrites
\beq
& & i\int_{[0,-i\beta]} ds\,L(q(s),\dot{q}(s))\nonumber\\ 
& & \hspace{1.5cm}=\,\int_0^{\beta} d\tau\,\left[\frac{1}{2}m(i\dot{q}_E(\tau))^2-V(q_E(\tau))\right]\nonumber\\
& & \hspace{1.5cm}=-\int_0^{\beta} d\tau\,\left[\frac{1}{2}m\dot{q}^2_E(\tau)+V(q_E(\tau))\right].
\eeq
We have thus found that the partition function (\ref{eq:pos}) rewrites
\beq
Z=\int_{q_E(0)=q_E(\beta)} {\cal D}q_E\,e^{-S_E[q_E]}\,,
\eeq
in terms of the Euclidean action
\beq
S_E[q]\equiv \int_0^\beta d\tau\,\left[\frac{1}{2}m\dot{q}^2_E(\tau)+V(q_E(\tau))\right].
\eeq
In the following, we omit the subscript $E$ since we work exclusively with the Euclidean formulation.

\subsection{Euclidean YM Functional Integral}
Through a slightly more technical procedure, see for instance \cite{Laine:2016hma} and also Appendix A of the present lectures, the above derivation can be extended to SU(N) YM theory in equilibrium. The partition function of a thermal bath of gluons writes in terms of an Euclidean functional integral over the gluon field:
\beq
Z=\int_{p.b.c.} {\cal D}A\,e^{-S[A]}\,,\label{eq:EYM}
\eeq
where $S[A]$ denotes an Euclidean version of YM theory:
\beq
S[A]=\int_x\,\frac{1}{4g^2}F_{\mu\nu}^a(x)F_{\mu\nu}^a(x)\,,\label{eq:YM}
\eeq
with
\beq
F_{\mu\nu}^a(x)\equiv \partial_\mu A_\nu^a(x)-\partial_\nu A_\mu^a(x)+f^{abc} A^b_\mu(x) A^c_\nu(x)\,,
\eeq
the non-abelian {\it field-strength tensor.}

Here, $f^{abc}$ stands for the SU(N) {\it structure constants.} Recall that the latter are related to the generators $t^a$ of the SU(N) Lie algebra as $\smash{[t^a,t^b]=if^{abc}t^c}$. Sometimes, it will be convenient to see the gluon field or the field-strength tensor as elements of the Lie algebra, that is $\smash{A_\mu\equiv A_\mu^a t^a}$ and $\smash{F_{\mu\nu}\equiv F_{\mu\nu}^at^a}$ respectively. In this case
\beq
F_{\mu\nu}(x)=\partial_\mu A_\nu(x)-\partial_\nu A_\mu(x)-i[A_\mu(x),A_\nu(x)]
\eeq
and the YM action rewrites
\beq
S[A]=\int_x\,\frac{1}{2g^2}{\rm tr}\,F_{\mu\nu}(x)F_{\mu\nu}(x)\,,
\eeq
where, as it is conventional, we have taken $\smash{{\rm tr}\,t^at^b=\delta^{ab}/2}$. For most of these lectures, the focus will be put on the SU($2$) case with generators $t^i\equiv\sigma_i/2$ given in terms of the Pauli matrices
\beq
\sigma_1=\left(\begin{array}{cc}
0 & 1\\
1 & 0
\end{array}\right)\,, \quad \sigma_2=\left(\begin{array}{cc}
0 & -i\\
i & 0
\end{array}\right)\,, \quad \sigma_3=\left(\begin{array}{cc}
1 & 0\\
0 & -1
\end{array}\right).\label{eq:Pauli}
\eeq
The generalization to the SU($3$) case, more relevant to QCD is discussed in the final chapters.

Note finally that $\smash{\int_x\equiv \int_0^\beta d\tau \int d\vec{x}}$ is to be understood as an integral over $[0,\beta]\times\mathds{R}^3$, sometimes referred to as {\it Euclidean space-time.} This space is compact along the temporal direction, and, as it was already the case in the quantum mechanical example, the functional integral in Eq.~(\ref{eq:EYM}) is to be taken over gluon fields that obey periodic boundary conditions ({\it p.b.c.}) at the boundaries of this interval:
\beq
A_\mu^a(\tau=0,\vec{x})=A_\mu^a(\tau=\beta,\vec{x})\,.
\eeq
Boundary conditions are crucial in the context of thermal quantum field theory calculations. They allow in particular to distinguish between bosons and fermions, the latter obeying anti-periodic boundary conditions. Note that it is often convenient to extend the fields beyond the compact time interval $[0,\beta]$, in which case we require them to obey the {\it periodicity condition}
\beq
A_\mu^a(\tau+\beta,\vec{x})=A_\mu^a(\tau,\vec{x})\,.\label{eq:periodic}
\eeq
These conditions will be pivotal when discussing symmetries in the next section.

\subsection{The Polyakov Loop}
Once the partition function is known, the system can be probed by computing {\it thermal averages} of the form
\beq
\langle{\cal O}[A]\rangle\equiv\frac{1}{Z}\int_{p.b.c.} {\cal D}A\,e^{-S[A]}\,{\cal O}[A]\,,\label{eq:O}
\eeq
where ${\cal O}[A]$ denotes a functional of the gauge field. In principle, ${\cal O}[A]$ should be gauge-invariant but we shall come back to this question below since there is a subtle point at finite temperature.

\begin{figure}[t]
\begin{center}
\includegraphics[height=0.4\textheight]{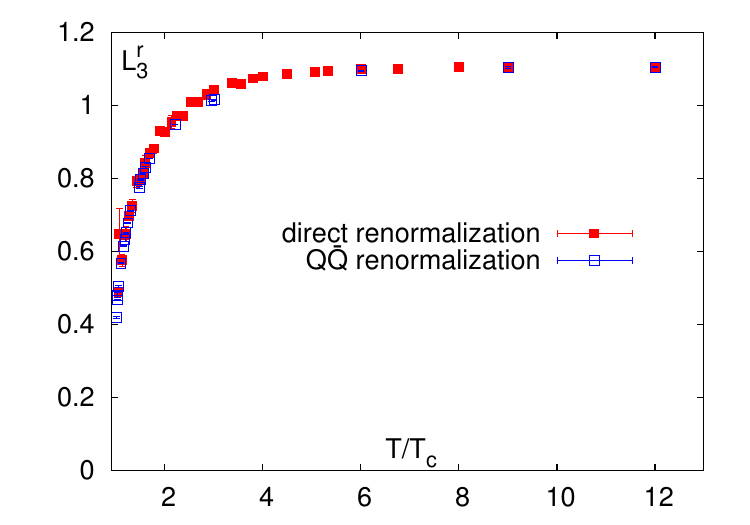}
\caption{Renormalized SU($3$) Polyakov loop as computed on the lattice \cite{Gupta:2007ax}.}\label{fig:lattice}
\end{center}
\end{figure}

A particularly interesting example is obtained by choosing a color-traced temporally-ordered Wilson line
\beq
{\cal O}[A]\to\Phi[A]\equiv\frac{1}{N}{\rm tr}\,\underbrace{{\cal P}\,\exp\left\{i\int_0^\beta d\tau\,A_0^a(\tau,\vec{x})t^a\right\}}_{\equiv L[A]}\,,
\eeq
whose thermal average defines the {\it Polyakov loop} $\smash{\ell\equiv\langle\Phi[A]\rangle}$. The relevance of this thermal average is that it provides an {\it order parameter} allowing one to distinguish between a confined and deconfined phase \cite{Polyakov:1978vu,Svetitsky:1985ye}. Indeed, we can write
\beq
Z\ell=\frac{1}{N}{\rm tr}\,\int_{p.b.c.}{\cal D}A\,e^{-S[A]}{\cal P}\,\exp\left\{i\int_0^\beta d\tau\,A_0^a(\tau,\vec{x})t^a\right\}\equiv Z_{\rm charge}.\label{eq:interpretation}
\eeq
Very schematically, if we forget for the moment the subtleties related to the path ordering and the trace, the right-hand side of this equation corresponds to the partition function $Z_{\rm charge}$ of a thermal bath of gluons in the presence of a static color charge $t^a$ that couples to $\int_0^\beta d\tau A_0^a(\tau,\vec{x})$. The actual derivation which explains the presence of the color trace and the time-ordering is given in Appendix A.

Coming back to Eq.~(\ref{eq:interpretation}) and recalling that the free-energy $F$ relates to $Z$ as $\smash{\ln Z=-\beta F}$, we deduce that
\beq
\ln \ell =\ln Z_{\rm charge}-\ln Z=-\beta(F_{\rm charge}-F)\equiv -\beta\Delta F\,,
\eeq
where $\smash{\Delta F\equiv F_{\rm charge}-F}$ is the energy cost for bringing the color charge into the bath of gluons. Equivalently, one can write $\smash{\ell=e^{-\beta\Delta F}}$. From here, we can contemplate two possibilities:
\begin{itemize}
\item[$-$] if the bath of gluons is in a confining state, then $\Delta F=\infty$ and $\ell=0$;
\item[$-$] if the bath of gluons is in a deconfined state, then $\Delta F<\infty$ and $\ell>0$.
\end{itemize}
Thus the Polyakov loop plays indeed the role of an order parameter distinguishing between a confined and a deconfined phase.

Being a particular  Wilson line, the Polyakov loop is particularly adapted for an evaluation on the lattice where it has been verified that it is indeed an order parameter in the case of YM theories, see Fig.~\ref{fig:lattice} for an illustration in the SU($3$) case.  The SU($2$) lattice simulations show a similar behavior with the difference that the transition is second order in this case. 

In the continuum, the Polyakov loop is not so easily computed and it might be useful to identify equivalent but simpler order parameters. A crucial prerequisite is that these alternative order parameters should probe the same symmetry as the Polyakov loop. This symmetry which we now discuss goes by the name of {\it center symmetry} \cite{Pisarski:2002ji}.

\section{Center Symmetry}
Usually, that an order parameter vanishes in one phase relates to the explicit -- or Wigner-Weyl -- realization of a symmetry. Similarly, the transition to another phase is interpreted in terms of the breaking -- or Nambu-Goldstone realization -- of that symmetry. Let us now see what is the symmetry associated to the Polyakov loop.

\subsection{Twisted Gauge Transformations}
Just as its Minkowskian analog, the Euclidean YM action (\ref{eq:YM}) is invariant under transformations of the form
\beq
A_\mu^U(x)\equiv U(x) A_\mu(x)U^\dagger(x)+iU(x)\partial_\mu U^\dagger(x)\,,\label{eq:transfo}
\eeq
where we recall our notation $\smash{A_\mu(x)\equiv A_\mu^a(x)t^a}$.  Now, since only periodic gauge fields should be considered in the evaluation of thermal averages, one should restrict to transformations (\ref{eq:transfo}) that preserve the condition (\ref{eq:periodic}). An obvious choice is that of transformations that are themselves periodic:
\beq
U(\tau+\beta,\vec{x})=U(\tau,\vec{x})\,.\label{eq:g0}
\eeq
But, as it can be trivially checked (do it!), a slightly more general choice is that of transformations that are periodic modulo a phase:
\beq
U(\tau+\beta,\vec{x})=e^{i\phi}U(\tau,\vec{x})\,.\label{eq:k}
\eeq
The phase is not totally arbitrary since both $U(\tau+\beta,\vec{x})$ and $U(\tau,\vec{x})$ are elements of SU(N) and thus of determinant $1$. From this, one deduces that
\beq
1={\rm det}\,U(\tau+\beta,\vec{x})=(e^{i\phi})^N{\rm det}\,U(\tau,\vec{x})=e^{iN\phi}\,,
\eeq
and thus $\smash{\phi=2\pi k/N}$, with $\smash{k=0,\dots,N-1}$. We refer to these transformations as {\it twisted gauge transformations.} They form a group denoted ${\cal G}$ in what follows.\\

\noindent{$\diamond\diamond\diamond\,\,${\bf Problem 1.1:} Show that these are the only possible transformations of the form (\ref{eq:transfo}) that preserve the periodicity condition (\ref{eq:periodic}). \underline{Tip:} express the periodicity of $A^U_\mu(x)$ as an equation for $\smash{Z(\tau,\vec{x})\equiv U^\dagger(\tau,\vec{x})U(\tau+\beta,\vec{x})}$, valid for an arbitrary configuration $A_\mu(x)$ obeying (\ref{eq:periodic}).}$\,\,\diamond\diamond\diamond$\\

The {\it periodic gauge transformations} (\ref{eq:g0}) form a subgroup of ${\cal G}$ denoted ${\cal G}_0$. The distinction between ${\cal G}$ and ${\cal G}_0$ will be crucial in what follows. We also introduce the set ${\cal G}_k$ of {\it $k$-twisted transformations} whose associated phase in Eq.~(\ref{eq:k}) is $e^{i2\pi k/N}$. Except for ${\cal G}_0$, none of these sets has a group structure. However, we note that, given an element $\smash{U\in {\cal G}_k}$ and and element $\smash{U'\in {\cal G}_{k'}}$, one has $\smash{UU'\in {\cal G}_{k+k'[N]}}$. This defines a group structure on the finite set $\{{\cal G}_0,\dots,{\cal G}_{N-1}\}$ which is isomorphic to the group $\mathds{Z}/N\mathds{Z}$ of relative integers modulo $N$, or to the group $\{1,e^{i2\pi/N},\dots,e^{i2\pi (N-1)/N}\}$ formed by the $N^{\rm th}$ roots of unity.

As an illustration, let us construct some examples of twisted transformations within SU($2$) YM theory. In this case, we have $\smash{k=0}$ or $\smash{k=1}$ and the phase in Eq.~(\ref{eq:k}) is $e^{i\pi k}$, so either $1$ or $-1$. For simplicity, consider transformations of the form 
\beq
U(\tau)=\exp\left\{i\theta\frac{\tau}{\beta}\frac{\sigma_3}{2}\right\},
\eeq
which are such that
\beq
U(\tau+\beta)=U(\beta)U(\tau),\label{eq:rhs}
\eeq
and let us look for values of $\theta$ such that $\smash{U(\beta)=\mathds{1}}$ in which case $U\in {\cal G}_0$, or $\smash{U(\beta)=-\mathds{1}}$ in which case $U\in {\cal G}_1$. This can actually be figured out without any calculation if one observes that 
\beq
U(\beta)=\exp\left\{i\theta\frac{\sigma_3}{2}\right\},\label{eq:Uofbeta}
\eeq
represents (in the sense of group representation theory) a rotation of angle $\theta$ around direction $3$ in the space of states of a spin $1/2$ particle. If we choose $\smash{\theta=\pm 2\pi}$, this is a full rotation in configuration space but, as it is well known, the wave function of a spin $1/2$ particle gets multiplied by a factor $-1$. We have thus found that the choice $\smash{\theta=\pm 2\pi}$ corresponds to a twisted gauge transformation with $k=1$. If we choose instead $\theta=\pm 4\pi$, one rotates twice around direction $3$ in configuration space and the wave function is restored to its original form, thus corresponding to a periodic gauge transformation, with $k=0$.\\

\noindent{$\diamond\diamond\diamond\,\,${\bf Problem 1.2:} Obtain the same results by brute-force calculating $U(\beta)$, see Eq.~(\ref{eq:Uofbeta}). \underline{Tip:} recall that $\sigma_3$ is diagonal, with diagonal elements $\pm 1$.}$\,\,\diamond\diamond\diamond$

\subsection{Relation to the Polyakov Loop}
Let us now investigate the relation of the twisted gauge transformations to the Polyakov loop. First of all, we note that the Wilson line $L[A]$ is such that
\beq
L[A^U]=U(\beta)L[A]U^\dagger(0)\,.\label{eq:LU}
\eeq

\vglue1mm

\noindent{$\diamond\diamond\diamond\,\,${\bf Problem 1.3:} Show this identity. \underline{Tip:} consider each side of the identity as a function of $\beta$ and show that they obey the same first order ordinary differential equation and the same initial condition.}$\,\,\diamond\diamond\diamond$\\

Consider now the traced Wilson line $\Phi[A]$. Using the cyclicality of the trace, it follows from Eq.~(\ref{eq:LU}) that
\beq
\Phi[A^U] & \!\!=\!\! & \frac{1}{N}{\rm tr}\,L[A^U]\nonumber\\
& \!\!=\!\! & \frac{1}{N}{\rm tr}\,U(\beta)L[A]U^\dagger(0)\nonumber\\
& \!\!=\!\! & \frac{1}{N}{\rm tr}\,U^\dagger(0)U(\beta)L[A]\,.
\eeq
Now, if the transformation $U$ is a $k$-twisted gauge transformation, then $\smash{U^\dagger(0)U(\beta)=e^{i2\pi k/N}}$ and we finally arrive at
\beq
\Phi[A^U]=\frac{e^{i2\pi k/N}}{N}{\rm tr}\,L[A]=e^{i2\pi k/N}\Phi[A]\,.\label{eq:Phi}
\eeq
This shows that the functional $\Phi[A]$ transforms simply under $k$-twisted gauge transformations.

Let us know see how this impacts the Polyakov loop, that is the thermal average of the traced Wilson line $\Phi[A]$. We shall implicitly assume that the symmetry associated to $k$-twisted gauge transformations is realized explicitly, that is in the Wigner-Weyl sense. In practice, this means that one can use a $k$-twisted gauge transformation as a change of variables under the functional integral that defines the thermal average. Since these transformations do not change the boundary conditions, it follows that
\beq
\ell & \!\!=\!\! & \frac{1}{Z}\int_{p.b.c.}{\cal D}A\,e^{-S[A]}\,\Phi[A]\nonumber\\
& \!\!=\!\! & \frac{1}{Z}\int_{p.b.c.}{\cal D}A^U\,e^{-S[A^U]}\,\Phi[A^U]\nonumber\\
& \!\!=\!\! & e^{i2\pi k/N}\frac{1}{Z}\int_{p.b.c.}{\cal D}A\,e^{-S[A]}\,\Phi[A]=e^{i2\pi k/N}\ell\,,\label{eq:ll}
\eeq
where, in the last step, we have used the invariance of the measure ${\cal D}A$ and the action $S[A]$, together with the transformation rule for $\Phi[A]$, Eq.~(\ref{eq:Phi}). Comparing the left and right-hand sides of Eq.~(\ref{eq:ll}), we deduce that $\smash{\ell=0}$ and thus, the confining phase is related to the explicit realization of the symmetry associated to twisted gauge transformations.\\

\noindent{$\diamond\diamond\diamond\,\,${\bf Problem 1.4:} In the case where the symmetry is explicitly broken, the expression for the Polyakov loop does not makes sense without adding to the action an infinitesimal source term $j=\int_{\vec{x}}\Phi[A](\vec{x})$ with $\smash{j=\rho e^{i\theta}\in\mathds{C}}$ and $\smash{\rho\to 0}$. Then, the Polyakov loop should more precisely be written}
\beq
\ell_{\rho,\theta}\equiv \frac{1}{Z}\int_{p.b.c.}{\cal D}A\,e^{-S[A]+\rho e^{i\theta}\int_{\vec{x}}\Phi[A](\vec{x})}\,\Phi[A]\,.\nonumber
\eeq 
How does the invariance under $k$-twisted transformations constrains $\ell_{\rho,\theta}$? Discuss what happens in the limit $\smash{\rho\to 0}$ when the limit is regular (that is independent on the phase $\theta$) or irregular (that is dependent on the phase $\theta$). $\,\,\diamond\,\diamond\,\diamond$

\subsection{The Center Symmetry Group}
It seems that we have a little paradox: we have found a quantity that allows one to distinguish between physically distinct phases of a thermal bath of gluons but this quantity transforms non-trivially under gauge transformations, so it does not seem to be an observable.

The solution to this paradox is that only the periodic gauge transformations belonging to ${\cal G}_0$ should be considered as the true gauge transformations, that do not alter the state of the system. In particular, they do not transform the Polyakov loop. In contrast, $k$-twisted transformations with $k\neq 0$ need to be seen as physical transformations for they act non-trivially on the Polyakov loop. More precisely, two transformations with the same $k$ should seen as representing the same transformation since they transform the Polyakov loop in the same way. In other words, the actual physical symmetry group is nothing but the group $\{{\cal G}_0,{\cal G}_1,\dots,{\cal G}_{N-1}\}$ discussed above. This is the actual, physical center symmetry group. In more technical terms, this group is the quotient group ${\cal G}/{\cal G}_0$ that removes away the gauge redundancy that is present within ${\cal G}$ due to the presence of the subgroup ${\cal G}_0$ of (unphysical) gauge transformations.

\section{What Comes Next}
As already emphasized, in the context of continuum calculations, it is worth asking whether there could exist equivalent, but simpler, alternatives to the Polyakov loop.  In particular, continuum methods provide an easy access to gluon correlators, that is thermal averages of products of gluon fields. So, one may ask whether it is possible to construct such alternative order parameters from the lowest correlators. 

In these lectures, we shall discuss the possibility of using the gluon field average $\langle A_\mu\rangle$. This discussion will occupy us quite some time because, unlike what happens with the Polyakov loop, the notion of gluon average exists only within a gauge-fixed setting. Now, gauge fixing comes with its own bunch of problems, which we will need to analyze and address as best as we can.

Before dealing with such complicated matters, however, there is a simpler question that one can address and which will turn out to be relevant when dealing with the gluon field average. This question can be formulated as follows. Suppose that, to a good approximation, the system is described by its classical limit. This means that its state is characterized by a classical gauge-field configuration. Knowing this state, how do we determine whether the system is in a confined or deconfined state. In other words, how do we identify among all possible classical configurations, those that are confining?

\chapter{SU($2$) Confining Configurations}

{\bf \underline{Goal:}} Learn how to identify classical, confining gluon field configurations.\\

For simplicity, we shall consider the SU($2$) case since anyone reading these notes should have a good intuition of how the color algebra operates in this case. In particular, recall that an SU(2) transformation $U=e^{i\theta \vec{n}\cdot\vec{\sigma}/2}$ acts on an element $\vec{X}\cdot\vec{\sigma}/2$ of the algebra as
\beq
U\vec{X}\cdot\vec{\sigma}U^\dagger=\vec{X}'\cdot\vec{\sigma}\,,\label{eq:21}
\eeq
where
\beq
\vec{X}'={\cal R}(\theta;\vec{n})\vec{X}\,,\label{eq:22}
\eeq
and ${\cal R}(\theta,\vec{n})$ is the SO($3$) rotation of angle $\theta$ and axis $\vec{n}$. In more technical terms, SU($2$) transformations are represented on the Lie algebra as standard three-dimensional rotations. This is known as the {\it adjoint representation} of the SU($2$) group. Elements to deal with the SU($3$) case will be given in the final chapters.

\section{Physical Invariance and Gauge Fields}
Suppose that you are given a classical gauge field configuration describing the state of a system of gluons in some classical limit. The question is now: how do we test that this classical configuration describes a confining phase? As we have already argued, the confining phase has to do with the explicit, or Wigner-Weyl realization of center symmetry, so one may look for configurations that comply with center symmetry explicitly. The question is in fact more general since, given a classical gluon configuration, one could ask whether it complies with a given physical symmetry. What criterion should be applied?

\subsection{Examples from Classical Electrodynamics}
To gain some intuition, let us start by considering simpler examples, borrowed from classical electrodynamics. 

Take first the case of a constant and static electric field along the $z$-direction, $\smash{\vec{E}(t,x,y,z)=E\vec{e}_z}$. This physical field is clearly invariant under any -- temporal or spatial -- translation. Indeed
\beq
\vec{E}(t+s,x+u,y+v,z+w)=\vec{E}(t,x,y,z)\,.
\eeq
What about the corresponding scalar potential $\smash{V(t,x,y,z)=-Ez}$? Clearly, the latter is explicitly invariant under time translations or under spatial translations along $x$ or $y$:
\beq
V(t+s,x+u,y+v,z)=V(t,x,y,z)\,.
\eeq
But it is clearly not invariant -- at least not in the usual sense -- under translations along $z$:
\beq
V(t,x,y,z+w)=V(t,x,y,z)-Ew\,.
\eeq
However, the right-hand side of this last equation can be seen as the gauge-transformed of the original scalar potential:
\beq
V(t,x,y,z)+\frac{\partial}{\partial t}(-Ew t)\equiv V^\Lambda(t,x,y,z)\,,
\eeq
where $\smash{\Lambda(t,x,y,z)=-Ewt}$\,. All in all, we have found that
\beq
V(t+s,x+u,y+v,z+w)=V^\Lambda(t,x,y,z)\,,
\eeq
so the scalar potential is translation invariant {\it modulo a gauge transformation.}

Similar considerations apply to the case of a constant and static magnetic field along the $z$-direction, $\smash{\vec{B}(t,x,y,z)=B\vec{e}_z}$. This physical field is clearly invariant under any, temporal or spatial, translation:
\beq
\vec{B}(t+s,x+u,y+v,z+w)=\vec{B}(t,x,y,z)\,.
\eeq
Now, the corresponding vector potential writes $\vec{A}(t,x,y,z)=(\vec{r}\times\vec{B})/2=B(y\vec{e}_x-x\vec{e}_y)/2$.\\

\noindent{$\diamond\diamond\diamond\,\,${\bf Problem 2.1:} Show this relation. \underline{Tip:} recall that $\vec{B}=\vec{\nabla}\times\vec{A}$.}$\,\,\diamond\diamond\diamond$\\

\noindent{Clearly, this vector potential is explicitly invariant under time translations or under spatial translations along $z$:}
\beq
\vec{A}(t+s,x,y,z+w)=\vec{A}(t,x,y,z)\,.
\eeq
But it is clearly not invariant -- at least not in the usual sense -- under translations along $x$ or $y$:
\beq
\vec{A}(t,x+u,y+v,z)=\vec{A}(t,x,y,z)+\frac{B}{2}(v\vec{e}_x-u\vec{e}_y)\,.
\eeq
However, the right-hand side of this last equation can be seen as the gauge-transformed of the original vector potential:
\beq
\vec{A}(t,x,y,z)-\vec{\nabla}\left[\frac{B}{2}(uy-vx)\right]\equiv\vec{A}^\Lambda(t,x,y,z)\,,
\eeq
where $\smash{\Lambda(t,x,y,z)=B(uy-vx)/2}$. All in all, we have found that
\beq
\vec{A}(t+s,x+u,y+v,z+w)=\vec{A}^\Lambda(t,x,y,z)\,,
\eeq
so the vector potential is also translation invariant {\it modulo a gauge transformation.}

\subsection{Invariance Modulo Gauge Transformations}
On general grounds, because gauge fields contain unphysical components that can be changed at will by means of gauge transformations, the ability of a gauge field to comply with a physical symmetry needs always to be understood modulo possible gauge transformations.

Coming back to our finite temperature set-up, recall that the actual gauge transformations are the transformations within ${\cal G}_0$, corresponding to periodic gauge transformations. Therefore, invariance under a certain physical transformation $\smash{A_\mu\to A_\mu^T}$ should be expressed as \cite{Reinosa:2019xqq,vanEgmond:2023lnw}
\beq
\exists U_0\in {\cal G}_0\,, \quad A_\mu^T=A_\mu^{U_0}\,.\label{sec:T}
\eeq
That is, upon physically transforming the classical gauge field configuration, we should retrieve a gauge-equivalent configuration, but not necessarily the original configuration.

In particular, in the case of center transformations $\smash{U\in {\cal G}}$, the confining or center-invariant configurations will be such that
\beq
\exists U_0\in {\cal G}_0\,, \quad A_\mu^U=A_\mu^{U_0}\,.\label{eq:center}
\eeq
But a similar criterion holds for other symmetries. For instance, in the case of charge conjugation $\smash{A_\mu\to A_\mu^C\equiv -A_\mu^{\rm t}}$, the configurations that are compatible with charge conjugation correspond to the condition
\beq
\exists U_0\in {\cal G}_0\,, \quad A_\mu^C=A_\mu^{U_0}\,.\label{eq:C}
\eeq
The goal of this chapter is to analyze the conditions (\ref{eq:center}) and (\ref{eq:C}) further in the SU($2$) case. The discussion in the SU($3$) is postponed to the last chapter.

\subsection{Constant, Temporal, Abelian Configurations}
It will not be necessary, however, to find all possible configurations obeying (\ref{eq:center}) or (\ref{eq:C}) but just particular candidates. A particularly convenient choice is to restrict to constant, temporal and Abelian configurations. In the SU(2) case, they read
\beq
A_\mu(x)=T\delta_{\mu0}\,r \frac{\sigma_3}{2}\,,\label{eq:form2}
\eeq
with $\sigma_3$ the diagonal Pauli matrix, see Eq.~(\ref{eq:Pauli}). The explicit factor of $T$ in Eq.~(\ref{eq:form2}) has been introduced to make the variable $r$ dimensionless. By restricting to configurations of the form (\ref{eq:form2}), the problem simplifies considerably since the space of configurations that we consider is isomorphic to the real axis. We want to find which points of this real axis correspond to confining configurations obeying (\ref{eq:center}).\\

\noindent{$\diamond\diamond\diamond\,\,${\bf Problem 2.2:} Consider the traced Wilson line $\Phi[A]$ defined in the previous chapter. Show that, for backgrounds of the form (\ref{eq:form2}), one has}
\beq
\Phi[A]=\cos\left(r/2\right).
\eeq 
\underline{Tip:} Note that the path-ordering that enters the definition of $\Phi[A]$ can be ignored since the considered configurations are constant. $\,\,\diamond\diamond\diamond$

\section{Anatomy of a Gauge Transformation}
The criterion (\ref{eq:center}) rewrites
\beq
\exists U_0\in {\cal G}_0\,, \quad (A_\mu^U)^{U_0}=A_\mu\,.
\eeq
A strategy to find the invariant configurations $A_\mu$ can be then formulated as follows: construct a gauge transformation $\smash{U_0\in {\cal G}_0}$ such that its combination with the center transformation $\smash{U\in {\cal G}}$ gives a transformation $A_\mu\to (A_\mu^U)^{U_0}$ on field space that possesses fixed-points. In order to see how to construct such a gauge transformation $U_0$, we first need to understand a bit deeper the structure of the group ${\cal G}_0$ of gauge transformations. Since we have restricted to configurations of the form (\ref{eq:form2}), we shall also restrict to gauge transformations that preserve this form.

\subsection{Weyl Transformations}
\begin{figure}[h]
\begin{center}
\includegraphics[height=0.045\textheight]{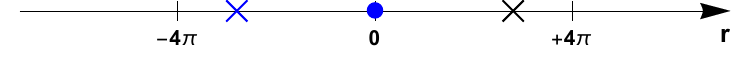}
\caption{Weyl transformation $\smash{r\to -r}$. The two crosses represent two configurations connected by the Weyl transformation.}\label{fig:Weyl2}
\end{center}
\end{figure}

We can first look for global gauge transformations (that is color rotations) that preserve the form (\ref{eq:form2}), known as {\it Weyl transformations.} In the SU($2$) case, a general such transformation $e^{i\theta\vec{n}\cdot\vec{\sigma}/2}$ acts on the Lie algebra as a rotation of angle $\theta$ around direction $\vec{n}$, Eqs.~(\ref{eq:21})-(\ref{eq:22}). The only such transformations that preserve the form (\ref{eq:form2}) and thus direction $3$ are either the identity or any rotation of angle $\pi$ whose axis $\vec{n}$ is orthogonal to direction $3$. The identity does not do anything non-trivial to (\ref{eq:form2}). In contrast, the other transformations act on (\ref{eq:form2}) as $r\to -r$, that is a point reflection with respect to the origin, the blue dot in Fig.~\ref{fig:Weyl2}.

\subsection{Winding Transformations}
\begin{figure}[h]
\begin{center}
\includegraphics[height=0.054\textheight]{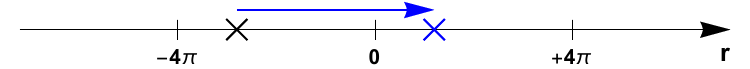}
\caption{Winding transformation $\smash{r\to r+4\pi}$. The two crosses represent two configurations connected by the winding transformation.}\label{fig:Winding2}
\end{center}
\end{figure}

As for local gauge transformations, we shall limit ourselves to the transformations
\beq
U(\tau)=\exp\left\{i\theta\frac{\tau}{\beta}\frac{\sigma_3}{2}\right\},
\eeq
with $\smash{\theta=\pm 4\pi}$ already discussed in the previous chapter. It is easily seen (check it!) that they act on configurations of the form (\ref{eq:form2}) as $r\to r\pm 4\pi$, that is a translation by $\pm 4\pi$. The translation by $4\pi$ is represented in Fig.~\ref{fig:Winding2}.

\vglue-2mm
\noindent{Obviously, by iterating various of these transformations, one obtains similar transformations with $\smash{\theta=4\pi n}$ any multiple of $4\pi$, which we refer to as {\it winding transformations.} They act on configurations of the form (\ref{eq:form2}) as $r\to r+4\pi n$.}\\

\noindent{$\diamond\diamond\diamond\,\,${\bf Problem 2.3:} Show that the most general transformation that preserves the form (\ref{eq:form2}) is either a winding transformation or a combination of a winding transformation with a Weyl transformation.} \underline{Tip:} Write the condition expressing that the form (\ref{eq:form2}) is preserved by the transformation and use that it applies to any component $\mu$ and any value of $r$ in (\ref{eq:form2}). $\,\,\diamond\diamond\diamond$

\subsection{Weyl Chambers}
\begin{figure}[h]
\begin{center}
\includegraphics[height=0.043\textheight]{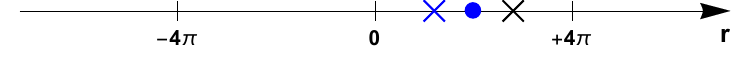}
\caption{Point reflection $r\to 2\pi-r$ obtained by combining a Weyl transformation $r\to-r$ and the winding transformation $r\to r+4\pi$.}\label{fig:Weyl22}
\end{center}
\end{figure}

By combining the above transformations, one generates new elements of ${\cal G}_0$. In particular, by combining a Weyl transformation and a winding transformation, one obtains $r\to 4\pi n-r$, that is a point reflection with respect to $2\pi n$, with $n\in\mathds{Z}$, see Fig.~\ref{fig:Weyl22} for an illustration in the case $n=1$.

\begin{figure}[h]
\begin{center}
\includegraphics[height=0.05\textheight]{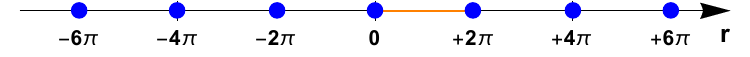}
\caption{Weyl chambers obtained by finding all point reflections that belong to ${\cal G}_0$. We have singled out one particular Weyl chamber (in orange) for later purpose.}\label{fig:chambers}
\end{center}
\end{figure}

Note that we can completely forget about the winding transformations $r\to r+4\pi n$ since the latter can be generated from combining two point reflections with respect to $2\pi k$ and $2\pi (k+n)$:
\beq
r\to  4\pi k-r\to 4\pi (k+n)-(4\pi k-r)=r+4\pi n\,.
\eeq
More importantly, the point reflections with respect to $2\pi n$ divide the $r$-axis into intervals $[2\pi n,2\pi (n+1)]$ which are all physically equivalent to each other since connected by genuine gauge transformations. These intervals are known as {\it Weyl chambers,} see Fig.~\ref{fig:chambers}.

\section{Symmetries}
We are now fully equipped to investigate physical symmetries in terms of gauge fields, and in particular to identify the gauge-field configurations that are invariant (modulo gauge transformations) under a given physical transformation.

\subsection{Center Symmetry}
Consider first the case of center transformations. We have seen in the previous chapter that one example of non-trivial center transformation is given by
\beq
U(\tau)=\exp\left\{i\theta\frac{\tau}{\beta}\frac{\sigma_3}{2}\right\},
\eeq
with $\smash{\theta=\pm 2\pi}$ this time. They act on configurations of the form (\ref{eq:form2}) as $r\to r\pm 2\pi$, that is translations by $\pm 2\pi$ (not to be mistaken with the winding transformations that correspond to translations by $\pm 4\pi$).

\begin{figure}[t]
\begin{center}
\includegraphics[height=0.07\textheight]{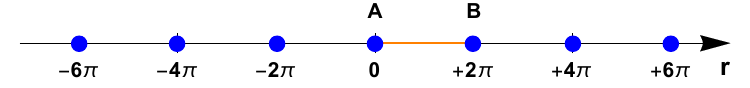}\\

\vglue5mm

\includegraphics[height=0.07\textheight]{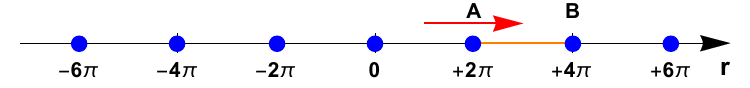}\\
\vglue6mm

\includegraphics[height=0.07\textheight]{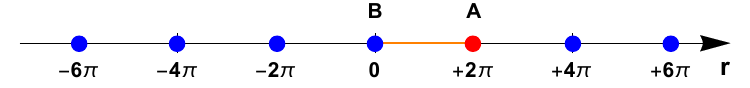}
\caption{Center transformation as a mapping of a given Weyl chamber ``AB'' into itself.}\label{fig:center}
\end{center}
\end{figure}

Let us take for instance $\smash{\theta=2\pi}$. The corresponding transformation is a translation by $2\pi$, red arrow in Fig.~\ref{fig:center}. In particular, it transforms the Weyl chamber $[0,2\pi]$ into $[2\pi,4\pi]$. But we can now use gauge transformations to fold this latter Weyl chamber into the original one. Since point reflections with respect to the edges of the Weyl chambers correspond to gauge transformations, this is easily done. Here, we just need to use the point reflection with respect to $2\pi$, red dot in Fig.~\ref{fig:center}. All together, this corresponds to the transformation
\beq
r\to r+2\pi\to 4\pi-(r+2\pi)=2\pi-r\,,
\eeq
that is a point reflection with respect to $\pi$.

\subsection{Confining Configurations}
 We have thus shown that, by using an appropriate gauge transformation, the center transformation appears as a transformation $\smash{r\to 2\pi-r}$ of the Weyl chamber $[0,2\pi]$ into itself, see Fig.~\ref{fig:center}. This transformation possesses a fixed point $r=\pi$, which is thus a confining configuration $\smash{A_\mu=T\delta_{\mu_0}\,\pi\sigma_3/2}$. 

\begin{figure}[h]
\begin{center}
\includegraphics[height=0.045\textheight]{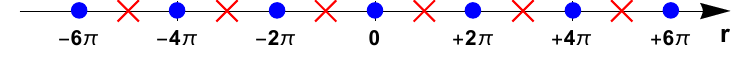}
\caption{Confining configurations (red crosses) within each Weyl chamber.}\label{fig:confining}
\end{center}
\end{figure}

Similarly, one shows (do it!) that any $\smash{r=\pi+2\pi n}$ corresponds to a center-invariant configuration, red crosses in Fig.~\ref{fig:confining}.\\

\noindent{$\diamond\diamond\diamond\,\,${\bf Problem 2.4:} Verify that $\Phi[A]$ vanishes on these configurations. \underline{Tip:} recall the formula derived in Problem 2.2.}$\,\,\diamond\diamond\diamond$

\subsection{Charge Conjugation}

We can proceed similarly with charge conjugation. Charge conjugation is defined as $\smash{A_\mu^C=-A_\mu^{\rm t}}$. For configurations of the form (\ref{eq:form2}) it corresponds to $\smash{r\to -r}$ because $\smash{\sigma_3^{\rm t}=\sigma_3}$. However, this is nothing but the Weyl transformation which is part of ${\cal G}_0$. We thus arrive at the conclusion that any configuration of the form (\ref{eq:form2}) is invariant under charge conjugation modulo a gauge transformation. 

This result is in fact more general because for any configuration, charge conjugation writes $(A_\mu^1,A_\mu^2,A_\mu^3)\to (-A_\mu^1,A_\mu^2,-A_\mu^3)$ since $\smash{\sigma_{1,3}^{\rm t}=\sigma_{1,3}}$ while $\smash{\sigma_2^{\rm t}=-\sigma_2}$. Thus, charge conjugation always corresponds to a color rotation, more precisely the rotation by an angle $\pi$ around direction $2$. This is a speficity of SU($2$) which will not apply anymore in the SU($3$) case.\\

\noindent{$\diamond\diamond\diamond\,\,${\bf Problem 2.5:} Evaluate $\Phi[A^C]$ and compare it to $\Phi[A]$ in the case of a configuration of the form (\ref{eq:form2}).\underline{Tip:} recall the formula derived in Problem 2.2.}$\,\,\diamond\diamond\diamond$

\chapter{Thermal Gluon Average}

{\bf \underline{Goal:}} Discuss under which circumstances the thermal gluon average is an order parameter for the confinement/deconfinement transition.\\

So far, we have restricted our discussion to the classical level, learning how to tell apart classical gluon configurations that are confining from those which are not. In the quantum world, however, we have access to thermal averages and so it is important to wonder which of them can help distinguishing between the two phases of the system. In the continuum, one has easy access to correlation functions, corresponding to thermal averages of products of gluon fields $\langle A_{\mu_1}\cdots A_{\mu_n}\rangle$. One question that we would like to address in this chapter is whether these correlation functions can be used as order parameters. For simplicity, we focus on the simplest of them, namely the one-gluon-field average $\langle A_\mu\rangle$.

One difficulty pops up immediately because, unlike the thermal average of a gauge-invariant functional
\beq
\langle{\cal O}[A]\rangle\equiv\frac{\int_{p.b.c.} {\cal D}A\,e^{-S[A]}\,{\cal O}[A]}{\int_{p.b.c.} {\cal D}A\,e^{-S[A]}}\,,
\eeq
which is defined directly in terms of the gauge-invariant (Euclidean) Yang-Mills action $S[A]$, the thermal average of $A_\mu$ makes sense only if one first fixes the gauge. In what follows, we recall the usual Faddeev-Popov gauge-fixing procedure that is used to give a meaning to the correlation functions and discuss the new problems that arise.

\section{Faddeev-Popov (FP) Procedure}
In order to define the thermal average of a product of gluon fields, one needs to restrict the functional integral to gluon configurations that obey a certain condition, known as a {\it gauge-fixing condition.} In what follows, we consider the {\it Landau gauge} defined by the condition
\beq
\partial_\mu A_\mu=0\,.
\eeq
To restrict to gauge field configurations that fulfil this condition, one inserts a {\it unity} under the functional integral in the form
\beq
1=\int_{{\cal G}_0}{\cal D}U_0\,\,J[A^{U_0}]\,\delta(\partial_\mu A^{U_0}_\mu)\,,\label{eq:unity}
\eeq
where $\int_{{\cal G}_0} {\cal D}U_0$ designates a formal integration over the group of gauge transformations and $\delta(\partial_\mu A^{U_0}_\mu)$ is a functional Dirac $\delta$ which selects a configuration obeying the Landau gauge condition along the gauge orbit of $A_\mu$. Finally, 
\beq
J[A^{U_0}]={\rm det}\,\left(\frac{\delta(\partial_\mu A^{U_0}_\mu)}{\delta U_0}\right),
\eeq
is the Jacobian that accounts for the change from the variable $U_0$ to the variable $\partial_\mu A_\mu$. The identity (\ref{eq:unity}) is a functional generalization of the well known identity
\beq
1=\int_{-\infty}^\infty dx\,f'(x)\delta(f(x))\,.
\eeq
We are here skipping some important subtleties (as most presentations of the Faddeev-Popov procedure do). We shall come back to them below.

\subsection{Inserting Unity}
Plugging Eq.~(\ref{eq:unity}) under the functional integral, the thermal average (\ref{eq:O}) of a gauge-invariant functional becomes
\beq
\langle{\cal O}\rangle=\frac{\int_{{\cal G}_0}{\cal D}U_0 \int_{p.b.c.} {\cal D}A\,e^{-S[A]}\,{\cal O}[A]\,J[A^{U_0}]\,\delta(\partial_\mu A^{U_0}_\mu)}{\int_{{\cal G}_0}{\cal D}U_0\int_{p.b.c.} {\cal D}A\,e^{-S[A]}J[A^{U_0}]\,\delta(\partial_\mu A^{U_0}_\mu)}\,.
\eeq
Now, because ${\cal D}A$, $S[A]$ and ${\cal O}[A]$ are gauge invariant, one can rewrite this as
\beq
\langle{\cal O}\rangle=\frac{\int_{{\cal G}_0}{\cal D}U_0 \int_{p.b.c.} {\cal D}A^{U_0}\,e^{-S[A^{U_0}]}\,{\cal O}[A^{U_0}]\,J[A^{U_0}]\,\delta(\partial_\mu A^{U_0}_\mu)}{\int_{{\cal G}_0}{\cal D}U_0\int_{p.b.c.} {\cal D}A^{U_0}\,e^{-S[A^{U_0}]}J[A^{U_0}]\,\delta(\partial_\mu A^{U_0}_\mu)}\,,
\eeq
and, because everything now depends only on $A^{U_0}$, one can make this transformed field the new integration variable via a change of variables
\beq
\langle{\cal O}\rangle=\frac{\int_{{\cal G}_0}{\cal D}U_0 \int_{p.b.c.} {\cal D}A\,e^{-S[A]}\,{\cal O}[A]\,J[A]\,\delta(\partial_\mu A_\mu)}{\int_{{\cal G}_0}{\cal D}U_0\int_{p.b.c.} {\cal D}A\,e^{-S[A]}J[A]\,\delta(\partial_\mu A_\mu)}\,.
\eeq
Doing so, we note that the volume of the gauge group has now factorized and cancels between the numerator and the denominator leading finally to an alternative (equivalent) definition of the thermal average of a gauge-invariant functional:
\beq
\langle{\cal O}\rangle\equiv\frac{\int_{p.b.c.} {\cal D}A\,e^{-S[A]}\,{\cal O}[A]\,J[A]\,\delta(\partial_\mu A_\mu)}{\int_{p.b.c.} {\cal D}A\,e^{-S[A]}J[A]\,\delta(\partial_\mu A_\mu)}\,.
\eeq
Since the integrands are not anymore gauge-invariant (thanks to the gauge-fixing factors $J[A]$ and $\delta(\partial_\mu A_\mu)$), we can now extend this definition of the thermal average to functionals that are not gauge-invariant. In particular, we define the gluon thermal average as
\beq
\langle A_\mu \rangle\equiv\frac{\int_{p.b.c.} {\cal D}A\,e^{-S[A]}\,A_\mu\,J[A]\,\delta(\partial_\mu A_\mu)}{\int_{p.b.c.} {\cal D}A\,e^{-S[A]}J[A]\,\delta(\partial_\mu A_\mu)}\,.
\eeq

\subsection{Auxiliary Fields}
In practice, it is not simple to perform calculations in this form because $J[A]\delta(\partial_\mu A_\mu)$ is highly non-local. It is convenient to introduce an equivalent local formulation to the price of introducing some auxiliary fields. In particular, the {\it Nakanishi-Lautrup field} $h$ allows one to write
\beq
\delta(\partial_\mu A_\mu)\propto\int {\cal D}h\,e^{i\int_x h^a(x)\partial_\mu A_\mu^a(x)}\,,
\eeq
which generalizes the well known formula of Fourier analysis. As for the factor $J[A]$, we first consider the following result:\\

\noindent{$\diamond\diamond\diamond\,\,${\bf Problem 3.1:} Show that, for an infinitesimal gauge transformation $U_0=\mathds{1}+i\theta^at^a$, one has}
\beq
A^{U_0}_\mu=A_\mu+D_\mu[A](\theta^at^a)\,,
\eeq 
with $D_\mu[A]\equiv\partial_\mu-i[A_\mu,\,\,]$ the {\it covariant derivative} in the adjoint representation. Deduce that $J[A]={\rm det}\,\partial_\mu D_\mu[A]$ which is known as {\it Faddeev-Popov determinant}. $\,\,\diamond\diamond\diamond$\\

With this result in mind, we can now introduce {\it ghost and antighost Grassmanian fields} $c$ and $\bar c$ to rewrite the Faddeev-Popov determinant as a gaussian Grassmanian integral:
\beq
J[A]={\rm det}\,\partial_\mu D_\mu[A]=\int{\cal D}c{\cal D}\bar c\,e^{\int_x\bar c^a(x)\partial_\mu D_\mu^{ab}[A]c^b(x)}\,,
\eeq
where $D_\mu^{ab}[A]$ denotes the matrix representation of the covariant derivative $\smash{D_\mu[A]=\partial_\mu-i[A_\mu,\,\,]}$ in the basis $t^a$, namely $\smash{D_\mu^{ab}[A]=\partial_\mu\delta^{ab}+f^{acb}A_\mu^c}$. All together, one arrives at the following FP prescription to evaluate the thermal gluon average in the Landau gauge
\beq
\langle A_\mu\rangle_{\rm Landau}\equiv\frac{\int_{p.b.c.} {\cal D}[A,c,\bar c,h]\,e^{-S_{\rm Landau}[A,c,\bar c,h]}\,A_\mu}{\int_{p.b.c.} {\cal D}[A,c,\bar c,h]\,e^{-S_{\rm Landau}[A,c,\bar c,h]}}\,,
\eeq
with
\beq
S_{\rm Landau}[A,c,\bar c,h]=S[A] & \!\!\!+\!\!\! & \int_x \bar c^a(x)\partial_\mu D_\mu^{ab}[A]c^b(x)\nonumber\\
& \!\!\!+\!\!\! & \int_x ih^a(x)\partial_\mu A_\mu^a(x)\,,
\eeq
known as the {\it Faddeev-Popov (FP) action} associated to the Landau gauge.

This {\it gauge-fixed formulation} is not void of difficulties however:
\begin{itemize}
\item[$-$] First, it is not always compatible with the symmetries. In particular, it does not necessarily makes explicit the center-symmetry of the problem, which is annoying if one wants to study the dynamical breaking of that symmetry. We discuss this problem and a possible fix in the next two sections. 

\item[$-$] Second, it is well known since the work of Gribov that the FP action is valid strictly speaking in the ultraviolet and any application at lower energies requires in principle a -- to date unknown -- extension of the Faddeev-Popov prescription. We discuss a possible phenomenological take on this question in the last section of this chapter.
\end{itemize}


\section{Background Landau Gauges}
To understand the problem with center symmetry, it is convenient to work within a more general class of gauges than the Landau gauge, namely {\it the family of background Landau gauges}
\beq
D_\mu[\bar A](A_\mu-\bar A_\mu)=0\,.
\eeq
This class of gauges is defined in terms of the covariant derivative $D_\mu[\bar A]\equiv\partial_\mu -i[\bar A_\mu,\,\,]$ in the presence of a {\it background configuration.} Each background configuration represents a different gauge and in what follows, as a convenient short-cut, we shall identify a given choice of gauge with the choice of background. In particular, the gauge $\smash{\bar A=0}$ is nothing but the Landau gauge treated in the previous section. 

By repeating the same argumentation as in the previous section, one arrives at the following Faddeev-Popov gauge-fixed action
\beq
S_{\bar A}[A,c,\bar c,h]=S[A] & \!\!\!+\!\!\! & \int_x \bar c^a(x)D_\mu^{ab}[\bar A] D_\mu^{bc}[A]c^c(x)\nonumber\\
& \!\!\!+\!\!\! & \int_x ih^a(x)D_\mu^{ab}[\bar A] (A_\mu^b(x)-\bar A_\mu^b(x))\,.\label{eq:SAbar}
\eeq


\noindent{$\diamond\diamond\diamond\,\,${\bf Problem 3.2:} Use the Faddeev-Popov procedure presented above to derive this action.} $\,\,\diamond\diamond\diamond$\\

\subsection{Background Symmetry}
Recall that the covariant derivative is taylor-made such that it transforms covariantly under gauge transformations. More precisely:\\

\noindent{$\diamond\diamond\diamond\,\,${\bf Problem 3.3:} Given a field $X$ leaving on the algebra, show that}
\beq
D_\mu[A^U](UXU^\dagger)=U(D_\mu[A]X)U^\dagger\,.
\eeq
This is true of course for any $A$, so it works for $D_\mu[\bar A]$ as well. $\,\,\diamond\diamond\diamond$\\

From this result, it is easily shown that
\beq
S_{\bar A^U}[A^U,c^U,\bar c^U,h^U]=S_{\bar A}[A,c,\bar c,h]\,,\label{eq:bg_sym}
\eeq
with $\smash{c^U\equiv UcU^\dagger}$, $\smash{\bar c^U\equiv U\bar cU^\dagger}$ and $\smash{h^U\equiv UhU^\dagger}$. Let us check it. To do so, it is convenient to rewrite the action (\ref{eq:SAbar}) in a way that does not depend on the particular basis of generators $t^a$. Using that ${\rm tr}\,t^at^b=\delta^{ab}/2$, we then find
\beq
S_{\bar A}[A,c,\bar c,h]=S[A] & \!\!\!+\!\!\! & 2\int_x {\rm tr}\,\bar c(x) D_\mu[\bar A] D_\mu[A]c(x)\nonumber\\
& \!\!\!+\!\!\! & 2\int_x {\rm tr}\,ih(x)D_\mu[\bar A] (A_\mu(x)-\bar A_\mu(x))\,.
\eeq
Then, using the result of Problem 3.3 and the cyclicality of the trace, the identity (\ref{eq:bg_sym}) is easily checked.

This identity is usually referred to as {\it background symmetry.} It is important to stress, however, that, despite its name, this is not really a symmetry for it does not leave invariant the FP action associated to a given gauge $\bar A$. Rather, it connects the FP actions associated to two different gauges, $\bar A$ and $\bar A^U$.

\subsection{Thermal Gluon Average}
The background symmetry (\ref{eq:bg_sym}) leaves its imprint on the thermal gluon average:
\beq
\langle A_\mu\rangle_{\bar A}=\frac{\int_{p.b.c.} {\cal D}[A,c,\bar c,h]\,e^{-S_{\bar A}[A,c,\bar c,h]}\,A_\mu}{\int_{p.b.c.} {\cal D}[A,c,\bar c,h]\,e^{-S_{\bar A}[A,c,\bar c,h]}}\,.
\eeq
Indeed, upon using Eq.~(\ref{eq:bg_sym}), we can write
\beq
\langle A_\mu\rangle_{\bar A}=\frac{\int_{p.b.c.} {\cal D}[A,c,\bar c,h]\,e^{-S_{\bar A^U}[A^U,c^U,\bar c^U,h^U]}\,A_\mu}{\int_{p.b.c.} {\cal D}[A,c,\bar c,h]\,e^{-S_{\bar A^U}[A^U,c^U,\bar c^U,h^U]}}\,.
\eeq
But now, upon using the invariance of the measure under the change of variables $(A_\mu,c,\bar c,h)\to (A_\mu^{U^\dagger},c^{U^\dagger},\bar c^{U^\dagger},h^{U^\dagger})$, we arrive at
\beq
\langle A_\mu\rangle_{\bar A}=\frac{\int_{p.b.c.} {\cal D}[A,c,\bar c,h]\,e^{-S_{\bar A^U}[A,c,\bar c,h]}\,A^{U^\dagger}_\mu}{\int_{p.b.c.} {\cal D}[A,c,\bar c,h]\,e^{-S_{\bar A^U}[A,c,\bar c,h]}}=\langle A^{U^\dagger}_\mu\rangle_{\bar A^U}\,.
\eeq
This rewrites more conveniently as
\beq
\langle A_\mu\rangle^U_{\bar A}=\langle A_\mu\rangle_{\bar A^U}\,,\label{eq:bg}
\eeq
which shows that, as one applies a transformation $U$ to the thermal gluon average in a given gauge $\bar A$, one obtains again the thermal gluon average, but within another gauge $\bar A^U$. 

This change of gauge as one applies the transformation is problematic because it implies that the thermal gluon average cannot be used in general as an order parameter for the symmetry. This is because, the symmetry does not constrain the thermal average to acquire any particular, confining, configuration. This is very different from what happens with gauge-invariant quantities such as the Polyakov for which the symmetry imposes a constraint
\beq
\ell=e^{i2\pi k/N}\ell
\eeq
which implies that $\ell=0$ if the symmetry is not broken.

\section{Center-Symmetric Landau Gauges}
Interestingly enough, there exist certain choices of background gauges for which the thermal gluon average is constrained by the symmetry and thus becomes an order parameter allowing one to distinguish between confined and deconfined phases.

\subsection{Definition}
Suppose indeed that we choose a gauge corresponding to a confining background configuration $\bar A_c$. Remember that such a configuration obeys
\beq
\forall U\in {\cal G}\,, \quad \exists U_0\in {\cal G}_0\,, \quad \bar A^{U_0U}=\bar A_c\,.
\eeq
Thus, in that gauge and using Eq.~(\ref{eq:bg}), we find
\beq
\langle A_\mu\rangle^{U_0U}_{\bar A_c}=\langle A_\mu\rangle_{\bar A_c^{U_0U}}=\langle A_\mu\rangle_{\bar A_c}\,.\label{eq:constraint}
\eeq
which is a symmetry constraint for the thermal gluon average $\langle A_\mu\rangle_{\bar A_c}$ in the considered gauge $\smash{\bar A=\bar A_c}$. In this case, $\langle A_\mu\rangle_{\bar A_c}$ becomes an order parameter for center-symmetry, equivalent to the Polyakov, but more easily tractable in the continuum. We refer to background Landau gauges corresponding to $\smash{\bar A=\bar A_c}$ as {\it center-symmetric Landau gauges} \cite{vanEgmond:2021jyx}.

\subsection{Constant, Temporal and Abelian Backgrounds}
If, as we did in the classical analysis, we restrict to backgrounds of the form
\beq
\bar A_\mu(x)=T\delta_{\mu0}\,\bar r \frac{\sigma_3}{2}\,,\label{eq:form22}
\eeq
the center-symmetric gauges correspond to the values $\smash{\bar r=\pi+2\pi n\equiv \bar r_c}$ located at the center of the Weyl chambers.

Moreover, in any gauge of the form (\ref{eq:form22}), we have the following result:\\

\noindent{$\diamond\diamond\diamond\,\,${\bf Problem 3.4:} Show that}
\beq
\langle A_\mu(x)\rangle_{\bar A}=T\delta_{\mu0}\, r \frac{\sigma_3}{2}\,.\label{eq:form222}
\eeq
\underline{Tip:} Apply the general identity (\ref{eq:bg}) to constant color rotations of the form $\smash{U=e^{i\theta\sigma_3/2}}$. $\,\,\diamond\diamond\diamond$\\

With this result in mind, it is easily seen that the constraint (\ref{eq:constraint}) in the gauge $\smash{\bar r=\bar r_c}$ now becomes $\smash{r=\bar r_c}$. So, in that gauge, the thermal average $r$ becomes an order parameter, equal to $\bar r_c$ in the confining phase, and departing from $\bar r_c$ in the deconfined phase. In other words, it plays the role of an order parameter for the confinement/deconfinement transition.

\section{Infrared Gauge Fixing}

Let us reconsider the Faddeev-Popov gauge-fixing procedure in the Landau gauge.

\subsection{Gribov Copy Problem}
 Recall that the procedure is based on inserting a unity in the form (\ref{eq:unity}) which is a particular case of the well known formula
\beq
1=\int_{-\infty}^\infty dx\,f'(x)\delta(f(x))\,.
\eeq
What we did not mention above is that this identity is valid only under specific conditions. Namely, the equation $\smash{f(x)=0}$ should have only one solution $\smash{x=x_0}$, and the latter should be such that $\smash{f'(x_0)>0}$. In the presence of multiple solutions $x_i$ with no definite sign of $f'(x_i)$, we have instead
\beq
\int_{-\infty}^\infty dx\,f'(x)\delta(f(x))=\sum_i \frac{f'(x_i)}{|f(x_i)|}=\sum_i {\rm sign}\,\big(f(x_i)\big)\,,
\eeq
which can then not be used as a unity.

The same is true for the Faddeev-Popov procedure. The unity that we used earlier is only justified if the equation $\smash{\partial_\mu A_\mu^{U_0}=0}$ has a unique solution $U_0(A)$ for each $A_\mu$. We know since the work of Gribov that this assumption is incorrect and that there are actually infinitely many solutions $U_{0,i}(A)$ known as {\it Gribov copies.} This invalidates a priori the whole Faddeev-Popov procedure. The usual consensus is that:
\begin{itemize}
\item[$-$] the Faddeev-Popov procedure is justified at high energies since one explores a small patch of gluon configurations under the functional integral, and, therefore, one should not be sensitive to Gribov copies;

\item[$-$] at low energies, there is no reason to think that the FP procedure is correct. In fact there is good evidence from the lattice that it is not.
\end{itemize}

\subsection{Lattice Results}

It was shown by Kugo and Ojima that the FP action predits a specific behavior of the ghost and gluon two-point correlation functions in the Landau gauge, known as {\it scaling behavior,} characterized by an enhancement of the ghost propagator at low momenta with respect to its tree-level value and a vanishing of the gluon propagator in the same limit.

\begin{figure}[t]
  \centering
  \includegraphics[width=0.45\linewidth]{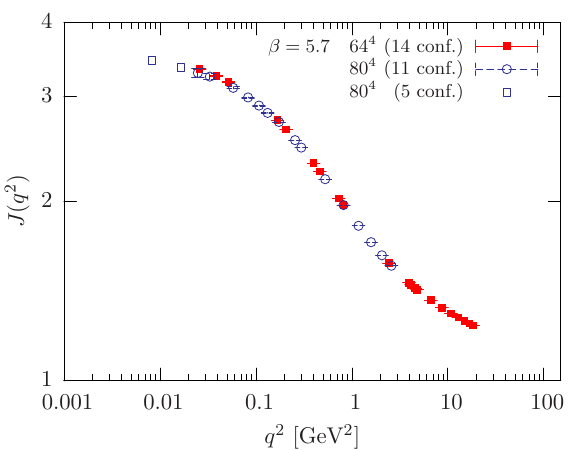} \quad\includegraphics[width=0.45\linewidth]{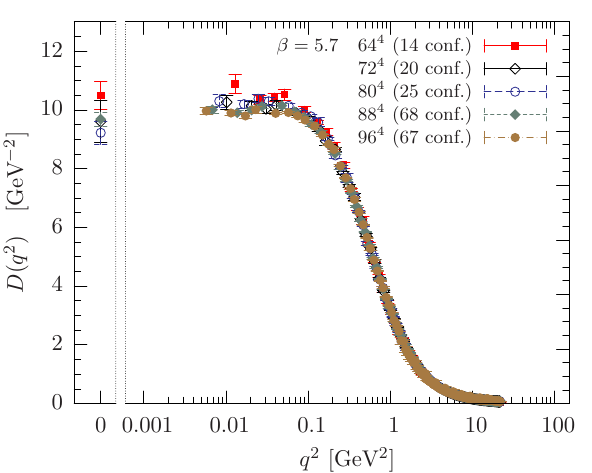}
 \caption{Left: Landau gauge ghost dressing function (propagator divided by its free counterpart) as computed on the lattice. Right: Landau gauge gluon propagator \cite{Bogolubsky:2009dc}.}\label{fig:props}
\end{figure}

The lattice finds instead that the ghost propagator behaves essentially as the tree-level one in the infrared, while the gluon saturates to a finite non-zero value, what is nowadays known as the {\it decoupling behavior,} see Fig.~\ref{fig:props}. This clearly shows that the FP action cannot be a good starting point to tackle low energy properties and that it needs to be replaced by something else.

Of course, finding what this something else is is a complicated matter because it relates to the Gribov copy problem which remains an open problem to date. One can try to tackle the Gribov copies partially. This leads to the Gribov-Zwanziger procedure and the Gribov-Zwanziger action \cite{Vandersickel:2012tz}, but this is not the topic of the present lectures. Another possibility is to follow a more phenomenological approach: since the FP action is incomplete in the infrared, one can try to propose extensions by adding new operators and trying to constrain their couplings using lattice data/experiments.

\subsection{Curci-Ferrari Model}
The Curci-Ferrari model is one example of such phenomenological approach. It is based on the observation made on the lattice that the ghost propagator remains pretty similar to its tree-level counterpart while the gluon propagator develops a screening mass at low momenta. The model then amounts to adding to the FP action, a mass terms for the gluon field:
\beq
S_{\rm Landau}[A,c,\bar c,h] & \!\!\!=\!\!\! & \int_x\,\frac{1}{4g^2}F_{\mu\nu}^a(x)F_{\mu\nu}^a(x)\nonumber\\
& \!\!\!+\!\!\! & \int_x \frac{1}{2}m^2(A_\mu^a(x))^2\nonumber\\
& \!\!\!+\!\!\! & \int_x \bar c^a(x)\partial_\mu D_\mu^{ab}[A]c^b(x)\nonumber\\
& \!\!\!+\!\!\! & \int_x ih^a(x)\partial_\mu A_\mu^a(x)\,.
\eeq
It is important to realize that this is not the same as ``massive Yang-Mills'' a.k.a. {\it Proca theory} which would correspond to the first two lines of the previous action. Proca theory is not only an intentional modification of a fundamental theory but it also leads to a non-renormalizable theory. 

On the other hand, the Curci-Ferrari model should be seen as a phenomenological account of the, yet to be understood, extension of the Landau gauge-fixed action beyond the Faddeev-Popov approximation. Moreover, the Curci-Ferrari model is renormalizable, meaning that, even though it is a model, there is only one additional parameter, the Curci-Ferrari mass, to be fixed in addition to the fundamental coupling parameter. This mass can be fixed for instance by fitting the ghost and gluon propagators, as computed within the model, to available lattice data for the Landau-gauge YM propagators. In what follows, we shall employ the parameter values obtained by fitting the propagators computed in a scheme where the renormalized CF mass $m$ is defined from the zero-momentum gluon propagator $G(Q\to 0)\equiv 1/m^2$. In the SU($2$) case, these values are
\beq
m\simeq 680\,{\rm MeV} \quad {\rm and} \quad g\simeq 7.5\,,
\eeq
while in the SU($3$) case, we should take instead
\beq
m\simeq 510\,{\rm MeV} \quad {\rm and} \quad g\simeq 4.9\,.
\eeq
For a review of applications of the CF model as a model for Landau gauge-fixing in the infrared, see Ref.~\cite{Pelaez:2021tpq}.

As before, if we want to properly account for center-symmetry, we need to generalize the model to background Landau gauges. The natural generalization is
\beq
S_{\bar A}[A,c,\bar c,h] & \!\!\!=\!\!\! & \int_x\,\frac{1}{4g^2}F_{\mu\nu}^a(x)F_{\mu\nu}^a(x)\nonumber\\
& \!\!\!+\!\!\! & \int_x \frac{1}{2}m^2(A_\mu^a(x)-\bar A_\mu^a(x))^2\nonumber\\
& \!\!\!+\!\!\! & \int_x ih^a(x)D_\mu^{ab}[\bar A] (A_\mu^b(x)-\bar A_\mu^b(x))\nonumber\\
& \!\!\!+\!\!\! & \int_x \bar c^a(x)D_\mu^{ab}[\bar A] D_\mu^{bc}[A]c^c(x)\,.\label{eq:action}
\eeq
This choice of mass terms makes sure that the background symmetry (\ref{eq:bg_sym}) is still valid, as it can be easily checked. In particular, if we choose the background to be confining $\smash{\bar A=\bar A_c}$, the starting action $S_{\bar A_c}[A,c,\bar c,h]$ is center-invariant, a minimum pre-requisite for the study of the spontaneous breaking of center symmetry and, thus, of the confinement/deconfinement transition.

\chapter{Effective Action}

{\bf \underline{Goal:}} Learn how to evaluate the thermal gluon average within the Curci-Ferrari model.\\

 A convenient way to do so is through the effective action which we first recall in the simpler set-up of a scalar theory before evaluating it within the CF model.

\section{Scalar Theory}

Consider the (Euclidean) scalar theory 
\beq
S[\varphi]=\int_x \left\{\frac{1}{2}\partial_\mu\varphi(x)\partial_\mu\varphi(x)+\frac{1}{2}m^2\varphi^2(x)+\frac{\lambda}{4!}\varphi^4(x)\right\}.
\eeq
This model has an obvious $Z_2$ symmetry $\smash{\varphi\to-\varphi}$, so if we were to compute the thermal field average naively
\beq
\langle\varphi(x)\rangle=\frac{1}{Z}\int_{p.b.c.} {\cal D}\varphi\,e^{-S[\varphi]}\,\varphi(x)\,,
\eeq
we would necessarily obtain $0$ since $\smash{\langle\varphi\rangle=-\langle\varphi\rangle}$ by symmetry. The subtlety, however, is that the system can never be thought in complete isolation. There is always a tiny perturbation that breaks the symmetry explicitly, say a tiny external source $J$ coupled linearly to the field. In the presence of this perturbation, the thermal field average writes instead
\beq
\langle\varphi(x)\rangle_J=\frac{1}{Z}\int_{p.b.c.} {\cal D}\varphi\,e^{-S[\varphi]+\int_x J(x)\varphi(x)}\,\varphi(x)\,,
\eeq
and the symmetry imposes now $\smash{\langle\varphi\rangle_J=-\langle\varphi\rangle_{-J}}$. Accordingly, we can distinguish two phases corresponding to two different realizations of the symmetry:
\begin{itemize}
\item[$-$] Wigner-Weyl realization: in this case, the zero-source limit does not depend on the way the limit is taken. The symmetry identity in the zero-source limit gives $\smash{\langle\varphi\rangle_{J\to 0}=-\langle\varphi\rangle_{J\to 0}}$ and thus $\smash{\langle\varphi\rangle_{J\to 0}=0}$;

\item[$-$] Nambu-Goldsone realization: in this case, the zero-source limit depends on the way the limit is taken. The best we can achieve is $\langle\varphi\rangle_{J\to 0^+}=-\langle\varphi\rangle_{J\to 0^-}$ but neither of these limits needs to be $0$.
\end{itemize}

\subsection{Effective Action}
A convenient way to distinguish between the two realizations of the symmetry is to introduce the functional
\beq
W[J]\equiv\ln \int_{p.b.c.} {\cal D}\varphi\,e^{-S[\varphi]+\int_x J(x)\varphi(x)}\,,
\eeq
know as the {\it generating functional for connected correlators.} Indeed, its functional derivatives with respect to the source $J$ give access to connected thermal averages of products of fields. For instance, the thermal field average is given by the first such derivative
\beq
\frac{\delta W[J]}{\delta J(x)}=\langle\varphi(x)\rangle_J\equiv \frac{\int_{p.b.c.} {\cal D}\varphi\,e^{-S[\varphi]+\int_x J(x)\varphi(x)}\,\varphi(x)}{\int_{p.b.c.} {\cal D}\varphi\,e^{-S[\varphi]+\int_x J(x)\varphi(x)}}\,,\label{eq:1pt}
\eeq
while the second derivative gives the two-point correlator 
\beq
\frac{\delta^2 W[J]}{\delta J(x)\delta J(y)}=\langle\varphi(x)\varphi(y)\rangle_J-\langle\varphi(x)\rangle_J\langle\varphi(y)\rangle_J\,,
\eeq
also known as {\it propagator.}

\begin{figure}[t]
  \centering
  \includegraphics[width=0.45\linewidth]{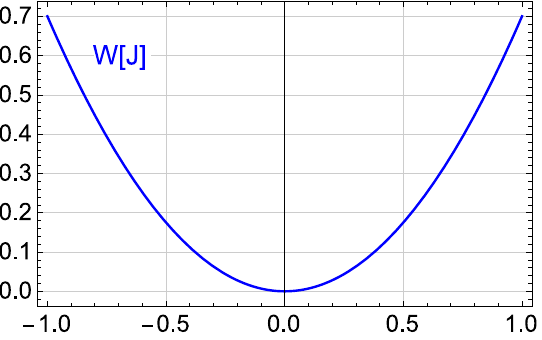} \quad\includegraphics[width=0.45\linewidth]{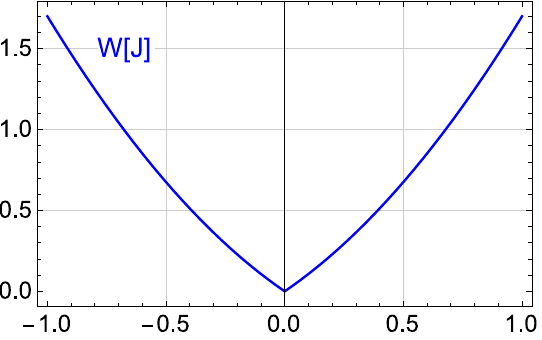}
 \caption{Wigner-Weyl and Nambu-Goldstone realization of $Z_2$ symmetry in terms of $W[J]$.}\label{fig:cusp}
\end{figure}

\begin{figure}[t]
  \centering
  \includegraphics[width=0.45\linewidth]{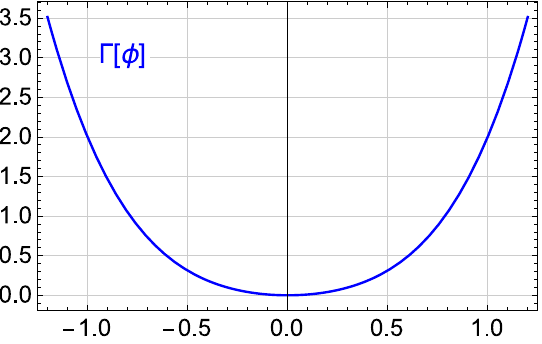} \quad\includegraphics[width=0.45\linewidth]{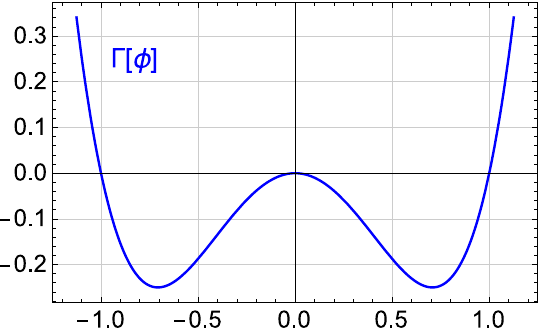}
 \caption{Wigner-Weyl and Nambu-Goldstone realization of $Z_2$ symmetry in terms of $\Gamma[\phi]$.}\label{fig:cusp_2}
\end{figure}

According to Eq.~(\ref{eq:1pt}), the broken phase appears as the phase where $W[J]$ becomes irregular, with its first functional derivative at $\smash{J=0}$ depending on the way $\smash{J=0}$ is approached. This is represented in Fig.~\ref{fig:cusp}. An even more convenient interpretation is obtained in terms of the {\it effective action} $\Gamma[\phi]$ corresponding to the Legendre Transform of $W[J]$. To define it, one first considers Eq.~(\ref{eq:1pt}) as a functional $\phi[J]$ associating a certain thermal field average $\phi$ to a given source $J$. One then considers the functional inverse $J[\phi]$ which, given a thermal field average $\phi$ allows one to retrieve back the source $J$ that generated this thermal average. The effective action is then defined as
\beq
\Gamma[\phi]\equiv -W[J[\phi]]+\int_x J[\phi](x)\phi(x)\,,
\eeq
and it is a well known result that (check it!)
\beq
\frac{\delta\Gamma}{\delta\phi(x)}=J(x)\,.
\eeq
From this latter equation, it follows that the zero-source limit corresponds to the extremization (in the present case minimization) of the effective action $\Gamma[\phi]$. Another nice feature is that (for linearly realized symmetries), the effective action reflects the symmetries of the problem. In the present scalar model, we have $\smash{\Gamma[\phi]=\Gamma[-\phi]}$ which implies in particular that, if $\phi_{\rm min}$ is a minimum, so is $-\phi_{\rm min}$. Then, when the minimum is unique, it has to be $0$ by symmetry. This corresponds to the Wigner-Weyl realization of the symmetry. When the minimum is not unique, the various minima do not need to be zero but are connected to each other by the symmetry. This is the Nambu-Goldstone realization of the symmetry, see Fig.~\ref{fig:cusp_2}.

\subsection{One-loop Expression}
The recipe to obtain the effective action at one-loop order is quite simple. One writes the field $\varphi$ as a classical contribution $\phi$ plus a fluctuation $\delta\varphi$ and expands the action $S[\varphi]$ with respect to that fluctuation:
\beq
S[\phi+\delta\varphi] & \!\!=\!\! & S[\phi]+\int_x \delta\varphi^a(x)\left.\frac{\delta S}{\delta\varphi^a(x)}\right|_{\varphi=\phi}\nonumber\\
& \!\!+\!\! & \frac{1}{2}\int_x\int_y \delta\varphi^a(x)\delta\varphi^b(y)\left.\frac{\delta^2 S}{\delta\varphi^a(x)\delta\varphi^b(y)}\right|_{\varphi=\phi}+\cdots\label{eq:exp2}
\eeq
For later purpose, we have here extended our model to a multicomponent model where the scalar field $\varphi^a$ carries a ``color'' index:
\beq
S[\varphi]=\int_x \left\{\frac{1}{2}\partial_\mu\varphi_a(x)\partial_\mu\varphi_a(x)+\frac{1}{2}m^2\varphi_a(x)\varphi_a(x)+\frac{\lambda}{4!}(\varphi_a(x)\varphi_a(x))^2\right\}.
\eeq

The two important ingredients to be kept from the above expansion are the order $0$ term, corresponding to the action $S[\phi]$ associated to the classical configuration $\phi$, and the next-to-leading (or quadratic) term that gives access to the {\it free inverse propagator}
\beq
G_{0,ab}^{-1}[\phi](x,y)\equiv\left.\frac{\delta^2 S}{\delta\varphi^a(x)\delta\varphi^b(y)}\right|_{\varphi=\phi}
\eeq
in the presence of the classical configuration $\phi$. These two ingredients enter the one-loop expression of the effective action as
\beq
\Gamma[\phi]=S[\phi]+\frac{1}{2}\ln {\rm det}\,G_0^{-1}[\phi]\,,\label{eq:ea}
\eeq
where the determinant in this formula accounts for both the space-time and the color dependence.  For later purpose, we also note that, in the case of conjugated Grassmanian fields (such as quarks and antiquarks or ghosts and antighosts), the factor of $1/2$ should be replaced by $-1$.\\

\noindent{$\diamond\diamond\diamond\,\,${\bf Problem 4.1:} In the present scalar model, show that the free inverse propagator is}
\beq
G_{0,ab}^{-1}[\phi]=\left(-\partial^2_x+m^2+\frac{\lambda}{6}\phi^2(x)\right)\delta_{ab}\delta(x-y)+\frac{\lambda}{3}\phi_a(x)\phi_b(x)\delta(x-y)\,.\label{eq:G0m1}
\eeq
$\,\,\diamond\diamond\diamond$

\subsection{Effective Potential}
If the system is translationally invariant, it is possible to restrict to constant classical configurations $\smash{\phi(x)=\phi}$. For those configurations, the effective action scales like the volume $\beta L^3$ of the Euclidean space time $[0,\beta]\times\mathds{R}^3$. This means that we can write $\smash{\Gamma[\phi]=\beta L^3 V(\phi)}$, where $V(\phi)$ is known as the {\it effective potential.} 

The scaling with the volume is obvious for the first term in Eq.~(\ref{eq:ea}) since for constant configurations
\beq
S[\phi]=\int_x \left\{\frac{1}{2}m^2\phi^a\phi^a+\frac{\lambda}{4!}(\phi^a\phi^a)^2\right\}=\beta L^3 {\cal L}(\phi)\,,
\eeq
where ${\cal L}(\phi)$ is the Lagrangian density. As for the second term in Eq.~(\ref{eq:ea}), the scaling with the volume is less obvious but it can be worked out (this is actually a good exercice). One eventually arrives at the formula
\beq
V(\phi)={\cal L}(\phi)+\frac{1}{2}\int_Q \ln {\rm det}\,G_0^{-1}[\phi](Q)\,,\label{eq:pot_Q}
\eeq
where $G_{0,ab}^{-1}[\phi](Q)$ is the Fourier transform of $G_{0,ab}^{-1}[\phi](x,y)$ which actually becomes a function of $x-y$ when $\phi$ is constant.\\

\noindent{$\diamond\diamond\diamond\,\,${\bf Problem 4.2:} In the present scalar model, show that }
\beq
G_{0,ab}^{-1}[\phi]=\left(Q^2+m^2+\frac{\lambda}{6}\phi^2\right)\delta_{ab}+\frac{\lambda}{3}\phi_a\phi_b\,.
\eeq
$\,\,\diamond\diamond\diamond$\\

Note that a convenient way to read $G_{0,ab}^{-1}(Q)$ directly is to express the quadratic part in Eq.~(\ref{eq:exp2}) in terms of the Fourier components of the fields. Indeed, it is easily checked (do it) that
\beq
& & \frac{1}{2}\int_x\int_y \delta\varphi^a(x)\delta\varphi^b(y)G_{0,ab}^{-1}(x-y)\nonumber\\
& & \hspace{0.5cm}=\,\frac{1}{2}\int_{Q} \delta\varphi^a(-Q)G_{0,ab}^{-1}(Q)\delta\varphi^b(Q)\,.
\eeq
We shall exploit this below.

\section{Calculation within the CF Model}
Let us now apply the formalism to the model (\ref{eq:action}). As it is customary before considering the loop expansion, we rescale all fields by $g$ while redefining $m^2$ as $m^2/g^2$. The action reads then
\beq
S_{\bar A}[A,c,\bar c,h] & \!\!\!=\!\!\! & \int_x\,\frac{1}{4}F_{\mu\nu}^a(x)F_{\mu\nu}^a(x)\nonumber\\
& \!\!\!+\!\!\! & \int_x \frac{1}{2}m^2(A_\mu^a(x)-\bar A_\mu^a(x))^2\nonumber\\
& \!\!\!+\!\!\! & \int_x ih^a(x)D_\mu^{ab}[\bar A] (A_\mu^b(x)-\bar A_\mu^b(x))\nonumber\\
& \!\!\!+\!\!\! & \int_x \bar c^a(x)D_\mu^{ab}[\bar A] D_\mu^{bc}[A]c^c(x)\,,\label{eq:action2}
\eeq
with
\beq
F_{\mu\nu}^a(x)\equiv \partial_\mu A_\nu^a(x)-\partial_\nu A_\mu^a(x)+gf^{abc} A^b_\mu(x) A^c_\nu(x)\,,
\eeq
and
\beq
D_\mu^{ab}[A]=\partial_\mu\delta^{ab}+gf^{acb}A^c_\mu\,,
\eeq
and similarly for $D_\mu^{ab}[\bar A]$.

When expanding the action (\ref{eq:action2}) around a classical configuration of the fields, the fields $c$ and $\bar c$ can be treated as fluctuations since, being Grassmanian variables, they do not have any classical counterpart. Something similar occurs for the field $h$ but the reason here is that it appears linearly in the action (\ref{eq:action2}) and shifting it around a non-vanishing configuration would not affect the evaluation of the effective action. 

All in all, we only need to shift $A_\mu$. For simplicity we also denote by $A_\mu$ the classical configuration around which the field is varied while the fluctuation is denoted by $a_\mu$. The goal is then to extract the quadratic part of
\beq
S_{\bar A}[A+a,c,\bar c,h]
\eeq
with respect to the variables $a$, $c$, $\bar c$ and $h$. We shall from now one assume that both $\bar A$ and $A$ are constant, temporal and abelian:
\beq
\bar A_\mu(x)=\frac{T}{g}\delta_{\mu0}\,\bar r \frac{\sigma_3}{2} \quad {\rm and} \quad A_\mu(x)=\frac{T}{g}\delta_{\mu0}\,r \frac{\sigma_3}{2}\,.\label{eq:form2222}
\eeq
Recall that, eventually, $\bar r$ is to be chosen equal to a confining configuration and that $r$ should be obtained from minimizing the corresponding effective potential in that particular gauge.

\subsection{Quadratic Part}
In the ghost sector, the quadratic part is simply
\beq
\int_x \bar c^a(x)\bar D_\mu^{ab} D_\mu^{bc}c^c(x)\,,
\eeq
where, to slightly simplify the notation we have defined $\smash{D^{ab}_\mu\equiv D^{ab}_\mu[A]}$ and $\smash{\bar D^{ab}_\mu\equiv D^{ab}_\mu[\bar A]}$.\\

In order to identify the quadratic part in the gluon sector, we first consider the following result:\\

\noindent{$\diamond\diamond\diamond\,\,${\bf Problem 4.3:} Consider the field-strength tensor as a functional of $A$:
\beq
F^a_{\mu\nu}[A]=\partial_\mu A^a_\nu-\partial_\nu A^a_\mu+gf^{abc}A^b_\mu A^c_\nu\,.
\eeq
Show that
\beq
F^a_{\mu\nu}[A+a]=F^a_{\mu\nu}[A]+D^{ab}_\mu a^b_\nu-D^{ab}_\nu a^b_\mu+gf^{abc}a^b_\mu a^c_\nu\,,
\eeq
where $\smash{D_\mu^{ab}X^b\equiv\partial_\mu X^a+gf^{acb}A_\mu^c X^b}$. $\,\,\diamond\diamond\diamond$\\

In the present case (\ref{eq:form2222}), the first term $F_{\mu\nu}[A]$ in the decomposition of $F^a_{\mu\nu}[A+a]$ vanishes and the quadratic part originates in the squaring of the combined next two terms. The quadratic part in the gluon sector then reads
\beq
& &\int_x\frac{1}{2}\Big(D^{ab}_\mu a^b_\nu(x)  D^{ac}_\mu a^c_\nu(x)-D^{ab}_\mu a^b_\nu(x) D^{ac}_\nu a^c_\mu(x)+\frac{1}{2}m^2(a_\mu^a(x))^2\Big)\nonumber\\
& & +\,\int_x ih^a(x)\bar D_\mu^{ab} a_\mu^b(x)\,,
\eeq
where we stress that one should not forget the last term coupling $a_\mu$ to $h$ since it is also quadratic. To properly identify the quadratic form, we use the following result:\\

\noindent{$\diamond\diamond\diamond\,\,${\bf Problem 4.4:} Show that
\beq
\int_x (D_\mu^{ab}X^b(x))Y^a(x)=-\int_x X^a(x)(D_\mu^{ab}Y^b(x))\,.\nonumber
\eeq
\underline{Tip:} start by showing that 
\beq
\partial_\mu (X^a(x)Y^a(x))=(D_\mu^{ab}X^b(x))Y^a(x)+X^a(x)(D_\mu^{ab}Y^b(x))\,.\nonumber
\eeq 
$\,\,\diamond\diamond\diamond$\\

We then arrive at
\beq
& & \int_x\frac{1}{2}a^a_\mu(x)\Big(D^{ac}_\nu D^{cb}_\mu- \delta_{\mu\nu}(D^{ac}_\rho D^{cb}_\rho -m^2\delta^{ab})\Big)a^b_\nu(x)\nonumber\\
& & +\frac{1}{2}\int_x ih^a(x)\bar D_\nu^{ab} a_\nu^b(x)-\frac{1}{2}\int_x a_\mu^a(x) \bar D_\mu^{ab}(ih^b(x))\,.\label{eq:quad}
\eeq

\subsection{Color Structure}
Clearly the index structure in Eq.~(\ref{eq:quad}) is rather complicated since it is neither diagonal in the Lorentz indices nor in the color indices. We shall deal with the Lorentz structure below. Let us first deal with the color structure.

The problem originates in the presence of the covariant derivatives $\smash{D_\mu^{ab}=\partial_\mu\delta^{ab}+gf^{acb}A_\mu^c}$ and $\smash{\bar D_\mu^{ab}=\partial_\mu\delta^{ab}+gf^{acb}\bar A_\mu^c}$ which are not diagonal in color. However, this is nothing but the expression of the covariant derivatives $\smash{D_\mu=\partial_\mu-ig[A_\mu,\,\,]}$ and $\smash{\bar D_\mu=\partial_\mu-ig[\bar A_\mu,\,\,]}$ in a particular basis $t^a$. The question is now: can we find another basis where the covariant derivative becomes diagonal? At least in the SU($2$) case, we can answer positively. Indeed since $\smash{A_\mu\propto\sigma_3}$ and $\smash{\bar A_\mu\propto\sigma_3}$, we can use the basis $\{\sigma_3/2,\sigma_+/2,\sigma_-/2\}$ where 
\beq
\sigma_\pm\equiv\frac{\sigma_1\pm i\sigma_2}{\sqrt{2}}
\eeq 
are the ladder operators that appear in the classification of the representations of the SU($2$) algebra. 

The basis $\{\sigma_3/2,\sigma_+/2,\sigma_-/2\}$ is known as a {\it Cartan-Weyl basis} and it is convenient to denote it as $\{t^\kappa\}$ with $\smash{\kappa=0,+1,-1}$, $\smash{t^0=\sigma_3/2}$, $\smash{t^+=\sigma_+/2}$ and $\smash{t^-=\sigma_-/2}$. One great benefit of this basis is that it diagonalizes the action of the commutator $[\sigma_3/2,\,\,]$. Indeed, one has (check it!)
\beq
\left[\frac{\sigma_3}{2},t^\kappa\right]=\kappa t^\kappa\,.
\eeq
In particular, the action of $D_\mu$ in such basis writes:
\beq
D_\mu(X^\kappa t^\kappa) & = & \partial_\mu(X^\kappa t^\kappa)-ig[A_\mu,X^\kappa t^\kappa]\nonumber\\
& = & (\partial_\mu X^\kappa) t^\kappa-iT\delta_{\mu0}\,rX^\kappa \left[\frac{\sigma_3}{2}, t^\kappa\right]\nonumber\\
& = & (\partial_\mu X^\kappa) t^\kappa-iT\delta_{\mu0}\,r \kappa X^\kappa  t^\kappa\nonumber\\
& = & (\partial_\mu-iT\delta_{\mu0}\,r \kappa) X^\kappa  t^\kappa\equiv (D_\mu^\kappa X^\kappa) t^\kappa\,,
\eeq
and similarly of course for $\bar D_\mu(X^\kappa t^\kappa)$ with $r$ replaced by $\bar r$. All together, the quadratic part within a Cartan-Weyl color basis reads then
\beq
& & \int_x\frac{1}{2}a^{-\kappa}_\mu(x)\Big(D^\kappa_\nu D^\kappa_\mu- \delta_{\mu\nu}(D^\kappa_\rho D^\kappa_\rho -m^2)\Big)a^\kappa_\nu(x)\nonumber\\
& & +\frac{1}{2}\int_x ih^{-\kappa}(x)\bar D_\nu^\kappa a_\nu^\kappa(x)-\frac{1}{2}\int_x \bar  a_\mu^{-\kappa}(x) \bar D_\mu^\kappa (ih^\kappa(x))\,.
\eeq
The presence of the labels $-\kappa$ relates to the fact that ${\rm tr}\,t^\kappa t^\lambda\propto\delta^{\kappa(-\lambda)}$ rather than ${\rm tr}\,t^a t^b\propto\delta^{ab}$ for the usual cartesian bases. This will gain even more sense in the next section.

\subsection{Lorentz Structure}
Recall that, in order to evaluate the potential (\ref{eq:pot_Q}),  we need to identify the quadratic form in Fourier space. With the convention $\partial_\mu\to -iQ_\mu$, we find
\beq
& & \int_Q\frac{1}{2}a^{-\kappa}_\mu(-Q)\Big(\delta_{\mu\nu}(Q^\kappa_\rho Q^\kappa_\rho +m^2)-Q^\kappa_\nu Q^\kappa_\mu\Big)a^\kappa_\nu(Q)\nonumber\\
& & +\int_Q h^{-\kappa}(-Q)\bar Q_\nu^\kappa a_\nu^\kappa(Q)-\int_Q \bar  a_\mu^{-\kappa}(-Q) \bar Q_\mu^\kappa h^\kappa(Q)\,,
\eeq
where $\smash{Q^\kappa_\mu\equiv Q_\mu+T\delta_{\mu 0}\,r\kappa}$ and $\smash{\bar Q^\kappa_\mu\equiv Q_\mu+T\delta_{\mu 0}\,\bar r\kappa}$ are {\it generalized momenta} that combine the momentum $Q_\mu$ and the charge $\kappa$ into a single object. From this perspective, it seems natural that  $-\kappa$ appears in those Fourier components of momentum $-Q$. These generalized momenta involved shifted frequencies, $\smash{Q^\kappa=(\omega_n^\kappa,\vec{q})}$ and  $\smash{\bar Q^\kappa=(\bar \omega_n^\kappa,\vec{q})}$, with $\smash{\omega_n^\kappa\equiv\omega_n+r\kappa T}$ and $\smash{\bar\omega_n^\kappa\equiv\omega_n+\bar r\kappa T}$. In a certain sense, these frequency shifts play the role of (imaginary) chemical potentials for the color charge.

In matrix form, the quadratic form whose determinant we need to evaluate writes
\beq
\left(\begin{array}{cc}
\delta_{\mu\nu}(Q^\kappa_\rho Q^\kappa_\rho +m^2)-Q^\kappa_\mu Q^\kappa_\nu & \bar Q^\kappa_\mu\\
\bar Q^\kappa_\mu & 0
\end{array}\right).\label{eq:quadro}
\eeq
To evaluate its determinant, we use the following result:\\

\noindent{$\diamond\diamond\diamond\,\,${\bf Problem 4.5:} Consider a square matrix $M$ subdivided in four square blocks $A$, $B$, $C$ and $D$:
\beq
M=\left(\begin{array}{cc}
A & B\\
C & D
\end{array}\right).\nonumber
\eeq
Assuming that $A$ is invertible, verify that
\beq
M=\left(\begin{array}{cc}
A & 0\\
C & 1
\end{array}\right)\left(\begin{array}{cc}
1 & A^{-1}B\\
0 & D-CA^{-1}B
\end{array}\right).
\eeq
Deduce that ${\rm det}\,M={\rm det}\,A\times {\rm det}(D-CA^{-1}B)$. The combination $D-CA^{-1}B$ is known as {\it Schur complement.}
$\,\,\diamond\diamond\diamond$\\

In the present case, the block $A$ writes
\beq
A=\delta_{\mu\nu}(Q_\kappa^2 +m^2)-Q^\kappa_\mu Q^\kappa_\nu=(Q_\kappa^2 +m^2)P^\perp_{\mu\nu}(Q_\kappa)+m^2P^\parallel_{\mu\nu}(Q_\kappa)
\eeq
where we have introduced the usual orthogonal projectors
\beq
P^\parallel_{\mu\nu}(Q_\kappa)\equiv\frac{Q^\kappa_\mu Q^\kappa_\nu}{Q^2_\kappa} \quad {\rm and} \quad P^\perp_{\mu\nu}(Q_\kappa)\equiv\delta_{\mu\nu}-\frac{Q^\kappa_\mu Q^\kappa_\nu}{Q^2_\kappa}\,,
\eeq
but with respect to the generalized momentum $Q^\kappa_\mu$. The block $A$ is clearly invertible, with inverse $A^{-1}$ given by
\beq
A^{-1}=\frac{P^\perp_{\mu\nu}(Q_\kappa)}{Q_\kappa^2 +m^2}+\frac{P^\parallel_{\mu\nu}(Q_\kappa)}{m^2}
\eeq
and determinant
\beq
{\rm det}\,A=m^2(Q_\kappa^2 +m^2)^{d-1}\,,
\eeq
the power of $d-1$ coming from the fact that $P^\perp(Q_\kappa)$ corresponds to an eigenspace of dimension $d-1$. Finally, the Schur complement is just a number in this case and computed to be
\beq
D-CA^{-1}B & = & \frac{\bar Q^\kappa_\mu P^\perp_{\mu\nu}(Q_\kappa)\bar Q^\kappa_\nu}{Q_\kappa^2 +m^2}+\frac{\bar Q^\kappa_\mu P^\parallel_{\mu\nu}(Q_\kappa)\bar Q^\kappa_\nu}{m^2}\nonumber\\
& = & \frac{\bar Q_\kappa^2}{Q_\kappa^2 +m^2}+\frac{(\bar Q_\kappa\cdot Q_\kappa)^2}{Q^2_\kappa}\left[\frac{1}{m^2}-\frac{1}{Q_\kappa^2+m^2}\right]\nonumber\\
& = & \frac{1}{Q_\kappa^2 +m^2}\left[\bar Q_\kappa^2+\frac{(\bar Q_\kappa\cdot Q_\kappa)^2}{m^2}\right].
\eeq
Here, we have introduced the notation $X^\kappa\cdot Y^\kappa\equiv X^\kappa_\mu Y^\kappa_\mu$ when a sum over $\mu$ is implied. Putting all the pieces together, the determinant of the quadratic form (\ref{eq:quadro}) writes
\beq
(Q_\kappa^2 +m^2)^{d-2}\left[m^2\bar Q_\kappa^2+(\bar Q_\kappa\cdot Q_\kappa)^2\right]\,.\label{eq:det_g}
\eeq

\section{Final Result}
The logarithm of the determinant (\ref{eq:det_g}) contributes to the effective potential with a factor 
$1/2$, see Eq.~(\ref{eq:ea}). Similarly, the quadratic part in the ghost sector gives the determinant $Q_\kappa\cdot\bar Q_\kappa$ whose logarithm contributes to the effective potential with a factor $-1$. All together we arrive at the formula
\beq
V_{\bar r}(r) & \!\!\!=\!\!\! & \frac{m^2T^2}{2g^2}(r-\bar r)^2-\sum_\kappa\int_Q\ln\big[\bar Q_\kappa\cdot{Q}_\kappa\big]\nonumber\\
& \!\!\!+\!\!\! & \frac{d-2}{2}\sum_\kappa\int_Q\ln\big[{Q}_\kappa^2+m^2\big]+\frac{1}{2}\sum_\kappa\int_Q\ln\left[(\bar Q_\kappa\cdot{Q}_\kappa)^2+m^2\bar Q_\kappa^2\right],\label{eq:1l}\nonumber\\
\eeq
where the first term is just the Lagrangian density for constant field and background of the form (\ref{eq:form2222}). We are being deliberately vague about the precise meaning of our notation $\int_Q$. This will be clarified in the next chapter where we deal with the practical evaluation of the effective potential.

Recall that $\bar r$ is a background that characterizes the chosen gauge within the class of background Landau gauges. We are particularly interested in the center-symmetric Landau gauge corresponding to the choice $\smash{\bar r=\bar r_c}$, where $\bar r_c$ denotes a confining configuration. On the other hand, the variable $r$ represents the gluon thermal average in the presence of an external source. Its value $r_{\rm min}$ in the limit of zero sources is obtained by minimizing the effective potential $V_{\bar r_c}(r)$ with respect to $r$ and plays, as we have seen, the role of an order parameter for the confinement/deconfinement transition, namely one should have $\smash{r_{\rm min}=\bar r_c}$ in the low temperature, confining phase, and $\smash{r_{\rm min}\neq\bar r_c}$ in the high temperature, deconfined phase.\\

\noindent{$\diamond\diamond\diamond\,\,${\bf Problem 4.6:} Check that $V_{\bar r_c}(r)$ as given by Eq.~(\ref{eq:1l}) for $\smash{\bar r=\bar r_c=\pi}$ is symmetric under Weyl transformations $\smash{r\to -r}$ and under the twisted transformations $\smash{r\to r+2\pi}$. It is thus invariant under $\smash{r\to 2\pi-r}$. \underline{Tip:} employ appropriate changes of variables in the Matsubara sums.}
$\,\,\diamond\diamond\diamond$\\

\chapter{Matsubara Sum-Integrals}

{\bf \underline{Goal:}} Learn how to evaluate one-loop integrals at finite temperature.\\

In order to pursue our investigation of the effective potential, we now need to determine the one-loop integrals present in Eq.~(\ref{eq:1l}). These integrals are similar to the usual vacuum integrals with the important difference that the frequencies take discrete values ($n\in\mathds{Z}$)
\beq
\omega_n\equiv \frac{2\pi}{\beta} n\,,\label{eq:Mat}
\eeq
known as {\it Matsubara frequencies.} The presence of discrete frequencies relates to the fact that the associated time variable spans a compact time interval $[0,\beta]$ over which the fields satisfy periodic boundary conditions. For such functions, the appropriate Fourier decomposition is not a Fourier integral but rather a Fourier series with discrete frequencies as given by (\ref{eq:Mat}). 

We can now give a more precise meaning to the integral notation $\int_Q$ appearing in Eq.~(\ref{eq:1l}). It stands for
\beq
\int_Q \equiv \frac{1}{\beta}\sum_{n\in\mathds{Z}}\int\frac{d^{d-1}q}{(2\pi)^{d-1}}\,,\label{eq:sum2}
\eeq
also known as a {\it Matsubara sum-integral.} In this chapter, we discuss a rather generic situation where it is possible to analytically evaluate the discrete sum in the sum-integral (\ref{eq:sum2}) and apply the method to extract the deconfinement temperature as predicted by the Curci-Ferrari model in the SU($2$) case. We also explain what to do in those cases where the sum has no simple analytical expression.

\section{The Contour Trick}
Suppose we are interested in evaluating a generic sum
\beq
F\equiv \frac{1}{\beta}\sum_{n\in\mathds{Z}} f(i\omega_n)\,,\label{eq:sum}
\eeq
with $\omega_n$ given in Eq.~(\ref{eq:Mat}). In what follows, we assume that $f(z)$ is a function defined over the complex plane, everywhere analytic except for a finite number of simple poles $z_i$, which should of course be distinct from the $i\omega_n$ (otherwise the sum is not even defined in the first place). We also assume that $f(z)$ goes to $0$ fast enough and uniformly enough as $|z|\to\infty$.

\subsection{General Formula}
Consider a circular contour $C_N$ centered at the origin, with radius $R_N=2\pi/\beta (N+1/2)$, so in between two consecutive Matsubara frequencies see Fig.~\ref{fig:contour}, and define the associated contour integral
\beq
I_N\equiv\int_{C_N}\frac{dz}{2\pi i}\,f(z)n(z)\,,\label{eq:contour_int}
\eeq
where
\beq
n(z)=\frac{1}{e^{\beta z}-1}
\eeq
is a complex version of the {\it Bose-Einstein factor.} Note that the latter is analytic, except for simple poles located precisely at $\smash{z=i\omega_n}$ with residue $1/\beta$ (verify it!). We also note for later purpose that $\smash{n(-z)=-1-n(z)}$ (show it!).

\begin{figure}[t]
  \centering
  \includegraphics[width=0.8\linewidth]{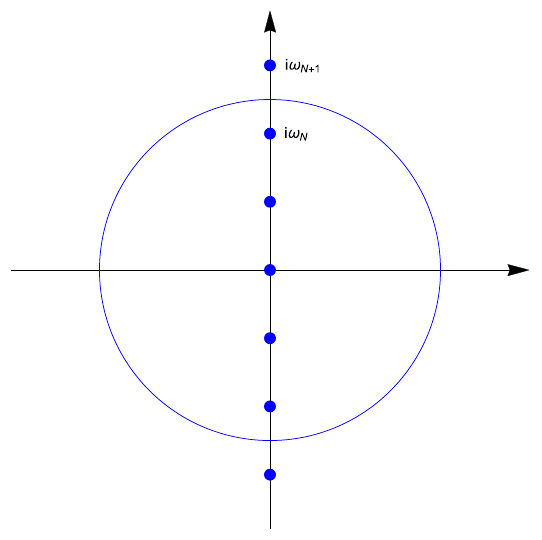}
 \caption{The integration contour $C_N$ in Eq.~(\ref{eq:contour_int}) and the Matsubara frequencies along the imaginary axis as poles of the complex Bose-Einstein factor $n(z)$.}\label{fig:contour}
\end{figure}

Let us now consider the integral $I_N$ the limit $\smash{N\to\infty}$ of infinite radius. There are two ways in which this limit can be evaluated. First of all, because we have assumed that $f(z)$ goes to $0$ fast enough and regularly enough as $|z|\to\infty$, we can write
\beq
\lim_{N\to\infty} I_N=0\,.\label{eq:limit}
\eeq
On the other hand, because $f(z)$ and $n(z)$ have only simple poles which do not coincide with each other by assumption, so is the case of $f(z)n(z)$. The residue of this function at a pole $i\omega_n$ is $f(i\omega_n)/\beta$ while the residue at a pole $z_i$ is $n(z_i) {\rm Res}\,f|_{z_i}$. The residue theorem then gives
\beq
I_N=\frac{1}{\beta}\sum_{n=-N}^N f(i\omega_n)+\sum_i n(z_i) {\rm Res}\,f|_{z_i}\,.
\eeq
In the limit $N\to\infty$, owing to Eqs.~(\ref{eq:sum}) and (\ref{eq:limit}), this rewrites
\beq
F=-\sum_i n(z_i) {\rm Res}\,f|_{z_i}\,,\label{eq:generic}
\eeq
which is an explicit result for the original sum (\ref{eq:sum}) in terms of the simple poles and residues of $f(z)$. Let us mention that the formula can be generalized to the case where $f(z)$ possesses multiple poles but we shall not need this more general formula since we shall always massage our expressions so that only simple poles are present.

\subsection{Examples}
Consider for instance the sum
\beq
\frac{1}{\beta}\sum_{n\in\mathds{Z}}\frac{1}{Q^2+m^2}\equiv\frac{1}{\beta}\sum_{n\in\mathds{Z}}\frac{1}{-(i\omega_n)^2+q^2+m^2}\,,
\eeq
corresponding to (we define $\varepsilon_q\equiv\sqrt{q^2+m^2}$)
\beq
f(z)=\frac{1}{-z^2+\varepsilon_q^2}=\frac{1}{2\varepsilon_q}\left[\frac{-1}{z-\varepsilon_q}+\frac{1}{z+\varepsilon_q}\right],
\eeq
with simple poles at $\smash{z=\pm\varepsilon_q}$ with residue $\mp 1/(2\varepsilon_q)$. Direct application of Eq.~(\ref{eq:generic}) then gives
\beq
\frac{1}{\beta}\sum_{n\in\mathds{Z}}\frac{1}{Q^2+m^2} & = & -\left[-\frac{1}{2\varepsilon_q}n(\varepsilon_q)+\frac{1}{2\varepsilon_q}n(-\varepsilon_q)\right]=\frac{1+2n(\varepsilon_q)}{2\varepsilon_q}\,.
\eeq
Combining this result with a spatial momentum integral, we arrive at
\beq
\int_Q\frac{X(q)}{Q^2+m^2}=\int\frac{d^{d-1}q}{(2\pi)^{d-1}}\frac{X(q)}{2\varepsilon_q}+\int\frac{d^{d-1}q}{(2\pi)^{d-1}}X(q)\frac{n(\varepsilon_q)}{\varepsilon_q}\,,\label{eq:vac_th}
\eeq
where $X(q)$ denotes a generic function depending only on the spatial momentum, which we assume to grow at most polynomially at large $q$. We have here split the result into a {\it finite-temperature part} (second term of Eq.~(\ref{eq:vac_th})) which goes to $0$ as $\smash{T\to 0}$ due to the presence of $n(\varepsilon_q)$ and a {\it zero-temperature or vacuum part} (first term of Eq.~(\ref{eq:vac_th})) which survives the $\smash{T\to 0}$ limit. This type of splitting is very useful since the finite-temperature part is UV-finite thanks to the presence of the thermal factor which cuts off the large momentum tails. Possible UV-divergences are only present in the vacuum part.\\

In what follows, we shall need the generalization of this formula in the presence of a shifted momentum $\smash{Q_\kappa=Q+r\kappa T}$. Its derivation is left as an exercise:\\

\noindent{$\diamond\diamond\diamond\,\,${\bf Problem 5.1:} Show that
\beq
\int_Q\frac{X(q)}{Q^2_\kappa+m^2}=\int\frac{d^{d-1}q}{(2\pi)^{d-1}}\frac{X(q)}{2\varepsilon_q}+{\rm Re}\int\frac{d^{d-1}q}{(2\pi)^{d-1}}X(q)\frac{n(\varepsilon_q-ir\kappa T )}{\varepsilon_q}\,,\nonumber
\eeq
where ${\rm Re}$ denotes the real part. Note that the zero-temperature piece does not depend on $r$ and thus coincides with the zero-temperature piece in the $\smash{r=0}$ case. This is also true for the finite-temperature piece for the color mode $\smash{\kappa=0}$. In contrast, for $\smash{\kappa=\pm 1}$, the finite-temperature piece does depend on $r$ and it is here quite clear that $\pm r T$ plays the role of an imaginary chemical potential. When $r$ is taken to correspond to a confining configuration $\smash{r=\bar r_c=\pi}$  show that the formula becomes
\beq
\int_Q\frac{X(q)}{Q^2_\pm+m^2}\to\int\frac{d^{d-1}q}{(2\pi)^{d-1}}\frac{X(q)}{2\varepsilon_q}-\int\frac{d^{d-1}q}{(2\pi)^{d-1}}X(q)\frac{f(\varepsilon_q)}{\varepsilon_q}\,.\nonumber
\eeq
with $f(z)\equiv 1/(e^{\beta z}+1)$ the Fermi-Dirac distribution! $\,\,\diamond\diamond\diamond$

\section{Application}
In the SU($2$) case, the transition is known to be second order, that is, it occurs continuously. In terms of the effective potential $V_{\bar r_c}(r)$, this means that one moves smoothly from a situation where there is only one minimum located at $\smash{r=\bar r_c=\pi}$, to a situation with two minima connected by center symmetry $r\to 2\pi-r$ while $\smash{r=\bar r_c}$ has turned into a maximum. Thus, the transition occurs at the temperature at which the extremum at $\smash{r=\bar r_c}$ changes nature, or, in other words, the temperature at which the curvature of the effective potential at $\smash{r=\bar r_c}$ vanishes, see Fig.~\ref{fig:curvature}. The equation determining the transition temperature is thus
\beq
V''_{\bar r_c}(r=\bar r_c)=0\,.\label{eq:tc}
\eeq
Let us now evaluate this equation in more detail.

\begin{figure}[t]
  \centering
  \includegraphics[width=0.6\linewidth]{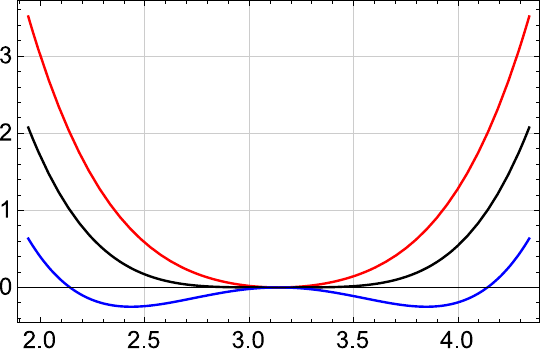}
 \caption{The continuous transition of a $Z_2$ symmetric theory. The transition occurs because the curvature of the potential at the $Z_2$-symmetric point changes sign. Because the transition is continuous, the curvature needs to vanish at the transition. This would not be true in the case of a first order transition, in which case the vanishing of the curvature defines the spinodals below and above the transition.}\label{fig:curvature}
\end{figure}

\subsection{Curvature}
Taking a first derivative with respect to $r$ in Eq.~(\ref{eq:1l}), we find
\beq
V'_{\bar r}(r) & \!\!\!=\!\!\! & \frac{m^2T^2}{g^2}(r-\bar r)+\frac{d-2}{2}\sum_\kappa\int_Q\frac{2\kappa T \omega_n^\kappa}{{Q}_\kappa^2+m^2}\nonumber\\
& \!\!\!-\!\!\! & \sum_\kappa\int_Q\frac{\kappa T \bar\omega_n^\kappa}{\bar Q_\kappa\cdot{Q}_\kappa}+\frac{1}{2}\sum_\kappa\int_Q\frac{2\kappa T \bar\omega_n^\kappa (\bar Q_\kappa\cdot{Q}_\kappa)}{(\bar Q_\kappa\cdot{Q}_\kappa)^2+m^2\bar Q_\kappa^2}.
\eeq
A second derivative gives
\beq
V''_{\bar r}(r) & \!\!\!=\!\!\! & \frac{m^2T^2}{g^2}+\sum_\kappa\int_Q\frac{(\kappa T)^2 (\bar\omega_n^\kappa)^2}{(\bar Q_\kappa\cdot{Q}_\kappa)^2}\nonumber\\
& \!\!\!+\!\!\! & \frac{d-2}{2}\sum_\kappa\int_Q\left[\frac{2(\kappa T )^2}{{Q}_\kappa^2+m^2}-\frac{4(\kappa T)^2 (\omega_n^\kappa)^2}{({Q}_\kappa^2+m^2)^2}\right]\nonumber\\
& \!\!\!+\!\!\! & \frac{1}{2}\sum_\kappa\int_Q\left[\frac{2(\kappa T)^2 (\bar\omega_n^\kappa)^2}{(\bar Q_\kappa\cdot{Q}_\kappa)^2+m^2\bar Q_\kappa^2}-\frac{4(\kappa T)^2 (\bar\omega_n^\kappa)^2(\bar Q_\kappa\cdot{Q}_\kappa)^2}{((\bar Q_\kappa\cdot{Q}_\kappa)^2+m^2\bar Q_\kappa^2)^2}\right].\nonumber\\
\eeq
This formula looks complicated but it can be simplified quite a bit. First, for later convenience, we bring a factor $T^2/g^2$ to the left-hand side. Next, we notice that only the modes $\smash{\kappa=\pm 1}$ contribute, and they actually contribute equally (as a change of variables $n\to -n$ allows one to show). Finally, we are eventually interested in the case $\smash{r=\bar r}$, see Eq.~(\ref{eq:tc}), in which case $\bar Q_\kappa\cdot Q_\kappa\to \bar Q_\kappa^2$ and $(\bar Q_\kappa\cdot{Q}_\kappa)^2+m^2\bar Q_\kappa^2\to \bar Q_\kappa^2(\bar Q_\kappa^2+m^2)$. We then arrive at
\beq
\frac{g^2}{T^2}V''_{\bar r}(\bar r) & \!\!\!=\!\!\! & m^2+g^2\left\{(d-2)\int_Q\left[\frac{2}{{\bar Q}_{+}^2+m^2}-\frac{4 (\bar \omega_n^{+})^2}{({\bar Q}_{+}^2+m^2)^2}\right]\right.\nonumber\\
& \!\!\!+\!\!\! & \left.\int_Q\left[\frac{ 2(\bar\omega_n^{+})^2}{\bar Q_{+}^4}+\frac{2(\bar\omega_n^{+})^2}{\bar Q_{+}^2(\bar Q_{+}^2+m^2)}-\frac{4 (\bar\omega_n^{+})^2}{(\bar Q_{+}^2+m^2)^2}\right]\right\}.\label{eq:temp2}
\eeq

\subsection{Renormalization}
Before evaluating the curvature, let us discuss its renormalization. 

The above expression contains one-loop sum-integrals. As we have seen, these sum-integrals typically split into a UV-finite finite-temperature piece and a possibly UV-divergent zero-temperature piece. These UV divergences should in principle be cancelled by the various renormalization factors/counterterms determined for instance from the evaluation of the primitively divergent correlation functions of the Curci-Ferrari model in a given renormalization scheme. Here, we can somehow short-circuit the determination of counterterms/elimination of divergences because the left-hand side of (\ref{eq:temp}) is nothing but the inverse two-point function in the zero-momentum limit.  To understand this in a simple set-up, let us go back to our scalar example. 

By construction, functional derivatives of the generating functional $W[J]$ with respect to the source $J$ give access to connected correlation functions:
\beq
\phi(x) & \!\!\!=\!\!\! & \frac{\delta W}{\delta J(x)}\,,\\
G(x,y) & \!\!\!=\!\!\! & \frac{\delta^2W}{\delta J(x)\delta J(y)}=\frac{\delta\phi(x)}{\delta J(y)}\,.
\eeq
The last equality shows that the propagator can be seen as the Jacobian of the transformation $J[\phi]$ that associates the one-point function $\phi$ to a given source $J$.
But we have also seen that the first derivative of the effective action  $\Gamma[\phi]$ allows one to retrieve back the source:
\beq
\frac{\delta\Gamma}{\delta\phi(x)}=J(x)\,.
\eeq
The second derivative is then
\beq
\frac{\delta^2\Gamma}{\delta\phi(x)\delta\phi(y)}=\frac{\delta J(x)}{\delta\phi(y)}\,,
\eeq
that is the Jacobian of the inverse transformation $J[\phi]$. Since $\phi[J]$ and $J[\phi]$ are inverses of each other, so should be the corresponding Jacobians. Then, we have
\beq
\delta(x-y)=\int_z \frac{\delta \phi(x)}{\delta J(z)}\frac{\delta J(z)}{\delta\phi(y)}=\int_z \frac{\delta^2\Gamma}{\delta\phi(x)\delta\phi(z)}G(z,y)\,,
\eeq
which shows that $\delta^2\Gamma/\delta\phi(x)\delta\phi(y)$ is nothing but the inverse propagator. Now, when we consider the potential $V(\phi)$, we are in fact evaluating the effective action for a constant field configuration $\smash{\phi(x)=\phi}$, that is $\smash{\Gamma[\phi]=\beta L^3 V(\phi)}$. The curvature of the potential is then
\beq
V''(\phi) & \!\!\!=\!\!\! & \frac{1}{\beta L^3}\int_x\int_y \left.\frac{\delta^2\Gamma}{\delta\phi(x)\delta\phi(y)}\right|_{\phi(\,\,)=\phi}\nonumber\\
& \!\!\!=\!\!\! & \frac{1}{\beta L^3}\int_x\int_y G^{-1}(x-y)\nonumber\\
& \!\!\!=\!\!\! & \frac{\int_y 1}{\beta L^3}\int_x G^{-1}(x)=G^{-1}(Q=0)\,.
\eeq
where we have used that $\smash{\int_y 1=\beta L^3}$ and $\smash{\int_x f(x)=f(Q=0)}$.

Coming back to our present problem, since there is a factor $T/g$ that connects the variable $r$ to the gluon field, see Eq.~(\ref{eq:form2222}), the left-hand side of Eq.~(\ref{eq:temp2}) including the factor $g^2/T^2$ is nothing but the inverse propagator $G^{-1}(Q)$ in the zero-momentum limit. Its zero-temperature part can then be seen as the renormalized mass in a scheme defined by the renormalization condition $\smash{G^{-1}_{T=0}(Q=0)=m^2}$. In practice, this means that, by choosing such type of renormalization scheme, one can manually set to $0$ the zero-temperature pieces of the one-loop sum-integrals in Eq.~(\ref{eq:temp2}) and keep only the finite-temperature pieces. For this reason, we shall also set $d=4$ in this formula:
\beq
\frac{T^2}{g^2}V''_{\bar r}(\bar r) & \!\!\!=\!\!\! & m^2+g^2\left\{\int_Q\frac{4}{{\bar Q}_{+}^2+m^2}\right.\nonumber\\
& \!\!\!+\!\!\! & \left.\int_Q\left[\frac{ 2(\bar\omega_n^{+})^2}{\bar Q_{+}^4}+\frac{2(\bar\omega_n^{+})^2}{\bar Q_{+}^2(\bar Q_{+}^2+m^2)}-\frac{12 (\bar\omega_n^{+})^2}{(\bar Q_{+}^2+m^2)^2}\right]\right\}_{\mbox{\tiny no $\smash{T=0}$ parts}}\!\!\!\!\!\!\!\!\!\!\!\!\!\!\!\!\!\!\!\!\!\!\!.\label{eq:inter}
\eeq
We are now ready to evaluate this expression using the contour trick presented above.

\subsection{Reduction to Basic Sum-Integrals}
Notice that the first sum-integral in Eq.~(\ref{eq:inter}) is precisely of the form treated in Problem 5.1. One strategy to deal with the other sum-integrals is to bring them to the same form. Take for instance the second sum-integral in the second line of Eq.~(\ref{eq:inter}). Using
\beq
\frac{1}{\bar Q_{+}^2(\bar Q_{+}^2+m^2)}=\frac{1}{m^2}\left[\frac{1}{\bar Q_{+}^2}-\frac{1}{\bar Q_{+}^2+m^2}\right],
\eeq
it becomes
\beq
\int_Q\frac{(\bar\omega_n^{+})^2}{\bar Q_{+}^2(\bar Q_{+}^2+m^2)}=\frac{1}{m^2}\int_Q\left[\frac{(\bar\omega_n^\kappa)^2}{\bar Q_{+}^2}-\frac{(\bar\omega_n^\kappa)^2}{\bar Q_{+}^2+m^2}\right].
\eeq
Then, noting that $\smash{(\bar\omega_n^\kappa)^2=\bar Q_\kappa^2-q^2=\bar Q_\kappa^2+m^2-q^2-m^2}$, we arrive at
\beq
\int_Q\frac{(\bar\omega_n^{+})^2}{\bar Q_{+}^2(\bar Q_{+}^2+m^2)}=\frac{1}{m^2}\int_Q\left[\frac{q^2+m^2}{\bar Q_{+}^2+m^2}-\frac{q^2}{\bar Q_{+}^2}\right].
\eeq
On the other hand, the third (and therefore the first) sum-integral in the second line of Eq.~(\ref{eq:inter}) can written as
\beq
\int_Q\frac{(\bar\omega_n^{+})^2}{(\bar Q_{+}^2+m^2)^2} & = & \int_Q\frac{1}{\bar Q_{+}^2+m^2}-\int_Q\frac{q^2+m^2}{(\bar Q_{+}^2+m^2)^2}\,.
\eeq
Then, noting that
\beq
\frac{1}{(\bar Q_{+}^2+m^2)^2}=-\frac{d}{dq^2}\frac{1}{\bar Q_{+}^2+m^2}=-\frac{1}{2q}\frac{d}{dq}\frac{1}{\bar Q_{+}^2+m^2}\,,
\eeq
we find, upon integrating by parts 
\beq
\int_Q\frac{(\bar\omega_n^{+})^2}{(\bar Q_{+}^2+m^2)^2} & = & \int_Q\frac{1}{\bar Q_{+}^2+m^2}+\frac{1}{2}\int_Q\frac{q^2+m^2}{q}\frac{d}{dq}\frac{1}{\bar Q_{+}^2+m^2}\nonumber\\
& = & \int_Q\frac{1}{\bar Q_{+}^2+m^2}-\frac{1}{2}\int_Q\left(3+\frac{m^2}{q^2}\right)\frac{1}{\bar Q_{+}^2+m^2}\nonumber\\
& = & -\frac{1}{2}\int_Q\frac{1}{\bar Q_{+}^2+m^2}-\frac{1}{2}\int_Q\frac{m^2}{q^2}\frac{1}{\bar Q_{+}^2+m^2}\,.
\eeq
Here, we have set $\smash{d=4}$ and we have not paid attention to boundary terms since we are only interested in the finite temperature pieces. Putting all the pieces together, we arrive at
\beq
\frac{T^2}{g^2}V''_{\bar r}(\bar r) & \!\!\!=\!\!\! & m^2+g^2\int_Q\left\{\left[6\frac{m^2}{q^2}+12+2\frac{q^2}{m^2}\right]\frac{1}{{\bar Q}_{+}^2+m^2}-\left[1+2\frac{q^2}{m^2}\right]\frac{1}{\bar Q_+^2}\right\}.\nonumber\\
\eeq

\subsection{Extracting $T_c$}
The condition (\ref{eq:tc}) defining $T_c$ is obtained by setting $\smash{\bar r=\bar r_c}$ in which case the sum-integrals can be expressed in terms of the Fermi-Dirac distribution, see Problem 5.1. 

Defining $\smash{t_c\equiv T_c/m}$ and rescaling the integrals by $m$, we find
\beq
\frac{2\pi^2}{g^2}=\int_0^\infty dq\,\frac{6+12q^2+2q^4}{\sqrt{q^2+1}e^{\sqrt{q^2+1}/t_c}+1}-\int_0^\infty dq\,\frac{q+2q^3}{e^{q/t_c}+1}\,.
\eeq
In the considered scheme, the fit of the lattice propagators at $T=0$ gives the best fitting parameters $\smash{g=7.5}$ and $\smash{m=680}$ MeV. With this value of $g$, we find $\smash{t_c=0.39}$ and thus $\smash{T_c=265}$ MeV, not so far from the lattice result of $295$ MeV.

\section{When the Contour Trick does not Work}
There exist situations for which the contour trick is not easily implemented. One example is that of the evaluation of the curvature of the potential for $r\neq\bar r$. In this case indeed there appears one denominator,
\beq
(\bar Q_\kappa\cdot{Q}_\kappa)^2+m^2\bar Q_\kappa^2\,,
\eeq
which depends quartically on the Matsubara frequencies with no obvious zeros. Even though roots of quartic polynomials are exactly known, evaluating the Matsubara sums in this way becomes a rather cumbersome exercise.

A different strategy consists in remarking that the above denominator is a quartic polynomial in the variable $q^2$ and can thus be written as
\beq
(q^2+M^2_{+,\kappa,n})(q^2+M^2_{-,\kappa,n})
\eeq
where $M^2_{\pm,\kappa,n}$, which are not necessarily real, depend on the Matsubara frequency $\omega_n$ and the charge $\kappa$.\\

\noindent{$\diamond\diamond\diamond\,\,${\bf Problem 5.2:} Show that
\beq
M^2_{\pm,\kappa,n} & = & \omega_n^\kappa\bar \omega_n^\kappa+\frac{m^2}{2}\nonumber\\
& \pm & \sqrt{\left(\omega_n^\kappa\bar \omega_n^\kappa+\frac{m^2}{2}\right)^2-(\bar\omega^\kappa_n)^2((\omega^\kappa_n)^2+m^2)}\nonumber\\
& = & \omega_n^\kappa\bar \omega_n^\kappa+\frac{m^2}{2}\pm\frac{m^2}{2}\sqrt{1+4\frac{\bar \omega_n^\kappa(\omega_n^\kappa-\bar \omega_n^\kappa)}{m^2}}\,,\nonumber
\eeq
where $\smash{\omega_n^\kappa=\omega_n+r\kappa\,T}$ and $\smash{\bar\omega_n^\kappa=\omega_n+\bar r\kappa\,T}$. $\,\,\diamond\diamond\diamond$

\begin{figure}[t]
  \centering
  \includegraphics[width=0.8\linewidth]{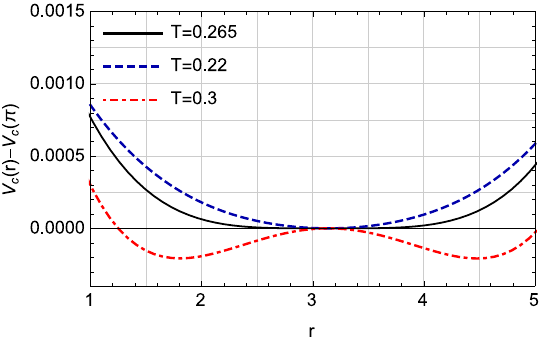}
 \caption{The potential $V_{\bar r_c}(r)$ in the SU(2) case.}\label{fig:su2_pot}
\end{figure}

This strategy can be applied in particular to deal with the effective potential (\ref{eq:1l}) which rewrites
\beq
V_{\bar r}(r) & \!\!\!=\!\!\! & \frac{m^2T^2}{2g^2}(r-\bar r)^2\nonumber\\
& \!\!\!-\!\!\! & T\sum_n\sum_\kappa\int_q\ln\big[q^2+M^2_{0,\kappa,n}\big]\nonumber\\
& \!\!\!+\!\!\! & \frac{d-2}{2}T\sum_\kappa\sum_n\int_q\ln\big[q^2+M^2_{\kappa,n}\big]\nonumber\\
& \!\!\!+\!\!\! & \frac{T}{2}\sum_\kappa\sum_n\int_q\ln\left[(q^2+M^2_{+,\kappa,n})(q^2+M^2_{-,\kappa,n})\right],
\eeq
with
\beq
M^2_{0,\kappa,n}\equiv \omega_n^\kappa\bar\omega_n^\kappa \quad {\rm and} \quad M^2_{\kappa,n}\equiv (\omega_n^\kappa)^2+m^2\,.
\eeq
The spatial momentum integrals can now all be performed analytically using the well known result
\beq
\int_q \ln[q^2+M^2]=-\frac{(M^2)^{(d-1)/2}}{(4\pi)^{(d-1)/2}}\Gamma\left(\frac{1-d}{2}\right)\,.
\eeq
One is then left with a $d$-regularized Matsubara sum to be performed numerically after possible divergences arsing from contributions of the form $\sum_n 1/n^{1+\#\epsilon}$ have been extracted. Figure \ref{fig:su2_pot} shows the so-obtained potential $V_{\bar r_c}(r)$ as a function of $r$ for various temperatures. We clearly observe a continuous confinement/deconfinement transition whose transition temperature we have already determined above.

\chapter{Extension to SU($3$)}
{\bf \underline{Goal:}} Learn how to extend the previous discussion to the SU($3$) case.\\

 The key notion is that of Cartan-Weyl bases and the associated roots which we briefly introduce in the first section. More details about roots are gathered in the Appendix.

\section{Roots and Cartan-Weyl Bases}

{\it Cartan-Weyl bases} are generalizations of the basis $\{\sigma_3/2,\sigma_+/2,\sigma_-/2\}$ which is so useful in the SU($2$) case due in particular to the relation
\beq
\left[\frac{\sigma_3}{2},\frac{\sigma_\pm}{2}\right]=\pm\frac{\sigma_\pm}{2}\,.\label{eq:roots2}
\eeq 
A Cartan-Weyl basis $\{t^j,t^\alpha\}$ consists of two types of generators: first, a set of generators $\{t^j\}$ that commute with each other and, second, a set of generators $\{t^\alpha\}$ which diagonalize the adjoint action of all the $t^j$:
\beq
[t^j,t^\alpha]=\alpha^j t^\alpha\,.
\eeq
To memorize these relations, one can adopt the following interpretation borrowed from Quantum Mechanics. The operators $[t^j,\,\,]$ form a set of commuting observables acting on a space of states as given by the Lie algebra of the group. In this space, one can look for states $t^\alpha$ that diagonalize simultaneously all the $[t^j,\,\,]$'s and which are, therefore, characterized by a set of well defined quantum numbers $\{\alpha^j\}$. This set of quantum numbers forms a vector $\alpha$ known as {\it root} which allows one to label the corresponding state $t^\alpha$. 

Note that the number of components of each root is equal to the number of commuting charges. In the SU($2$) case for instance, there is only one commuting charge $\smash{t^j=\sigma_3/2}$ and  the roots are just numbers. From Eq.~(\ref{eq:roots2}), they are found to be $\smash{\alpha=\pm 1}$, corresponding respectively to the states $\smash{t^\pm\equiv \sigma_\pm/2}$.  More generally, within the SU($N$) algebra, among the $N^2-1$ generators of the algebra, one can always find $N-1$ generators $t^j$ that commute with each other, so the roots are $(N-1)$-dimensional vectors.

Consider then SU($3$). A basis of the  algebra is $\{t^a\equiv\lambda_a/2\}$, given in terms of the Gell-Mann matrices
\beq
\lambda_1 & \!\!=\!\! & \left(
\begin{array}{ccc}
0 & 1 & 0\\
1 & 0 & 0\\
0 & 0 & 0
\end{array}
\right),\,\, \,\lambda_2=\left(
\begin{array}{ccc}
0 & -i & 0\\
i & 0 & 0\\
0 & 0 & 0
\end{array}
\right),\,\,\, \lambda_3=\left(
\begin{array}{ccc}
1 & 0 & 0\\
0 & -1 & 0\\
0 & 0 & 0
\end{array}
\right),\nonumber\\
\lambda_4 & \!\!=\!\! & \left(
\begin{array}{ccc}
0 & 0 & 1\\
0 & 0 & 0\\
1 & 0 & 0
\end{array}
\right),\,\,\, \lambda_5= \left(
\begin{array}{ccc}
0 & 0 & -i\\
0 & 0 & 0\\
i & 0 & 0
\end{array}
\right),\nonumber\\
\lambda_6 & \!\!=\!\! &\left(
\begin{array}{ccc}
0 & 0 & 0\\
0 & 0 & 1\\
0 & 1 & 0
\end{array}
\right),\,\,\, \lambda_7=\left(
\begin{array}{ccc}
0 & 0 & 0\\
0 & 0 & -i\\
0 & i & 0
\end{array}
\right),\,\,\, \lambda_8=\frac{1}{\sqrt{3}}\left(
\begin{array}{ccc}
1 & 0 & 0\\
0 & 1 & 0\\
0 & 0 & -2
\end{array}
\right)\!\,.\label{eq:gellmann}\nonumber\\
\eeq
There are two commuting generators, $t^3$ and $t^8$, so the roots are two-dimensional vectors. Two of the roots are obtained by noticing that $\{t^1,t^2,t^3\}$ generates the SU($2$) algebra while $t^8$ commutes with any element of this algebra. It follows that the combinations $t^\pm\equiv(t^1\pm it^2)/\sqrt{2}$ satisfy
\beq
[t^3,t^\pm]=\pm t^\pm \quad {\rm and} \quad [t^8,t^\pm]=0\,,
\eeq
and correspond thus to the roots $\alpha=(\pm 1,0)$.\\

\begin{figure}[t]
\begin{center}
\includegraphics[height=0.5\textheight]{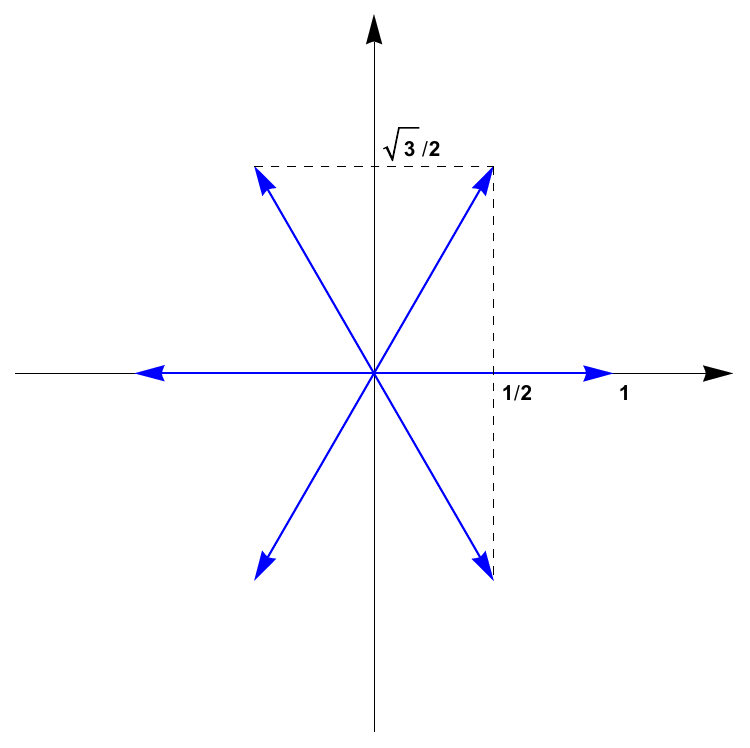}
\caption{Roots of the su($3$) algebra.}\label{fig:root}
\end{center}
\end{figure}

\noindent{$\diamond\diamond\diamond\,\,${\bf Problem 6.1} Find linear combinations $\tilde t^3$ and $\tilde t^8$ of $t^3$ and $t^8$ such that $\{t^4,t^5,\tilde t^3\}$ generates the SU($2$) algebra while $\tilde t^8$ commutes with any element of this algebra. Do the same after replacing $t^4$ and $t^5$ by $t^6$ and $t^7$ respectively.}$\,\,\diamond\diamond\diamond$\\

\noindent{$\diamond\diamond\diamond\,\,${\bf Problem 6.2} Use the results of Problem 6.1 to show that the other possible roots of SU($3$) are $(\pm 1/2,\sqrt{3}/2)$ and $(\pm 1/2,-\sqrt{3}/2)$ and construct the corresponding states from $t^4$, $t^5$, $t^6$ and $t^7$. \underline{Tip:} one can also apply some trial and error to guess the good combinations of these matrices.}$\,\,\diamond\diamond\diamond$\\

The collection of roots associated to an algebra forms the {\it root diagram} of that algebra. The root diagram of SU($3$) is represented in Fig.~\ref{fig:root} while the root diagram of SU($2$) can be read along the horizontal axis of that same figure. More details about roots can be found in Appendix B. In particular, one can show that the SU(N) roots are all of unit length, $\smash{\alpha^2=1}$. It is also generally true that the roots come by pairs, $\alpha$ and $-\alpha$.

\section{SU($3$) Gauge Transformations}

Recall that, in terms of classical gauge field configurations, invariance under some physical transformation $\smash{A_\mu\to A_\mu^T}$ needs always to be though modulo gauge transformations
\beq
\exists U_0\in {\cal G}_0\,, \quad A_\mu^T=A_\mu^{U_0}\,.
\eeq
So, in order to find configurations that obey this property, one needs first to identify gauge transformations $U_0\in {\cal G}_0$ such that the mapping $A_\mu\to  (A_\mu^T)^{U_0}$ possesses fixed-points in field space. The first step is to better understand the content of ${\cal G}_0$ as we did in the SU($2$) case. For simplicity, we restrict once again to configurations which are constant, temporal and Abelian:
\beq
A_\mu(x)=4\pi T\delta_{\mu0}\,x^j t^j\,.\label{eq:formn}
\eeq
As compared to the SU($2$) case, for later convenience, we have introduced an extra factor of $4\pi$ in the right-hand side of this equation. Since we restrict to configurations of this specific form, we need to restrict to gauge transformations that preserve this form.\\

\noindent{$\diamond\diamond\diamond\,\,${\bf Problem 6.3:} Consider the traced Wilson line $\Phi[A]$ defined in the chapter 2. Show that, for configurations of the form (\ref{eq:formn}), one has}
\beq
\Phi[A]=\frac{1}{3}\Bigg[e^{-i\frac{4\pi x_8}{\sqrt{3}}}+2e^{i\frac{2\pi x_8}{\sqrt{3}}}\cos\left(2\pi x_3\right)\Bigg].
\eeq
\underline{Tip:} Note that the path-ordering that enters the definition of $\Phi[A]$ can be ignored since the considered configurations are constant. $\,\,\diamond\diamond\diamond$

\begin{figure}[t]
\begin{center}
\includegraphics[height=0.4\textheight]{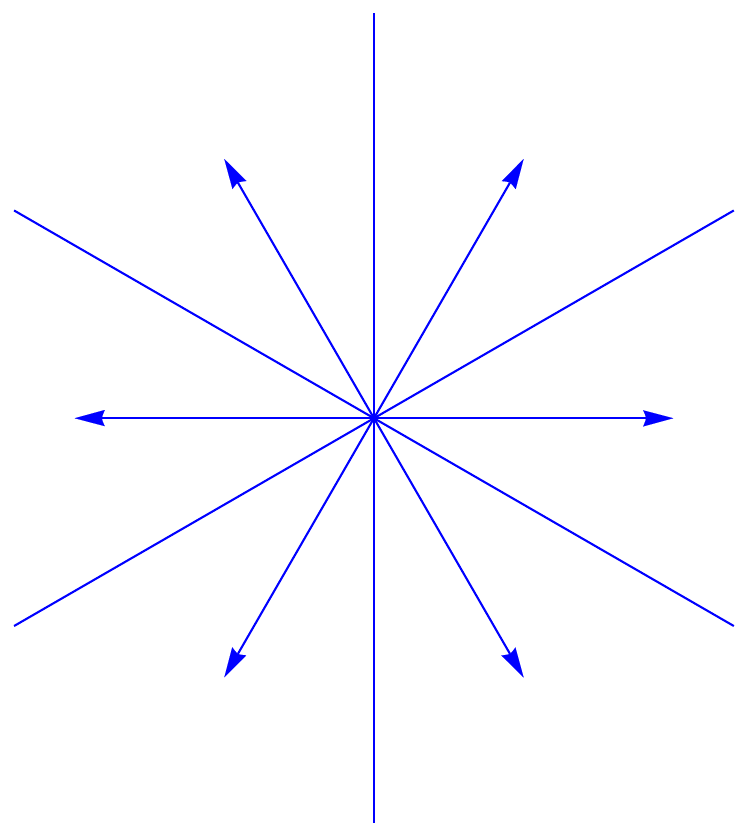}
\caption{Weyl reflections $W_\alpha$ and the associated pairs of roots $\alpha$ and $-\alpha$.}\label{fig:Weyl3}
\end{center}
\end{figure}

\subsection{Weyl Transformations}
We can first look for global gauge transformations (color rotations) that preserve this form. We saw in the SU($2$) case that the only non-trivial transformation that preserves direction $3$ of the algebra is $e^{i\pi \sigma_1/2}$ which acts on direction $3$ as a point reflection with respect to the origin. It is interesting to recast this transformation in terms of $t^\pm$. Since $t^\pm\equiv (\sigma_1\pm i\sigma_2)/2\sqrt{2}$, we have $\sigma_1/2=(t^++t^-)/\sqrt{2}$ and the transformation writes
\beq
e^{i\pi \sigma_1/2}=e^{i\frac{\pi}{\sqrt{2}}(t^++t^-)}\,.
\eeq
In the SU(N) case, it is generally true that, given a root $\alpha$,
\beq
W_\alpha\equiv e^{i\frac{\pi}{\sqrt{2}}(t^\alpha+t^{-\alpha})}\,,
\eeq
stabilizes the commuting part of the algebra. More precisely, one shows that
\beq
W_\alpha X^j t^j W^\dagger_\alpha=\left(X^j-2\frac{X\cdot\alpha}{\alpha^2}\alpha^j\right)t^j\,,
\eeq
see the Appendix. This corresponds to a reflection with respect to an hyperplane orthogonal to $\alpha$. Note that $\smash{W_\alpha=W_{-\alpha}}$ and thus the reflection is the same for both roots $\alpha$ and $-\alpha$. The Weyl reflections in the SU($3$) are represented in Fig.~\ref{fig:Weyl3} together with the corresponding roots.

\subsection{Winding Transformations}
In the SU($2$) case, we saw that the transformations
\beq
U(\tau)=\exp\left\{i\theta\frac{\tau}{\beta}\frac{\sigma_3}{2}\right\},
\eeq
with $\smash{\theta=\pm 4\pi}$ are genuine (in the sense of periodic) gauge transformations. Note that $\smash{\theta=4\pi\alpha}$ where $\alpha$ is an SU($2$) root. In fact the generalization of these transformations to the SU($N$) case is simply
\beq
U(\tau)=\exp\left\{i4\pi\frac{\tau}{\beta}\alpha^jt^j\right\},
\eeq
These transformations are periodic, $\smash{U(\beta)=U(0)}$, see the Appendix. Moreover, they act on configurations of the form (\ref{eq:formn}) as $\smash{x^j\to x^j+\alpha^j}$ that is a mere translation by the corresponding root. These transformations can also be seen in Fig.~\ref{fig:Weyl3}.

\begin{figure}[t]
\begin{center}
\includegraphics[height=0.2\textheight]{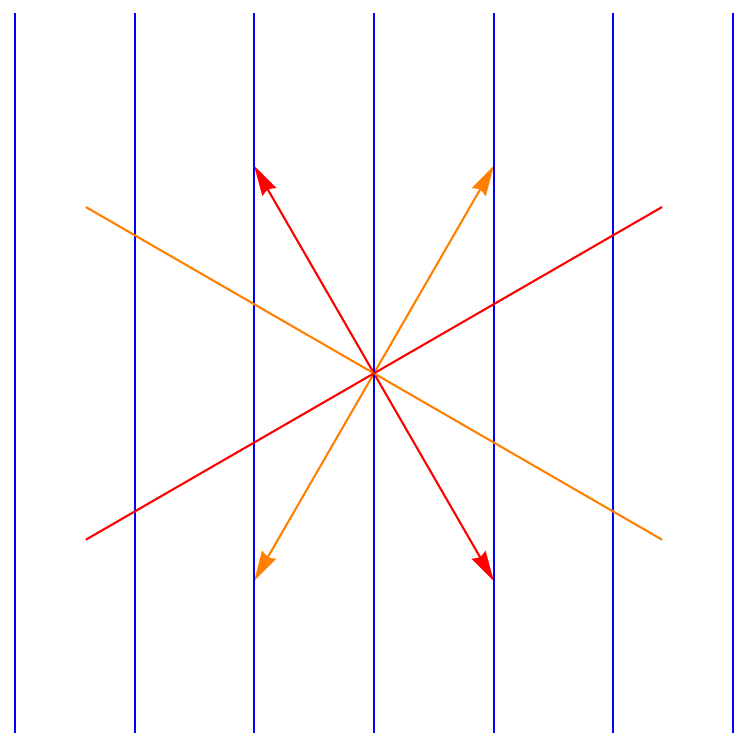}\quad\includegraphics[height=0.2\textheight]{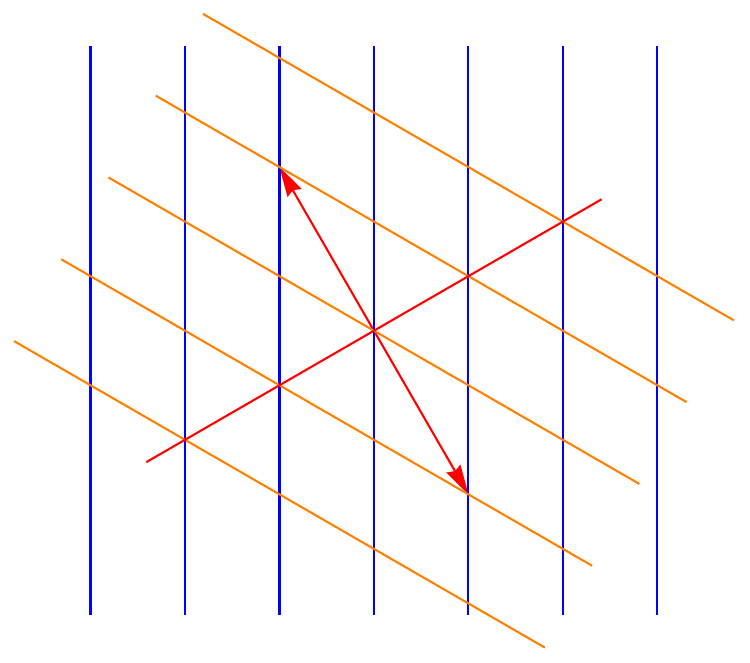}\quad\includegraphics[height=0.2\textheight]{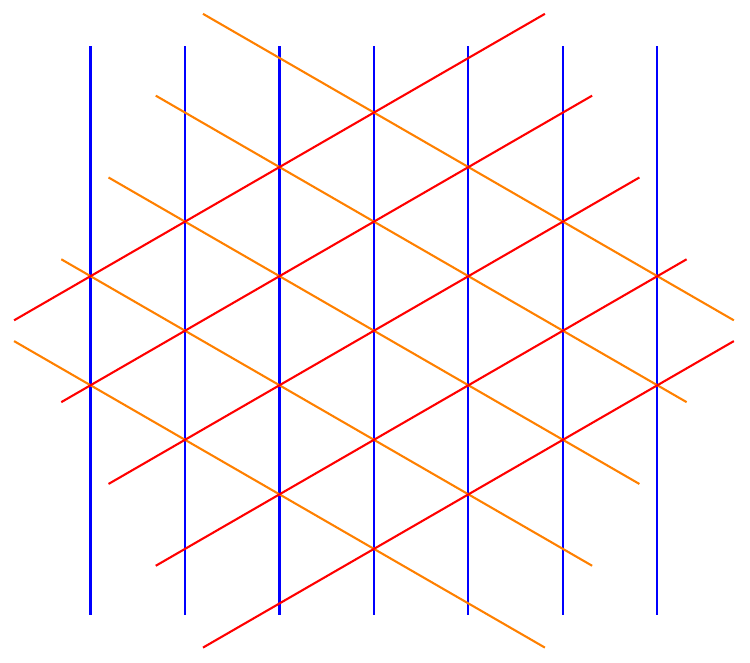}
\caption{For each root, one can combine the corresponding Weyl transformation and the winding transformations to generate a network of reflection symmetries that pave the plane $\mathds{R}^2$.}\label{fig:making}
\end{center}
\end{figure}

\subsection{Weyl Chambers}
For a given root, we can now consider the corresponding Weyl reflection orthogonal to a given root $\alpha$ and the two translations by $\alpha$ and $-\alpha$. Take for instance the transformations associated to the horizontal roots in Fig.~\ref{fig:Weyl3}. Similarly to what we did in the SU($2$) case, we can combine these transformations to generate new transformations. In particular one generates reflections with respect to hyperplanes orthogonal to the root and displaced by any multiple of half the root. In fact, we can completely forget about the translations and focus only on these displaced reflections. This gives the first picture in Fig.~\ref{fig:making}.

Repeating the same strategy for the two other root-directions, see Fig.~\ref{fig:making}, we obtain a paving of the plane $\mathds{R}^2$ into regions known as Weyl chambers, see the last figure of Fig.~\ref{fig:making}. These regions are physically equivalent since connected by genuine gauge transformations which correspond to reflections with respect to the edges of the Weyl chambers.\\

\noindent{$\diamond\diamond\diamond\,\,${\bf Problem 6.4} Show that the Weyl chambers form a lattice of points $s$ such that $s\cdot\alpha\in\mathds{Z}/2$. These vectors have a simple interpretation in terms of the {\it fundamental weights of the algebra,} see the Appendix B. $\,\,\diamond\diamond\diamond$\\

\section{Symmetries in the SU($3$) Case}
We are now ready to investigate physical transformations in terms of classical SU($3$) gauge-field configurations and to identify the invariant configurations modulo gauge transformations.

\subsection{Center Symmetry}
Let us consider center transformations first. We saw in the SU($2$) case that one particular example of non-trivial center transformations with $\smash{U(\beta)=-U(0)}$ is provided by
\beq
U(\tau)=\exp\left\{i\theta\frac{\tau}{\beta}\sigma_3/2\right\}.
\eeq
with $\smash{\theta=\pm 2\pi}$. Let us now see how this generalizes to the SU(N) case. Consider a transformation of the form
\beq
U(\tau)=\exp\left\{i4\pi\frac{\tau}{\beta}s^jt^j\right\}.
\eeq
We want $s^j$ to be chosen such that the transformation $U(\tau)$ preserves the periodicity of the gauge field (without being itself periodic necessarily). The action on the gauge field writes
\beq
A_\mu^U & = & e^{i4\pi\frac{\tau}{\beta}s^jt^j}A_\mu e^{-i4\pi\frac{\tau}{\beta}s^jt^j}+ie^{i4\pi\frac{\tau}{\beta}s^jt^j}\partial_\mu e^{-i4\pi\frac{\tau}{\beta}s^jt^j}\nonumber\\
& = & e^{i4\pi\frac{\tau}{\beta}s^j[t^j,\,\,]}A_\mu+\frac{4\pi}{\beta}\delta_{\mu0}s^jt^j\,,
\eeq
where we have used the well-known identity $\smash{e^XYe^{-X}=e^{[X,\,\,]}Y}$. The second term is obviously periodic, so the periodicity constraint comes actually from the first term and reads $e^{i4s^j[t^j,\,\,]}=\mathds{1}$. To make most of this condition, it is convenient to evaluate it on a Cartan-Weyl basis $\{t^j,t^\alpha\}$ that diagonalizes the defining action of the $t^j$'s. The only non-trivial conditions are obtained when acting on $t^\alpha$, namely $\smash{e^{i4\pi s^j\alpha^j}=1}$ and thus
\beq
s\cdot\alpha \in \frac{1}{2}\mathds{Z}\,.
\eeq
From Problem 6.3, we know that the solutions to this equation are nothing but the points of the lattice formed by the Weyl chambers. This is in agreement to what we found in the SU($2$) case where a given Weyl chamber was translated along one of its edges into the neighbouring one.

\begin{figure}[t]
\begin{center}
\includegraphics[height=0.2\textheight]{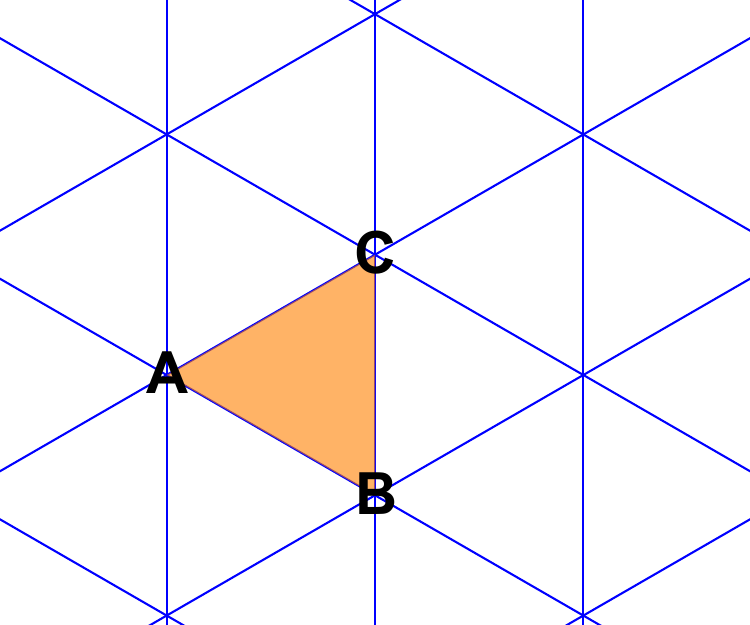}\quad\includegraphics[height=0.2\textheight]{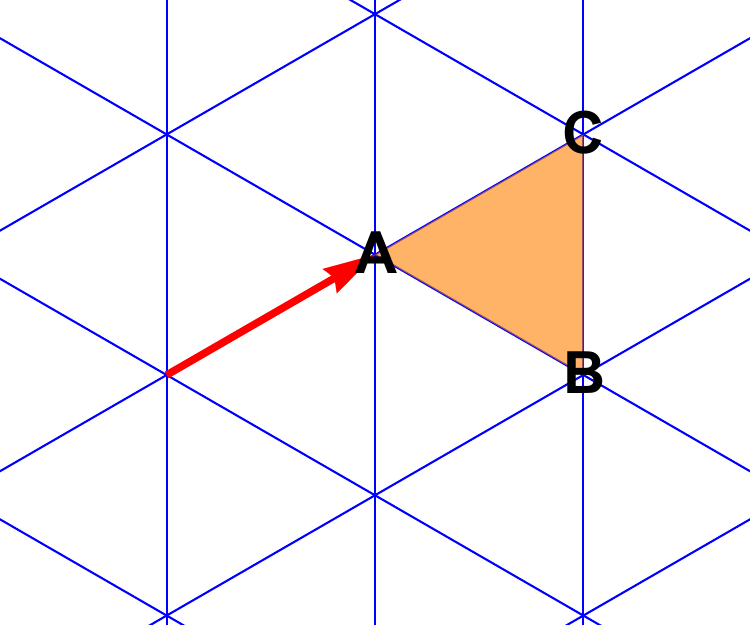}\\
\vglue4mm
\includegraphics[height=0.2\textheight]{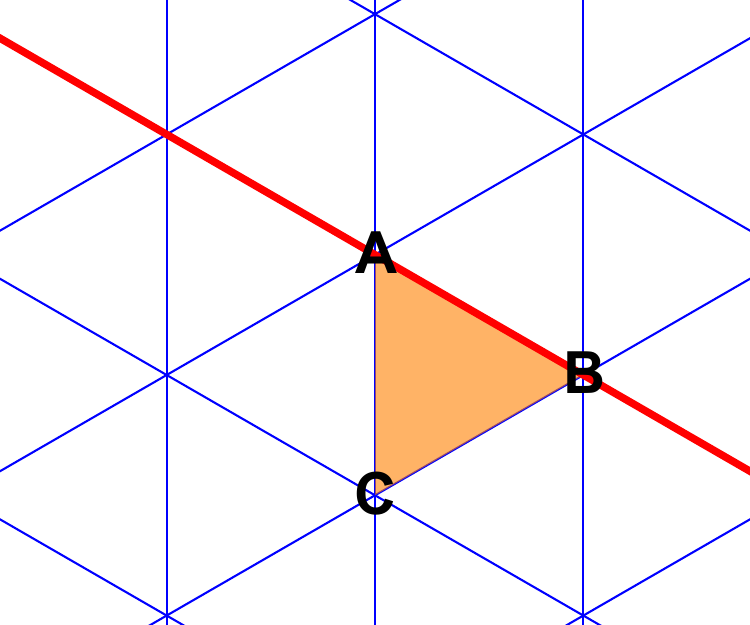}\quad\includegraphics[height=0.2\textheight]{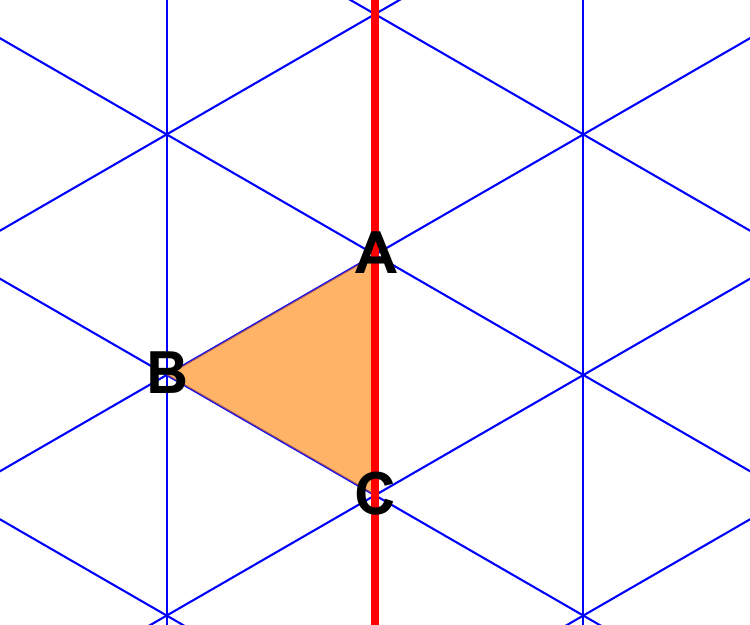}
\caption{Center transformation on a Weyl chamber, together with the gauge transformations that bring the Weyl chamber back to its original location. The combination of these transformations corresponds to a rotation of the Weyl chamber by an angle $2\pi/3$ leaving the center of the Weyl chamber as the only confining configuration in the Weyl chamber.}\label{fig:mv}
\end{center}
\end{figure}

\subsection{Confining Configurations}
Consider in particular a translation of a Weyl chamber along one of its edges, see the second plot of Fig.~\ref{fig:mv}. By using two reflections with respect to edges of Weyl chambers, that is two genuine gauge transformations, one can bring the displaced Weyl chamber back to its original location, see the last two plots of Fig.~\ref{fig:mv}. In the process, we observe that the original Weyl chamber has been rotated by an angle $2\pi/3$. Using the other possible translation, we find instead a rotation by angle $-2\pi/3$. We have thus shown that, up to gauge transformations, center transformations correspond to rotations of the Weyl chambers by $\pm 2\pi/3$. This means that the center of each Weyl chamber is center-invariant modulo gauge transformations and thus a confining configuration. For the considered Weyl chamber, this point is located at $(x_3,x_8)=(1/3,0)$ as a little geometry allows one to determine.\\

\noindent{$\diamond\diamond\diamond\,\,${\bf Problem 6.5:} Verify that $\Phi[A]$ vanishes on this configuration. \underline{Tip:} recall the formula derived in Problem 6.3.}$\,\,\diamond\diamond\diamond$


\subsection{Charge Conjugation}
We can proceed similarly to identify configurations that are charge conjugation invariant (modulo gauge transformations), see Fig.~\ref{fig:mv2}. Consider for instance the Weyl chamber depicted in the first plot of Fig.~\ref{fig:mv2}. Charge conjugation corresponds to a point symmetry of that Weyl chamber with respect to the origin, see the second plot of Fig.~\ref{fig:mv2}. By using a Weyl reflection, we can bring the so-transformed Weyl chamber back to its original location. In the process, we obtain the mirror image of the original Weyl chamber with respect to one its axes. In the considered Weyl chamber, this axis is $\smash{x_8=0}$ which corresponds thus to the location of the charge-invariant configurations (modulo gauge transformations).

Note that this is quite different from the SU($2$) case where all configurations were found to be charge conjugation invariant (modulo gauge transformations).\\

\noindent{$\diamond\diamond\diamond\,\,${\bf Problem 6.6:} Evaluate $\Phi[A^C]$ and compare it to $\Phi[A]$ in the case of a charge-conjugation-invariant configuration $\smash{x_8=0}$.\underline{Tip:} recall the formula derived in Problem 2.2.}$\,\,\diamond\diamond\diamond$

\begin{figure}[t]
\begin{center}
\includegraphics[height=0.17\textheight]{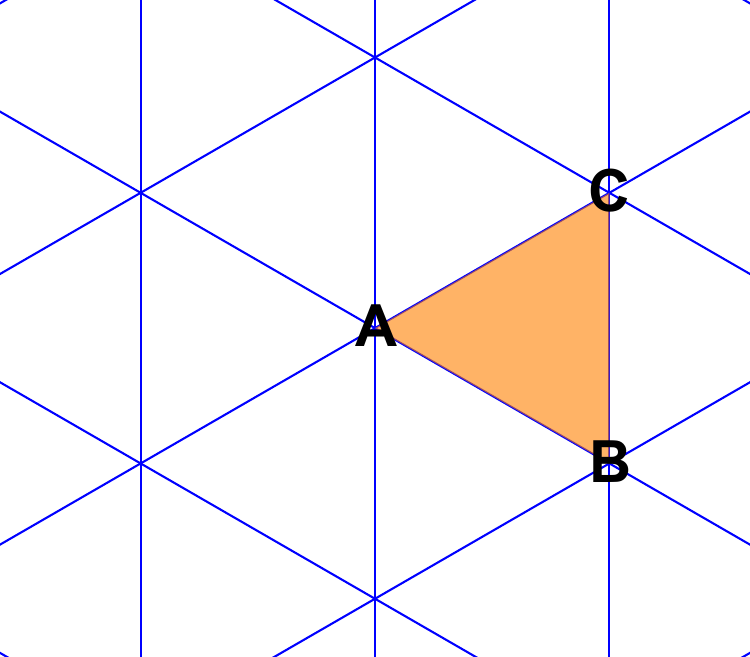}\quad\includegraphics[height=0.17\textheight]{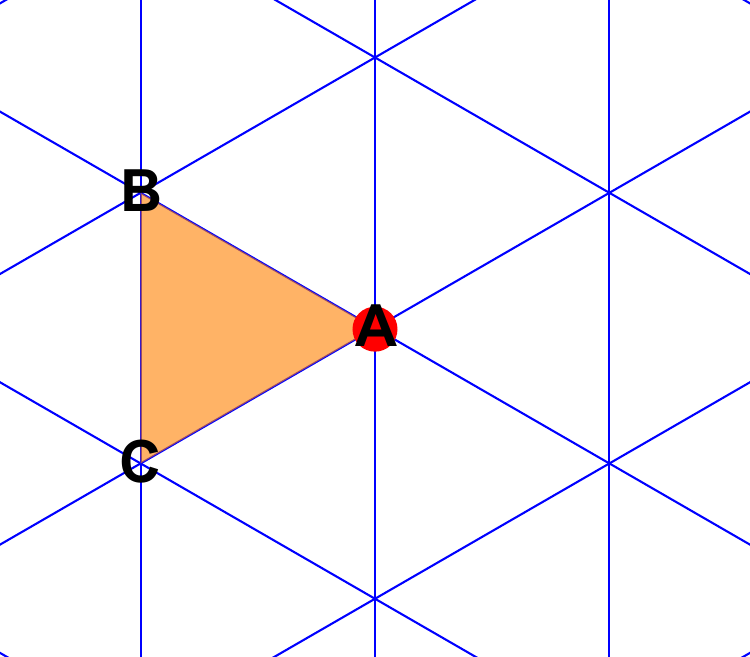}\quad\includegraphics[height=0.17\textheight]{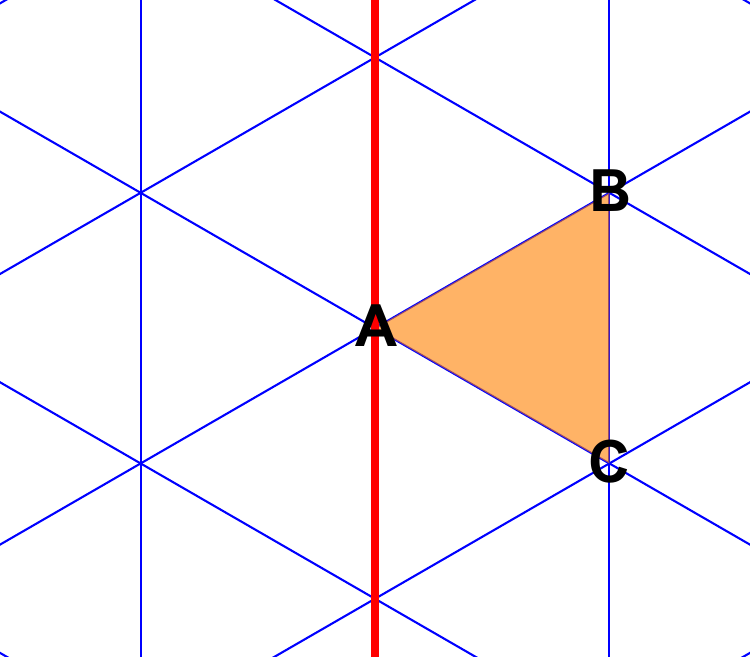}
\caption{Charge conjugation on a Weyl chamber, together with the gauge transformation that brings the Weyl chamber back to its original location. The final transformation corresponds to a reflection of the Weyl chamber with respect to one of its axes, so defining that axis as the location of the charge-invariant configurations in that Weyl chamber.}\label{fig:mv2}
\end{center}
\end{figure}

\section{The SU($3$) Effective Potential}
The Cartan-Weyl bases play also a central role in simplifying the evaluation of the effective potential. Recall that one problem in the SU($2$) case was the color structure which was not diagonal in the conventional cartesian basis due the presence of the covariant derivative
\beq
D_\mu(X^at^a) & = & \partial_\mu (X^a t^a)-i[A_\mu^bt^b,X^c t^c]\nonumber\\
& = & (\partial_\mu X^a) t^a-iA_\mu^b X^c[t^b,t^c]\nonumber\\
& = & (\partial_\mu X^a) t^a+f^{abc}A_\mu^b X^c t^a\,.
\eeq
In contrast, within a Cartan-Weyl basis $t^\kappa=\{\sigma_3/2,\sigma_+/2,\sigma_-/2\}$ with $\kappa=0,+1,-1$ such that $[t^3,t^\kappa]=\kappa t^\kappa$, and using the fact that the considered gauge fields are along direction $3$, one finds
\beq
D_\mu(X^\kappa t^\kappa) & = & \partial_\mu (X^\kappa t^\kappa)-iT\delta_{\mu 0}\,r X^\kappa[t^3,t^\kappa]\nonumber\\
& = & (\partial_\mu X^\kappa) t^\kappa-iT\delta_{\mu 0}\,r\kappa\, X^\kappa t^\kappa\nonumber\\
& = & (\partial_\mu-iT\delta_{\mu 0}\,r\kappa) X^\kappa t^\kappa\,.
\eeq
The generalization to the SU(N) case is straightforward. One first introduces a compact notation $t^\kappa$ for the Cartan-Weyl basis $\{t^j,t^\alpha\}$, with $\smash{\kappa=\alpha}$ when $\smash{t^\kappa=t^\alpha}$ and  $\smash{\kappa=0}$ when $\smash{t^\kappa=t^j}$. The reason for labelling the $t^j$ with $0$ is that they all commute with each other and have thus zero charge. One should pay attention to the fact that, beyond the SU($2$) case, this labelling is ambiguous for there are various states with zero-charge. To account for this degeneracy, we introduce various copies $0^{(j)}$ of $0$ such that $t^\kappa=t^j$ when $\kappa=0^{(j)}$. It should be clear, however, that any of these copies needs to be understood as the zero vector when it is not used as a label for the generator but appears instead in some algebraic expression. In particular, with this notation, we have $[t^j,t^\kappa]=\kappa^j t^\kappa$, and thus the action of the covariant derivative in a Cartan-Weyl basis writes (from now on we absorb factors of $4\pi$ in the notation $r^j\equiv 4\pi x^j$)
\beq
D_\mu(X^\kappa t^\kappa) & = & \partial_\mu (X^\kappa t^\kappa)-iT\delta_{\mu 0}\,r^j X^\kappa[t^j,t^\kappa]\nonumber\\
& = & (\partial_\mu X^\kappa) t^\kappa-iT\delta_{\mu 0}\,r^j\kappa^j\, X^\kappa t^\kappa\nonumber\\
& = & (\partial_\mu-iT\delta_{\mu 0}\,r\cdot\kappa) X^\kappa t^\kappa\,,
\eeq
where we have defined $r\cdot\kappa\equiv r^j\kappa^j$.

\begin{figure}[t]
  \centering
  \includegraphics[width=0.8\linewidth]{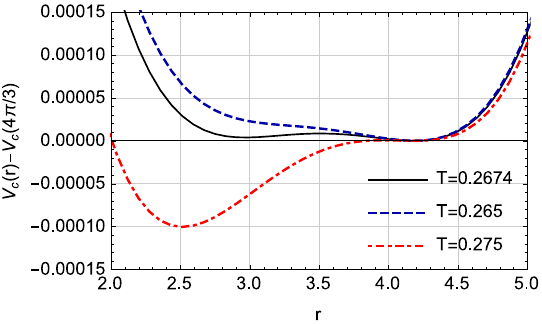}
 \caption{The potential $V_{\bar r_c}(r)$ in the SU(3) case.}\label{fig:su3_pot}
\end{figure}

In particular, this means that the expression for the SU(N) potential can be obtained by replacing $r\kappa$ by $r\cdot\kappa$ and $\bar r\kappa$ by $\bar r\cdot\kappa$ in the corresponding SU($2$) formula. One finds
\beq
V_{\bar r}(r) & \!\!\!=\!\!\! & \frac{m^2T^2}{2g^2}(r-\bar r)^2-\sum_\kappa\int_Q\ln\big[\bar Q_\kappa\cdot{Q}_\kappa\big]\nonumber\\
& \!\!\!+\!\!\! & \frac{d-2}{2}\sum_\kappa\int_Q\ln\big[{Q}_\kappa^2+m^2\big]+\frac{1}{2}\sum_\kappa\int_Q\ln\left[(\bar Q_\kappa\cdot{Q}_\kappa)^2+m^2\bar Q_\kappa^2\right],\nonumber\\
\eeq
with $\smash{Q_\kappa=Q+r\cdot \kappa T}$ and $\smash{\bar Q_\kappa=Q+\bar r\cdot\kappa T}$. This potential can be studied using the same techniques as those described in chapter 5. With the choice $\smash{\bar r=\bar r_c=\bar(4\pi/3,0)}$, the potential as a function of $r$ and for various temperatures is shown in Fig.~\ref{fig:su3_pot}. Note that, from charge conjugation invariance, we know that it is enough to plot the potential as a function of $r_3$ for $\smash{r_8=0}$. We find a first-order type confinement/deconfinement transition, in agreement with the lattice simulations. Using the values of the CF parameters obtained by fitting the lattice two-point functions in the presently considered renormalizaton scheme, we find a transition temperature $T_c\simeq 267$ MeV, quite close to the lattice value of $270$ MeV.\\

\appendix

\chapter{The Polyakov loop formula}
In the main text, we gave a hand-waving argument to justify the expression for the free-energy of a static color charge in a thermal bath of gluons in terms of the Polyakov loop. Here, we would like to sketch the actual derivation, explaining in particular the origin of the trace and the path-ordering.

The free-energy of a static color charge in a thermal bath of gluons is obtained from the associated partition function
\beq
Z={\rm Tr}\,e^{-\beta \hat{H}}\,,\label{eq:Z}
\eeq
where the Hamiltonian operator $\hat{H}$ and the appropriate trace ${\rm Tr}$ are determined from the canonical quantization of YM theory. More precisely, one should start from the classical Lagrangian in the presence of a static color charge density $\smash{\rho^a(\vec{x})=\rho^a\delta(\vec{x}-\vec{x}_0)}$, construct the classical Hamiltonian and promote it to an operator. The YM system is a constrained system, however, meaning that, in deriving the classical Hamiltonian, one obtains various constraints that need to be dealt with and which may restrict the quantum states that enter the trace. 

One such constraint is the fact that the momentum associated to the temporal component of the gauge field vanishes, so the latter is not a dynamical variable. A simple way to cope with this is to work in the temporal gauge $\smash{A^a_0=0}$. In this case, the Hamiltonian takes the form
\beq
H=\frac{1}{2}\int d^3x\,\Big(E^a_i(\vec{x})E^a_i(\vec{x})+B^a_i(\vec{x})B^a_i(\vec{x})\Big)\,,
\eeq
which can be immediately promoted into an Hamiltonian operator $\hat{H}$ after promoting the fields and associated momenta into operators as well. 

Another constraint comes from the non-Abelian Gauss law which reads classically
\beq
D_i^{ab}E^b_i(\vec{x})=\rho^a\delta(\vec{x}-\vec{x}_0)\,.
\eeq
At the quantum level, this constraint turns into a restriction on the states to be traced over. Naively, one could think of restricting to states $|\psi\rangle$ such that
\beq
\hat\rho^a(\vec{x})|\psi\rangle=\rho^a\delta(\vec{x}-\vec{x}_0)|\psi\rangle\,,\label{eq:GQ}
\eeq
with $\hat\rho^a(\vec{x})\equiv \hat D_i^{ab}\hat E^b_i(\vec{x})$ the color charge density operators.  However, integrating over space, one finds
\beq
\hat Q^a|\psi\rangle=\rho^a|\psi\rangle\,,
\eeq
and, because $\smash{[\hat Q^a,\hat Q^b]=if^{abc}\hat Q^c}$, this implies that
\beq
f^{abc}\rho^c|\varphi\rangle=0\,.
\eeq
Using $\smash{f^{abc}f^{abd}=N\delta_{cd}}$, this rewrites $\smash{\rho^d|\varphi\rangle=0}$, and, because at least one of the $\rho^d$'s is non-zero since we are consider a static source carrying color, this means that the only possible state is the null state, which is not normalizable and does no enter the evaluation of the trace.

The problem is of course that we have imposed too many restrictions on the states, because the charge density operators $\hat\rho^a(\vec{x})$ not all commute.  We should, instead, consider only those restrictions associated to commuting operators
\beq
\hat\rho^j(\vec{x})|\psi\rangle=\rho^j\delta(\vec{x}-\vec{x}_0)|\psi\rangle\,,\label{eq:W}
\eeq
where the labels $j$ refer to the Cartan subalgebra of commuting generators $t^j$. We have again $\smash{\hat Q^j|\psi\rangle=\rho^j|\psi\rangle}$, but, this time, because the charges $\hat Q^j$ commute with each other, the possible states do not need to vanish but only to belong to some representation $R$ of the color group in which case the $\rho^j$ are weights of that representation. In what follows, we assyme that the representation is irreducible.

The restriction to the states obeying (\ref{eq:GQ}) can be done by inserting the appropriate projector under the trace:
\beq
Z_\rho={\rm Tr}\,e^{-\beta \hat H}\int {\cal D}\theta^j\,\exp\left\{i\int d^3x\,\theta^j(\vec{x})(D_i^{jb}\hat E^b_i(\vec{x})-\rho^j\delta(\vec{x}-\vec{x}_0))\right\},\label{eq:2}
\eeq
where we note that there is no ambiguity in the writing of the projector since the charge density operators $D_i^{jb}\hat E^b_i(\vec{x})$, contrary to what would have happened in the case where we would have included all $D_i^{ab}\hat E^b_i(\vec{x})$.

What comes next is pretty standard and here we follow \cite{Laine:2016hma} for instance. The partition function (\ref{eq:2}) can be recast into a functional integral over the dynamical field variables $A_i^a(\vec{x})$ and their associated momenta $E_i^a(\vec{x})$ which get promoted into fields evolving in Euclidean time, $A_i^a(\tau,\vec{x})$ and $E_i^a(\tau,\vec{x})$, the former being periodic in $\tau$ with period $\beta$. In addition, there is an integration over the auxiliary variables $\theta^j(\vec{x})$ which also get promoted into fields evolving in Euclidean time, which we denote $A_4^j(\tau,\vec{x})$ in what follows for a reason that shall becomes clear below. 

The dependence in the fields $E_i^a(\tau,\vec{x})$ is quadratic, so the integration over $E_i^a(\tau,\vec{x})$ can be done analytically. One eventually arrives at
\beq
Z_{\rho}=\int {\cal D}[A_4^j,A_i^a]\,e^{-S_E[A_4^j,A_i^a]+i\rho^j\int_0^\beta d\tau\,A_4^j(\tau,\,\vec{x}_0)}\,.\label{eq:Zr}
\eeq
The action $S_E[A^a_4,A_i^a]$ is the Euclidean version of the YM action:
\beq
S_E[A^a_4,A_i^a]=\int_x\,\frac{1}{4}F_{\mu\nu}^a(\tau,\vec{x})F_{\mu\nu}^a(\tau,\vec{x})\,,
\eeq
with
\beq
F_{\mu\nu}^a=\partial_\mu A_\nu^a-\partial_\nu A_\mu^a+f^{abc}A_\mu^bA_\nu^c\,,
\eeq
and $\smash{\mu=4}$ or $\smash{\mu=i\in\{1,2,3\}}$.

At this point, various remarks are in order. First, the above formula does not look yet like the formula in the main text, involving the Polyakov loop. So, we still have to work a bit more. In fact, even in the absence of source (or when the source is colorless), the formula does not look like the one found in \cite{Laine:2016hma} because, here, only the color components of $A_4$ that are in the Cartan subalgebra are integrated over in (\ref{eq:Zr}). In fact, the starting point of \cite{Laine:2016hma} is a projector (\ref{eq:2}) where all charge density operators are considered which may seem contradicting what we have jus explained. However, we will now argue, first, that in the absence of source, the two approaches are perfectly equivalent as they can be connected by some manipulations, and, second that the same manipulations allow one to go from (\ref{eq:Zr}) to the well known formula involing the Polyakov loop.

Suppose first that $\rho$ in Eq.~(\ref{eq:Zr}) is a weight of the defining representation of SU(N). It is possible to check that this weight is connected to the other weights of the representation by certain geometrical transformations which are associated to particular color rotations known as Weyl transformations. More precisely, given another defining weight $\sigma$, one can find a Weyl transformation $W$ such that
\beq
\rho^j A_4^j(\tau,\,\vec{x}_0)=\sigma^j (A_4^W)^j(\tau,\vec{x}_0)\,.\label{eq:id3}
\eeq
Then, from Eq.~(\ref{eq:Zr}), and using the invariance of the integration measure and of the Euclidean action under color rotations, one arrives at the conclusion that $\smash{Z_\rho=Z_\sigma}$. This common value which we could denote $Z_{\rm def}$ can be conveniently written as
\beq
Z_{\rm def} & = & \frac{1}{N}\int {\cal D}A_4^j {\cal D}A_i^a\,e^{-S_E[A_4^j,A_i^a]}\sum_\rho e^{i\rho^j\int_0^\beta d\tau\, A^j_4(\tau,\,\vec{x}_0)}\nonumber\\
& = &  \frac{1}{N}\int {\cal D}A_4^j {\cal D}A_i^a\,e^{-S_E[A_4^j,A_i^a]}{\rm tr}\,e^{i\int_0^\beta d\tau\, A^j_4(\tau,\,\vec{x}_0)t^j}\,,
\eeq
where the $t^j$ are the generators of the defining representation. 

In a sense, this result is reassuring: based on the premise that color is not observable, one should not be able to distinguish the various color states of a representation by using an observable such as the free-energy. We mention, however, that the identity $\smash{Z_\rho=Z_\sigma}$ is not obvious beyond the fundamental representation because the relation (\ref{eq:id3}) is not guaranteed. In this case, the non-observability of color is rooted in the fact that the formula for the partition function should be extended in the presence of a matrix density which is proportionnal to the identity in the (color) space of the chosen representation. One then finds, in general
\beq
Z_R=\frac{1}{d_R}\int {\cal D}A_4^j {\cal D}A_i^a\,e^{-S_E[A_4^j,A_i^a]}{\rm tr}\,e^{i\int_0^\beta d\tau\, A^j_4(\tau,\,\vec{x}_0)t^j_R}\,,
\eeq
where the $t^j_R$ are the commuting generators in the chosen representation and $d_R$ the dimension of the latter. 

The above formulas still differ from the one,  Eq.~(\ref{eq:interpretation}), given in the main text. The two main differences are the absence of path-ordering and the fact that only the Cartan components of the temporal gluon field $A_4^j$ are integrated over. These two issues are in fact related as we now discuss. First, because $A^j_4t^j_R$ is in the Cartan subalgebra, it does not cost much to add a time-ordering
\beq
Z_R=\frac{1}{d_R}\int {\cal D}A_4^j {\cal D}A_i^a\,e^{-S_E[A_4^j,A_i^a]}{\rm tr}\,{\cal P}\,e^{i\int_0^\beta d\tau\, A^j_4(\tau,\,\vec{x}_0)t^j_R}\,.\label{eq:temp}
\eeq
Let us now argue that, up to a volume factor, this is nothing but
\beq
\frac{1}{d_R}\int {\cal D}A_4^a {\cal D}A_i^a\,e^{-S_E[A_4^a,A_i^a]}{\rm tr}\,{\cal P}\,e^{i\int_0^\beta d\tau\, A^a_4(\tau,\,\vec{x}_0)t^a_R}\,.\label{eq:latter}
\eeq
Since a gauge-field can always be gauge-transformed such that its temporal component sits in the Cartan subalgebra, Eq.~(\ref{eq:temp}) can be seen as a (partial) gauge-fixing of Eq.~(\ref{eq:latter}). Of course, this would require a more careful analysis, but this goes beyond the scope of these notes. Assuming that the argument goes through, it follows that
\beq
Z_R & = &  \frac{1}{{\rm Vol}}\frac{1}{d_R}\int {\cal D}A_4^a {\cal D}A_i^a\,e^{-S_E[A_4^a,A_i^a]}{\rm tr}\,{\cal P}\,e^{i\int_0^\beta d\tau\, A^a_4(\tau,\,\vec{x}_0)t^a_R}\,,
\eeq
where ${\rm Vol}$ is the volume of the gauge subgroup needed to map arbitary configurations onto configurations with temporal components in the Cartan subalgebra. The latter volume is infinite which is annoying. However, what one is interested in is the ratio to the partition function in the absence of source and then
\beq
\frac{Z_R}{Z_0}=e^{-\beta(f_R-f_0)}=\frac{\frac{1}{d_R}\int {\cal D}A_4^a {\cal D}A_i^a\,e^{-S_E[A_4^a,A_i^a]}{\rm tr}\,{\cal P}\,e^{i\int_0^\beta d\tau\, A^a_4(\tau,\,\vec{x}_0)t^a_R}}{\int {\cal D}A_4^a {\cal D}A_i^a\,e^{-S_E[A_4^a,A_i^a]}}\,.
\eeq
This is the well-known formula relating the Polyakov loop to the free-energy cost for bringing the color charge into the medium.

\chapter{Advanced Color Algebra}

Here, we discuss some useful notions related to the color algebra, including representations, weights and roots, see for instance Ref.~\cite{zuber} for more details.

\section{Representations of the su(N) Algebra}
In what follows, SU(N) denotes the group of special unitary matrices $U$ obeying
\beq
UU^\dagger=\mathds{1} \quad {\rm and} \quad {\rm det}\,U=1\,.
\eeq

\subsection{su(N) Algebra}
Since SU(N) is a Lie group, one can consider elements $\smash{U\approx 1+iX}$ that are infinitesimally close to the identity, in which case
\beq
X^\dagger=X \quad {\rm and} \quad {\rm tr}\,X=0\,.
\eeq
This defines a finite-dimensional vector space, denoted su(N), the elements of which are customary written as
\beq
X=X^a t^a\,,
\eeq
where $\smash{X^a\in\mathds{R}}$ and the $t^a$'s form a basis of {\it generators}. The Lie group nature of SU(N) implies that su(N) is in fact a Lie algebra. In particular, it is stable under commutation:
\beq
[t^a,t^b]=if^{abc}t^c\,.\label{eq:structure}
\eeq
The coefficients $\smash{f^{abc}\in\mathds{R}}$ are known as the {\it structure constants.}

To take a few examples, a basis of the su($2$) algebra can be constructed in terms of the Pauli matrices as $\smash{t^a\equiv \sigma_a/2}$, with $\smash{a=1,2,3}$ and where
\beq
\sigma_1=\left(
\begin{array}{cc}
0 & 1\\
1 & 0
\end{array}
\right), \quad \sigma_2=\left(
\begin{array}{cc}
0 & -i\\
i & 0
\end{array}
\right), \quad \sigma_3=\left(
\begin{array}{cc}
1 & 0\\
0 & -1
\end{array}
\right).\label{eq:pauli}
\eeq
As it is well known, $\smash{[t^a,t^b]=i\varepsilon^{abc}t^c}$ and thus $\smash{f^{abc}=\varepsilon^{abc}}$ in this case. As for the su($3$) algebra, a similar construction $\smash{t^a=\lambda_a/2}$ with $\smash{a=1,\dots,8}$, holds in terms of the Gell-Mann matrices
\beq
\lambda_1 & \!\!=\!\! & \left(
\begin{array}{ccc}
0 & 1 & 0\\
1 & 0 & 0\\
0 & 0 & 0
\end{array}
\right),\,\, \,\lambda_2=\left(
\begin{array}{ccc}
0 & -i & 0\\
i & 0 & 0\\
0 & 0 & 0
\end{array}
\right),\,\,\, \lambda_3=\left(
\begin{array}{ccc}
1 & 0 & 0\\
0 & -1 & 0\\
0 & 0 & 0
\end{array}
\right),\nonumber\\
\lambda_4 & \!\!=\!\! & \left(
\begin{array}{ccc}
0 & 0 & 1\\
0 & 0 & 0\\
1 & 0 & 0
\end{array}
\right),\,\,\, \lambda_5= \left(
\begin{array}{ccc}
0 & 0 & -i\\
0 & 0 & 0\\
i & 0 & 0
\end{array}
\right),\nonumber\\
\lambda_6 & \!\!=\!\! &\left(
\begin{array}{ccc}
0 & 0 & 0\\
0 & 0 & 1\\
0 & 1 & 0
\end{array}
\right),\,\,\, \lambda_7=\left(
\begin{array}{ccc}
0 & 0 & 0\\
0 & 0 & -i\\
0 & i & 0
\end{array}
\right),\,\,\, \lambda_8=\frac{1}{\sqrt{3}}\left(
\begin{array}{ccc}
1 & 0 & 0\\
0 & 1 & 0\\
0 & 0 & -2
\end{array}
\right)\!\,.\label{eq:gellmann}\nonumber\\
\eeq

\subsection{Representations}
Colored fields transform according to  finite-dimensional, unitary {\it representations} of the SU(N) group. For a field $\phi$ in a representation $R$, the possible color states are taken within a certain vector space ${\cal V}_R$ of dimension $d_R$, known as the {\it representation space.} The transformation of these color states under (infinitesimal) color transformations $U=1+i\theta^at^a$ writes
\beq
\phi'=\phi+i\theta^a t^a_R\phi\,,
\eeq
where the $t^a_R$'s denote a set of hermitian matrices acting on ${\cal V}_R$ which comply with the same commutation relations as the defining $t^a$'s:
\beq
[t^a_R,t^b_R]=if^{abc}t^c_R\,.\label{eq:rep}
\eeq
For instance, a quark field $\psi$ transforms in the fundamental or {\it defining representation,} $\smash{R={\rm def}}$. In this case, the representation space is $\smash{{\cal V}_{\rm def}=\mathds{C}^N}$ and $\smash{t^a_{\rm def}=t^a}$, so that
\beq
\psi'=\psi+i\theta^a t^a\psi\,.\label{eq:defining}
\eeq
That $t^a_{\rm def}$ obeys (\ref{eq:rep}) follows trivially from (\ref{eq:structure}).

Besides the defining representation (\ref{eq:defining}), the most useful one is certainly the {\it adjoint representation,} $\smash{R={\rm adj}}$. In this case, the representation space is the Lie algebra itself, $\smash{{\cal V}_{\rm adj}=su(N)}$, and $\smash{t^a_{\rm adj}=[t^a,\,\,]}$, such that
\beq
X'=X+i\theta^a t^a_{\rm ad} X=X+i\theta^a[t^a,X]\,.\label{eq:adjoint}
\eeq
Recall that gauge fields transform under the adjoint representation.\\

\noindent{$\diamond\diamond\diamond\,\,${\bf Problem B.1} Verify that $t^a_{\rm adj}$ complies with (\ref{eq:rep}). \underline{Tip:} use the Jacobi identity $\smash{[X,[Y,Z]]+[Y,[Z,X]]+[Z,[X,Y]]=0}$.}$\,\,\diamond\diamond\diamond$\\

For later purpose, it will be useful to bear in mind the finite versions of the infinitesimal transformations (\ref{eq:defining}) and (\ref{eq:adjoint}):
\beq
\psi' & \!\!\!=\!\!\! & e^{i\theta^a t^a}\psi\,,\nonumber\\
X' & \!\!\!=\!\!\! & e^{i[\theta^a t^a,\,\,]}X=e^{i\theta^a t^a}Xe^{-i\theta^a t^a}\,.
\eeq

\section{Weights and Roots}
To continue our study of color representations, it will be useful to adopt a quantum-mechanical language/notation from now on. For a given representation $R$, the representation space ${\cal V}_R$ can be seen as the space of color states which we denote $|\phi\rangle$ in what follows. The (hermitian) generators $t^a_R$ can be interpreted as observables that measure the color charges of these states. 

As usual in quantum mechanics, it is convenient to choose a collection of observables that commute with each other. This allows one to define a basis of states which all have well defined quantum numbers. Take for instance the case of a non-relativistic particle in a central potential. We like to introduce the basis of states $|E,\ell,m\rangle$ that simultaneously diagonalize $H$, $L^2$ and $L_z$:
\beq
H|E,\ell,m\rangle & \!\!\!=\!\!\! & E|E,\ell,m\rangle\,,\nonumber\\
L^2|E,\ell,m\rangle & \!\!\!=\!\!\! & \ell(\ell+1)|E,\ell,m\rangle\,,\label{eq:ex}\\
L_z|E,\ell,m\rangle & \!\!\!=\!\!\! & m|E,\ell,m\rangle\,.\nonumber
\eeq
Notice that the triplet $(E,\ell,m)$ which collects the various quantum numbers, is used to label the states of the basis. As we now describe, something similar can be done within any color representation.

\subsection{Weights of a Representation}
It is well known that one can choose the basis $\{t^a\}$ of the su(N) algebra  such that $\smash{d_C=N-1}$ of the generators commute with each other. These generators, denoted $t^j$ in what follows, generate an abelian subalgebra known as the {\it Cartan subalgebra.} The corresponding $t^j_R$'s in a given representation also commute with each other. They can then be diagonalized simultaneously as
\beq
t^j_R|\rho\rangle=\rho^j |\rho\rangle\,.
\eeq
Here $\rho$ is a vector in $\mathds{R}^{d_C}$ known as {\it $R$-weight} or {\it weight of the representation $R$.} It collects the various quantum numbers $\rho^j$ associated to each of the $t^j_R$ and allows one to label the various states, just as in (\ref{eq:ex}). In some cases, there might exist some degeneracy, with various linearly independent states producing the same quantum numbers. In this case, one needs to introduce one extra label $|\rho;\eta\rangle$ to lift the degeneracy. We will see an important example below.

As an example of weights, let us identify the defining weights of su(2). In this case, $\smash{d_C=1}$ and we can choose $\smash{t^i_{\rm def}=\sigma_3/2}$ which is already in diagonal form, see Eq.~(\ref{eq:pauli}). This gives immediately the two weights, which are numbers in this case, corresponding to the two diagonal elements, $1/2$ and $-1/2$, of $\sigma_3/2$. In the case of su(3), the commuting subalgebra is generated by $\lambda_3/2$ and $\lambda_8/2$ so that $\smash{d_C=2}$ and the weights are bidimensional vectors. There are three defining weights whose components can be read off from the diagonal components of $\lambda_3/2$ and $\lambda_8/2$. For instance, by looking at the first diagonal element of each of these matrices
\beq
\frac{\lambda_3}{2}=\left(
\begin{array}{ccc}
\boxed{\frac{1}{2}} & 0 & 0\\
0 & -\frac{1}{2} & 0\\
0 & 0 & 0
\end{array}
\right),\,\,\,\frac{\lambda_8}{2}=\left(
\begin{array}{ccc}
\boxed{\frac{1}{2\sqrt{3}}} & 0 & 0\\
0 & \frac{1}{2\sqrt{3}} & 0\\
0 & 0 & -\frac{1}{\sqrt{3}}
\end{array}
\right),
\eeq
one finds the weight
\beq
\left(\frac{1}{2},\frac{1}{2\sqrt{3}}\right)\,.
\eeq
The two other weights are obtained in a similar fashion by looking at the second and third diagonal elements. One finds the two other weights
\beq
\left(-\frac{1}{2},\frac{1}{2\sqrt{3}}\right) \quad {\rm and} \quad \left(0,-\frac{1}{\sqrt{3}}\right)\,.
\eeq
It is customary and quite useful to represent the weights in a diagram, see Fig.~\ref{fig:diag} for a representation of the su(3) defining weight diagram.\\

\begin{figure}[t]
\begin{center}
\includegraphics[height=0.5\textheight]{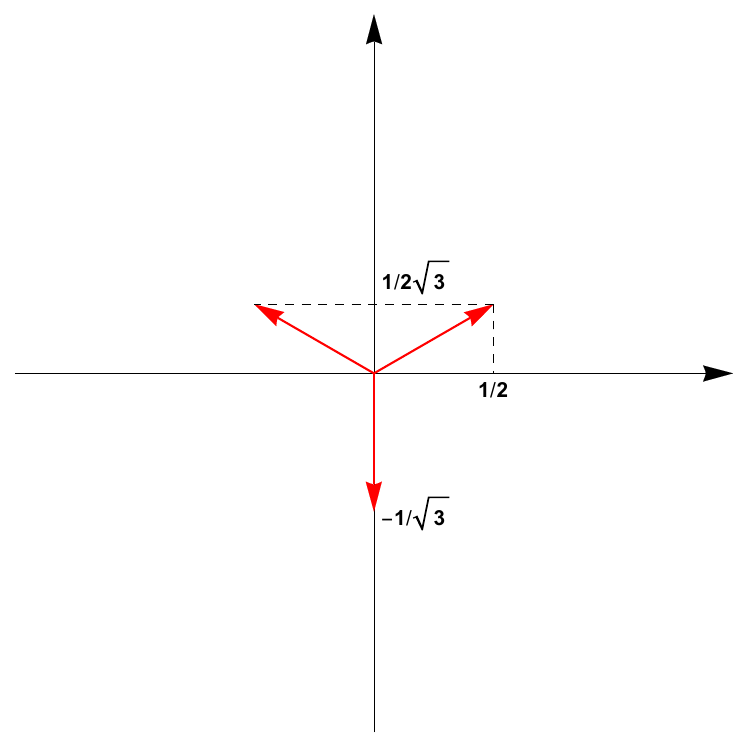}
\caption{Definining weights the su($3$) algebra.}\label{fig:diag}
\end{center}
\end{figure}

\noindent{$\diamond\diamond\diamond\,\,${\bf Problem B.2} Verify that the defining weights of SU(N) with $\smash{N=2}$ or $\smash{N=3}$ satisfy the relations}
\beq
\rho^2=\frac{1}{2}\left(1-\frac{1}{N}\right) \quad \mbox{and} \quad \rho\cdot\rho'=-\frac{1}{2N}\,, \quad \mbox{for } \rho\neq\rho'.
\eeq
These formulas are in fact valid for any value of $N$. Could you show it? \underline{Tip:} start by generalizing the Gell-Mann matrices to the SU(N) case.$\,\,\diamond\diamond\diamond$

\subsection{Roots of the Algebra}
Let us now consider the weights of the adjoint representation $t^a_{\rm adj}=[t^a,\,\,]$. We shall denote them by $\kappa$ such that
\beq
[t^j,t^\kappa]=\kappa^j t^\kappa\,.
\eeq
The adjoint weights can be of two types:

\begin{itemize}

\item[$\bullet$] Since the $t^j$ commute with each other, any $t^j$ is a state with vanishing color charges and therefore with vanishing adjoint weight. Because there are $N-1$ such linearly independent states sharing the same adjoint weight, we have here an example of degeneracy: if we want to label these states using their charge, we should keep track of the index $j$ by writing something like $\kappa=(0;j)$ or $\kappa=0^{(j)}$. This notation is useful when one wants to perform calculation in a compact way, as we have illustrated in chapter 6. Here, we shall stick with the notation $t^{j}$ for these states.

\item[$\bullet$] There are also non-vanishing adjoint weights, known as {\it roots} and denoted generically as $\alpha$. They are such that
\beq
[t^j,t^\alpha]=\alpha^j t^\alpha\,.\label{eq:ja}
\eeq
There is no degeneracy here meaning that there is a unique $t^\alpha$ associated to a given $\alpha$. It can also be shown that, when $\alpha$ is a root, then $-\alpha$ is also a root and $t^{-\alpha}$ can always be chosen to coincide with $(t^\alpha)^\dagger$ and such that
\beq
[t^\alpha,t^{-\alpha}]=\alpha^j t^j\,.\label{eq:aa}
\eeq
\end{itemize}

\noindent{$\diamond\diamond\diamond\,\,${\bf Problem B.3} Show indeed that $(t^\alpha)^\dagger$ has charges $-\alpha^j$. \underline{Tip:} apply the dagger on the appropriate equation. Show also that $[t^\alpha,t^{-\alpha}]$ has vanishing charges. \underline{Tip:} use the Jacobi identity.}$\,\,\diamond\diamond\diamond$\\

To illustrate the above general notions, consider the su($2$) algebra. In this case, $\smash{d_C=1}$ so the weights are one-dimensional vectors (numbers). The well known ladder operators $\smash{t_\pm\equiv (\sigma_1\pm i\sigma_2)/2\sqrt{2}}$ are such that
\beq
\left[\frac{\sigma_3}{2},t_\pm\right]=\pm t_\pm\,.
\eeq
There is thus one vanishing adjoint weight associated to $\sigma_3/2$ itself and two roots $+1$ and $-1$. In the case of su($3$), $\smash{d_C=2}$ so the weights are two-dimensional vectors and there are two vanishing weights, associated to $\lambda_3/2$ and $\lambda_8/2$. The SU($3$) roots have already been studied in chapter $3$. We here recall the corresponding root diagram represented in Fig.~\ref{fig:root}.\\

\begin{figure}[t]
\begin{center}
\includegraphics[height=0.5\textheight]{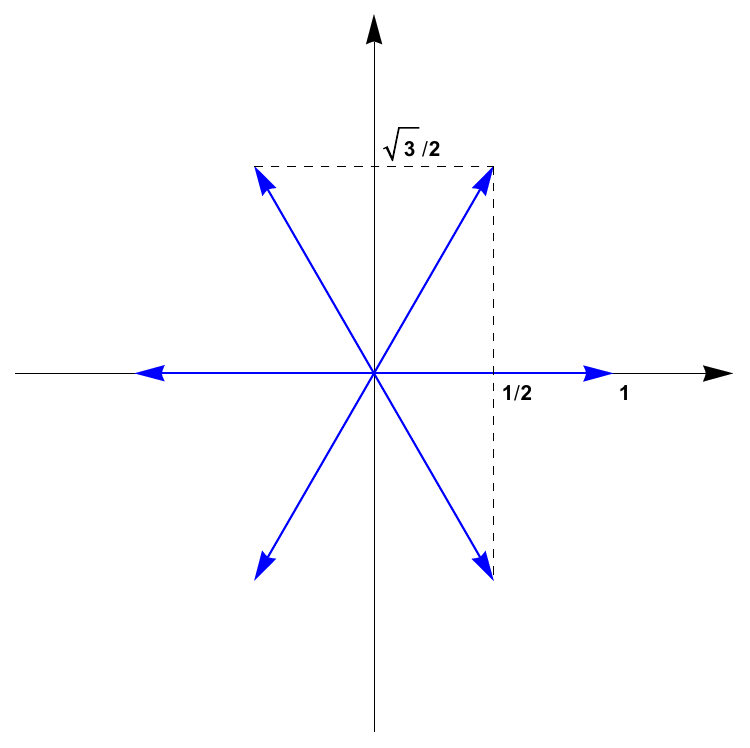}
\caption{Roots of the su($3$) algebra.}\label{fig:root2}
\end{center}
\end{figure}

The collection $\{t^j,t^\alpha\}$ forms a new basis of the su(N) algebra, known as {\it Cartan-Weyl basis} (more precisely, it is a basis of the complexified algebra). Any element of the original algebra can decomposed in this basis as
\beq
X=X^jt^j+X^\alpha t^\alpha\,,
\eeq
where a summation over $j$ and $\alpha$ is implied, and we recall that for each $\alpha$ in the sum there is also $-\alpha$. Since $X^\dagger=X$, $(t^j)^\dagger=t^j$ and $(t^\alpha)^\dagger=t^{-\alpha}$, we have $X^j\in\mathds{R}$, while $X^\alpha\in\mathds{C}$ with $(X^\alpha)^*=X^{-\alpha}$.\\

\subsection{Relation between Roots and Weights}

\begin{figure}[t]
\begin{center}
\includegraphics[height=0.4\textheight]{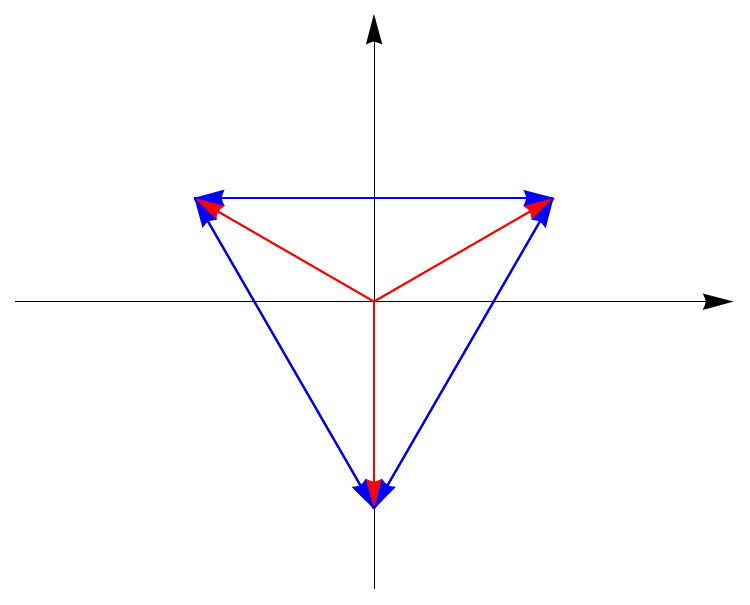}
\caption{Relation between the roots and the defining weights.}\label{fig:join}
\end{center}
\end{figure}

The roots and the defining weights are actually connected to each other, see Fig.~\ref{fig:join}.\\

\noindent{$\diamond\diamond\diamond\,\,${\bf Problem B.4} Consider SU(N) for $\smash{N=2}$ or $\smash{N=3}$. Verify that the roots are obtained by considering all possible differences of (non-equal) defining weights. This result is in fact valid for any value of $N$. Could you explain why? \underline{Tip:} think how the defining and adjoint representations are connected to each other.}$\,\,\diamond\diamond\diamond$\\ 

By combining this result with the one in Problem B.2, we can evaluate the scalar product of a root $\alpha$ and a defining weight $\rho$. Since the root writes as the difference of two weights, $\alpha\equiv\rho_\alpha-\bar\rho_\alpha$, with $\smash{\rho_\alpha\neq\bar\rho_\alpha}$we need to distinguish three cases:


\begin{itemize}
\item[$\bullet$] if $\smash{\rho\neq\rho_\alpha}$ and $\smash{\rho\neq\bar\rho_\alpha}$, then
\beq
\rho\cdot\alpha=\rho\cdot\rho_\alpha-\rho\cdot\bar\rho_\alpha=-\frac{1}{2N}-\left(-\frac{1}{2N}\right)=0\,;\nonumber
\eeq
\item[$\bullet$] if $\smash{\rho=\rho_\alpha}$ but $\smash{\rho\neq\bar\rho_\alpha}$, then
\beq
\rho\cdot\alpha=\rho\cdot\rho_\alpha-\rho\cdot\bar\rho_\alpha=\frac{1}{2}\left(1-\frac{1}{N}\right)-\left(-\frac{1}{2N}\right)=\frac{1}{2}\,;\nonumber
\eeq
\item[$\bullet$] if $\smash{\rho\neq\rho_\alpha}$ but $\smash{\rho=\bar\rho_\alpha}$, then
\beq
\rho\cdot\alpha=\rho\cdot\rho_\alpha-\rho\cdot\bar\rho_\alpha=\left(-\frac{1}{2N}\right)-\frac{1}{2}\left(1-\frac{1}{N}\right)=-\frac{1}{2}\,.\nonumber
\eeq
\end{itemize}
This shows that $\rho\cdot\alpha$ can only take three different values: $-1/2$, $0$ or $1/2$. One also shows in the same way that the su($n$) roots are all of unit length, $\alpha^2=1$ (do it!).

\section{Applications}
In the search for confining configurations, we needed to analyze more precisely the structure of the group ${\cal G}_0$ of periodic gauge transformations and more precisely the transformations that transform configurations of the form
\beq
A_\mu(x)=4\pi T\delta_{\mu0}\,x^j t^j\,,\label{eq:forma}
\eeq
into configurations of the same form, with possibly a different $x^j$.

\subsection{Color Rotation associated to a Root}
Let us first consider color rotations that involve the $t^\alpha$'s and see whether they stabilize the Cartan subalgebra generated by the $t^j$. Since the $t^\alpha$ do not commute in general, let us consider a rotation involving a single root
\beq
W_\alpha(z)\equiv e^{i(zt^\alpha+z^* t^{-\alpha})}
\eeq
The reason why we need to include $t^{-\alpha}$ is that the phase of the transformation needs to be an element of the su(N) algebra.

In principle, one can evaluate the action of $W_\alpha(z)$ on an arbitrary element of the algebra. Here, our main goal is to determine the action of the rotations on elements of the form $X=X^jt^j$. Forgetting $X^j$ for the moment, we then write
\beq
W_\alpha(z)t^j W^\dagger_\alpha(z) & = & e^{i(zt^\alpha+z^* t^{-\alpha})}t^j e^{-i(zt^\alpha+z^* t^{-\alpha})}\nonumber\\
& = & e^{i(zt^\alpha_{\tiny adj}+z^* t^{-\alpha}_{\tiny adj})}t^j\nonumber\\
& = & e^{i[zt^\alpha+z^* t^{-\alpha},\,\,]}t^j\,.
\eeq
To pursue the calculation, we expand the exponential
\beq
W_\alpha(z)t^j W^\dagger_\alpha(z)=\sum_{n=0}^\infty \frac{i^n}{n!}[zt^\alpha+z^* t^{-\alpha},\,\,]^nt^j\,,\label{eq:exp}
\eeq
where $[A,\,\,]^nB$ denotes the $n$-times nested commutator
\beq
[A,[A,[A,\dots [A,B]]]]\,,
\eeq
with $[A,\,\,]^0B=B$ and $[A,\,\,]^1B=[A,B]$.

To seek some guidance, let us evaluate the first iterations. Using Eqs.~(\ref{eq:ja}) and (\ref{eq:aa}), we find
\beq
{[}zt^\alpha+z^* t^{-\alpha},\,\,{]}^0t^j & = & t^j\,,\nonumber\\
{[}zt^\alpha+z^* t^{-\alpha},\,\,{]}^1t^j & = & \alpha^j(z^*t^{-\alpha}-zt^\alpha)\,,\\
{[}zt^\alpha+z^* t^{-\alpha},\,\,{]}^2t^j & = & 2\alpha^j|z|^2\alpha^k t^k\,.\nonumber
\eeq

\noindent{$\diamond\diamond\diamond\,\,${\bf Problem B.5} Show that}
\beq
{[}zt^\alpha+z^* t^{-\alpha},\,\,{]}^{2p+1}t^j & = & \alpha^j(2|z|^2)^p(z^*t^{-\alpha}-zt^\alpha)\,,p\geq 0\,,\nonumber\\
{[}zt^\alpha+z^* t^{-\alpha},\,\,{]}^{2p}t^j & = & \alpha^j(2|z|^2)^p\alpha^k t^k\,, p>0\,.
\eeq
\underline{Tip:} proceed by induction and recall that $\smash{\alpha^2=1}$. $\,\,\diamond\diamond\diamond$\\

Coming back to Eq.~(\ref{eq:exp}), we find
\beq
W_\alpha(z)t^j W^\dagger_\alpha(z) & = & t^j+\sum_{p=0}^\infty \frac{i^{2p+1}}{(2p+1)!}[zt^\alpha+z^* t^{-\alpha},\,\,]^{2p+1}t^j\nonumber\\
& & + \sum_{p=1}^\infty \frac{i^{2p}}{(2p)!}[zt^\alpha+z^* t^{-\alpha},\,\,]^{2p}t^j\nonumber\\
& = & t^j+\alpha^j\sum_{p=0}^\infty \frac{i^{2p+1}}{(2p+1)!}(2|z|^2)^p(z^*t^{-\alpha}-zt^\alpha)\nonumber\\
& & +\,\alpha^j\sum_{p=1}^\infty \frac{i^{2p}}{(2p)!}(2|z|^2)^p\alpha^k t^k\nonumber\\
& = & t^j+\alpha^j\sum_{p=0}^\infty \frac{i^{2p+1}}{(2p+1)!}(\sqrt{2}|z|)^{2p+1}\frac{z^*t^{-\alpha}-zt^\alpha}{\sqrt{2}|z|}\nonumber\\
& & +\,\alpha^j\sum_{p=1}^\infty \frac{i^{2p}}{(2p)!}(\sqrt{2}|z|)^{2p}\alpha^k t^k\,.
\eeq
We can now sum the series and we arrive at
\beq
W_\alpha(z)t^j W^\dagger_\alpha(z) & = & t^j+\alpha^j\big(\cos(\sqrt{2}|z|)-1\big)\,\alpha^k t^k\nonumber\\
& & +\,\alpha^j\sin(\sqrt{2}|z|)\,\frac{z^*t^{-\alpha}-zt^\alpha}{\sqrt{2}|z|}\,.
\eeq

\subsection{Weyl Transformations}
We note that for specific values of $|z|$ the previous transformation stabilizes the Cartan subalgebra. One possibility is to choose $|z|=\sqrt{2}\pi n$ for which the transformation becomes trivial $W_\alpha(z)t^j W^\dagger_\alpha(z)=t^j$. A more interesting possibility is $|z|=\sqrt{2}\pi (n+1/2)$, in which case the transformation reads
\beq
W_\alpha(z) t^j W^\dagger_\alpha(z)= t^j-2 \alpha^j\alpha^k t^k\,,
\eeq
For a generic element $X^jt^j$ of the Cartan subalgebra, this gives
\beq
W_\alpha X^jt^j W^\dagger_\alpha & = &  X^jt^j-2 X^j\alpha^j\alpha^k t^k\nonumber\\
& = & (X^j-2X^k\alpha^k\alpha^j)t^j\,,
\eeq
and so a transformation of the coordinates as
\beq
X^j\to X^j-2X^k\alpha^k\alpha^j\,.
\eeq 
We note that this transformation does not change the projection of $X$ orthogonal to $\alpha$, while the projection of $X$ along $\alpha$ is just multiplied by a sign:
\beq
X^j\alpha^j\to X^j\alpha^j-2X^k\alpha^k\alpha^2=-X^j\alpha^j\,.
\eeq
Thus it corresponds to a reflection with respect to an hyperplane orthogonal to $\alpha$. This generalizes what we have discussed in the main text for the SU($2$) case and which we claimed to be true in the SU($3$) case.

\subsection{Winding Transformations}
We can now look for local gauge transformations that preserve the form (\ref{eq:forma}). Consider in particular the transformations
\beq
U(\tau)=\exp\left\{i4\pi\frac{\tau}{\beta}s^jt^j\right\},
\eeq
see Sec.~6.2.2, which act on (\ref{eq:forma}) as $x^j\to x^j+s^j$ (show it!). If we want these transformations to correspond to genuine, that is periodic, gauge transformations, there are restrictions on the possible $s^j$. Indeed, because $\smash{U(0)=\mathds{1}}$, periodicity is tantamount to
\beq
U(\beta)=e^{i4\pi s^jt^j}=\mathds{1}\,.
\eeq
To make most of this condition, it is convenient to evaluate it on a basis $|\rho\rangle$ that diagonalizes the defining action of the $t^j$'s, that is $t^j|\rho\rangle=\rho^j|\rho\rangle$. This leads to the condition $\smash{\exp\left\{i4\pi s^j\rho^j\right\}=1}$, that is
\beq
s\cdot\rho \in \frac{1}{2}\mathds{Z}\,.
\eeq
We know already solutions to this condition thanks to the result discussed in Sec.~B.2.3: any root and more generally any linear combination of roots with integer coefficients, in line with what was announced in Sec.~6.2.2. It can be shown that this is actually the more general solution.

Let us finally mention that requiring $U(\tau)$ to preserve the periodicity of the fields without being itself necessarily periodic leads to the condition (as we have shown in Sec.~6.3.1)
\beq
s\cdot\alpha \in \frac{1}{2}\mathds{Z}\,,
\eeq
which is in a sense dual to the one above since it is solved for any linear combination of the weights with integer coefficients. Since, as we have seen also, this condition characterizes the vertices of the lattice of Weyl chamber, we then deduce that this lattice is generated by the defining weights, as one can simply verify it in the SU($2$) and SU($3$) cases treated in the main text.

\chapter{Solutions to the Problems}

\section{Polyakov Loop and Center Symmetry}

\noindent{{\bf Problem 1.1:} That the periodicity conditions are preserved by the transformation $U(\tau,\vec{x})$ means that, for any $A_\mu(\tau,\vec{x})$ such that $\smash{A_\mu(\tau+\beta,\vec{x})=A_\mu(\tau,\vec{x})}$, we should also find $\smash{A^U_\mu(\tau+\beta,\vec{x})=A^U_\mu(\tau,\vec{x})}$. Making this latter condition more explicit, we find
\begin{eqnarray}
& & U(\tau+\beta,\vec{x})A_\mu(\tau+\beta,\vec{x})U^\dagger(\tau+\beta,\vec{x})+iU(\tau+\beta,\vec{x})\partial_\mu U^\dagger(\tau+\beta,\vec{x})\nonumber\\
& & \hspace{1.0cm}=\,U(\tau,\vec{x})A_\mu(\tau,\vec{x})U^\dagger(\tau,\vec{x})+iU(\tau,\vec{x})\partial_\mu U^\dagger(\tau,\vec{x})\,.\nonumber
\end{eqnarray}
Using the assumed periodicity of $A_\mu(\tau,\vec{x})$ together with the unitarity of $U(\tau+\beta,\vec{x})$ which implies $U(\tau+\beta,\vec{x})\partial_\mu U^\dagger(\tau+\beta,\vec{x})=-(\partial_\mu U(\tau+\beta,\vec{x})) U^\dagger(\tau+\beta,\vec{x})$, we rewrite this conveniently as
\begin{eqnarray}
& & U(\tau+\beta,\vec{x})A_\mu(\tau,\vec{x})U^\dagger(\tau+\beta,\vec{x})-i(\partial_\mu U(\tau+\beta,\vec{x})) U^\dagger(\tau+\beta,\vec{x})\nonumber\\
& & \hspace{1.0cm}=\,U(\tau,\vec{x})A_\mu(\tau,\vec{x})U^\dagger(\tau,\vec{x})+iU(\tau,\vec{x})\partial_\mu U^\dagger(\tau,\vec{x})\,.\nonumber
\end{eqnarray}
Then, upon left multiplication by $U^\dagger(\tau,\vec{x})$ and right multiplication by $U(\tau+\beta,\vec{x})$, this rewrites
\begin{eqnarray}
& & Z(\tau,\vec{x})A_\mu(\tau,\vec{x})-iU^\dagger(\tau,\vec{x})\partial_\mu U(\tau+\beta,\vec{x})\nonumber\\
& & \hspace{1.0cm}=\,A_\mu(\tau,\vec{x})Z(\tau,\vec{x})+i(\partial_\mu U^\dagger(\tau,\vec{x}))U(\tau+\beta,\vec{x})\,,\nonumber
\end{eqnarray}
where $\smash{Z(\tau,\vec{x})\equiv U^\dagger(\tau,\vec{x})U(\tau+\beta,\vec{x})}$. In fact, this identity rewrites solely in terms of $Z(\tau,\vec{x})$:
 \beq
 \partial_\mu Z(\tau,\vec{x})-i[A_\mu(\tau,\vec{x}),Z(\tau,\vec{x})]=0\,.\nonumber
 \eeq 
 Now, since the latter identity is valid for any periodic gauge field, it should be valid for $\smash{A_\mu=0}$. From this we deduce that $Z$ is constant and that $[A_\mu,Z]=0$ for any periodic gauge field configuration. In particular this means that $Z$ should commute with any generator $t^a$ of the algebra, and, therefore, with any element of the SU(N) group. The only possibility for $Z$ is then to be of the form $e^{i\phi}\mathds{1}$ with $\phi=2\pi k/N$.}\\

\vglue2mm

\noindent{{\bf Problem 1.2:} 
Since
\beq
\sigma_3=\left(
\begin{array}{cc}
1 & 0\\
0 & -1
\end{array}\right),\nonumber
\eeq
we have
\beq
U(\beta)=\exp\left\{i\theta\frac{\sigma_3}{2}\right\}=\left(
\begin{array}{cc}
e^{i\theta/2} & 0\\
0 & e^{-i\theta/2}
\end{array}\right).\nonumber
\eeq
For $\smash{\theta=\pm 2\pi}$, we find indeed $\smash{U(\beta)=-\mathds{1}}$, while for $\smash{\theta=\pm 4\pi}$, we find $\smash{U(\beta)=\mathds{1}}$.\\

\vglue2mm

\noindent{{\bf Problem 1.3:} Owing to the path ordering, we have
\beq
\frac{d}{d\beta}L[A^U]=iA^U_0(\beta,\vec{x})L[A^U]\,.\nonumber
\eeq
On the other hand
\beq
& & \frac{d}{d\beta}U(\beta)L[A]U^\dagger(0)\nonumber\\
& &  \hspace{0.5cm}=\,\frac{dU(\beta)}{d\beta}L[A]U^\dagger(0)+iU(\beta)A_0(\beta,\vec{x})L[A]U^\dagger(0)\nonumber\\
& &  \hspace{0.5cm}=\,\left[\frac{dU(\beta)}{d\beta}U^\dagger(\beta)+iU(\beta)A_0(\beta,\vec{x})U^\dagger(\beta)\right]U(\beta)L[A]U^\dagger(0)\nonumber\\
& &  \hspace{0.5cm}=\,i\left[U(\beta)A_0(\beta,\vec{x})U^\dagger(\beta)+iU(\beta)\frac{dU^\dagger(\beta)}{d\beta}\right]U(\beta)L[A]U^\dagger(0)\nonumber\\
& &  \hspace{0.5cm}=\,iA^U_0(\beta,\vec{x})U(\beta)L[A]U^\dagger(0)\,.\nonumber
\eeq
This shows that $L[A^U]$ and $U(\beta)L[A]U^\dagger(0)$ obey the same first order differential equation with respect to $\beta$. Since they are both equal to $\mathds{1}$ for $\beta=0$, we deduce that $\smash{L[A^U]=U(\beta)L[A]U^\dagger(0)}$.\\

\pagebreak

\noindent{{\bf Problem 1.4:} Consider a change of variables in the form a $k$-twisted gauge transformation $A_\mu\to A_\mu^U$. We find
\beq
\ell_{\rho,\theta} & = & \frac{1}{Z}\int_{p.b.c.}{\cal D}A^U\,e^{-S[A^U]+\rho e^{i\theta}\int_{\vec{x}}\Phi[A^U](\vec{x})}\,\Phi[A^U]\nonumber\\
& = & \frac{e^{i2\pi k/N}}{Z}\int_{p.b.c.}{\cal D}A\,e^{-S[A]+\rho e^{i\theta} e^{i2\pi k/N}\int_{\vec{x}}\Phi[A](\vec{x})}\,\Phi[A]\,,\nonumber
\eeq 
where we have used the invariance of the measure ${\cal D}A$ and the action $S[A]$, together with the transformation rule for $\Phi[A]$. It follows that
\beq
\ell_{\rho,\theta}=e^{i2\pi k/N}\ell_{\rho,\theta+2\pi k/N}\,.\nonumber
\eeq
In the case where center-symmetry is realized explicitly, the limit of $\ell_{\rho,\theta}$ as $\smash{\rho\to 0}$ does not depend on $\theta$: $\smash{\lim_{\rho\to 0}\ell_{\rho,\theta}=\ell}$. The above identity becomes a constraint for $\ell$ which forces it to vanish. In contrast, in the case where center-symmetry is broken in the Nambu-Goldstone sense, the limit of $\ell_{\rho,\theta}$ as $\smash{\rho\to 0}$ still depends on $\theta$: $\smash{\lim_{\rho\to 0}\ell_{\rho,\theta}=\ell_\theta}$. In this case, the above identity does not impose that the various possible limits $\ell_{\theta}$ should vanish, but connects them to each other as
\beq
e^{-i2\pi k/N}\ell_{\theta}=\ell_{\theta+2\pi k/N}\,.\nonumber
\eeq

\section{SU($2$) Confining Configurations}

\noindent{{\bf Problem 2.1:} We have to compute the rotational of $\vec{A}=(\vec{r}\times\vec{B})/2=B(y\vec{e}_x-x\vec{e}_y)/2$. We find
\beq
(\vec{\nabla}\times\vec{A})_x & \!\!=\!\! & \frac{\partial A_y}{\partial z}-\frac{\partial A_z}{\partial y}=0\,,\nonumber\\
(\vec{\nabla}\times\vec{A})_y & \!\!=\!\! & \frac{\partial A_z}{\partial x}-\frac{\partial A_x}{\partial z}=0\,,\nonumber\\
(\vec{\nabla}\times\vec{A})_z & \!\!=\!\! & \frac{\partial A_x}{\partial y}-\frac{\partial A_y}{\partial x}=B/2-(-B/2)=B\,.\nonumber
\eeq
\\

\noindent{{\bf Problem 2.2:} Since the configuration is constant, the path-ordering can be ignored and we have
\beq
\Phi[A] & \!\!=\!\! & \frac{1}{2}{\rm tr}\,e^{i\beta A_0}=\frac{1}{2}{\rm tr}\,e^{ir\frac{\sigma_3}{2}}\nonumber\\
& \!\!=\!\! & \frac{1}{2}{\rm tr}\left(
\begin{array}{cc}
e^{ir/2} & 0\\
0 & e^{-ir_3/2}\\
\end{array}
\right)=\cos(r/2)\,.\nonumber
\eeq
\\

\noindent{{\bf Problem 2.3:} The looked for transformation $U(\tau,\vec{x})$ should be such that for any configuration of the form $\smash{A_\mu(\tau,\vec{x})=T\delta_{\mu0}\,r\sigma_3/2}$, the transformed configuration $A_\mu^U(\tau,\vec{x})$ should be also of this form, that is
\beq
A^U_\mu(\tau,\vec{x})=T\delta_{\mu0}\,\theta\frac{\sigma_3}{2}\,,\nonumber
\eeq
for some $\theta$. Making the left-hand side explicit, this rewrites
\beq
T\delta_{\mu0}\,rU(\tau,\vec{x})\frac{\sigma_3}{2}U^\dagger(\tau,\vec{x})+iU(\tau,\vec{x})\partial_\mu U^\dagger(\tau,\vec{x})=T\delta_{\mu0}\,\theta\frac{\sigma_3}{2}\,,\nonumber
\eeq
which should be true for any value of $\mu$ and any value of $r$, with $\theta$ depending on $r$. In particular, for $\smash{\mu=j\neq0}$, the condition reads $iU(\tau,\vec{x})\partial_j U^\dagger(\tau,\vec{x})=0$ which basically means that $U(\tau,\vec{x})$ depends only on $\tau$ and we denote it more simply as $U(\tau)$. Taking now $\smash{\mu=0}$ in the equation above, this transformation $U(\tau)$ is seen to obey the equation
\beq
T\,rU(\tau)\frac{\sigma_3}{2}U^\dagger(\tau)+iU(\tau)\frac{d}{d\tau}U^\dagger(\tau)=T\,\theta\frac{\sigma_3}{2}\,,\nonumber
\eeq
which rewrites as
\beq
\frac{d}{d\tau}U(\tau)=iT\left[\theta\frac{\sigma_3}{2}U(\tau)-U(\tau)r\frac{\sigma_3}{2}\right].\nonumber
\eeq
This equation should be valid for any value of $r$ with $\theta$ depending on $r$. In particular, for $\smash{r=0}$, we deduce that there is some $\theta$ such that
\beq
\frac{d}{d\tau}U(\tau)=iT\,\theta\frac{\sigma_3}{2}U(\tau)\,.\nonumber
\eeq
This equation is solved as
\beq
U(\tau)=We^{i\theta\frac{\tau}{\beta}\frac{\sigma_3}{2}},\nonumber
\eeq
where $W$ is a color rotation. Since both $U(\tau)$ and $e^{i\theta\frac{\tau}{\beta}\frac{\sigma_3}{2}}$ preserve direction $3$, we should have $W\sigma_3 W^\dagger=\sigma_3$ which implies that $W$ is either the identity or the Weyl transformation discussed in the main text.\\\\

\noindent{{\bf Problem 2.4:} Recall that $\Phi[A]=\cos(r/2)$. Thus, for $\smash{r=\pi+2\pi n}$, we find indeed
 \beq
\cos\left(\frac{\pi}{2}+n\pi\right)=(-1)^n\cos\left(\frac{\pi}{2}\right)=0.\nonumber
 \eeq
\\

\noindent{{\bf Problem 2.5:} For the considered configurations, we have
\beq
A^C_\mu(x)=T\delta_{\mu0}\,(-r)\frac{\sigma_3}{2}\,,\nonumber
\eeq
and thus
\beq
\Phi[A^C]=\cos((-r)/2)=\cos(r/2)=\Phi[A]\,.\nonumber
\eeq
This is expected since YM theory being charge conjugation invariant, there should be no way for a thermal bath of gluons to distinguish between a quark and an antiquark.\\

\section{Thermal Gluon Average}

\noindent{{\bf Problem 3.1:} To linear order in $\theta^a$, we can write
\beq
A^{U_0}_\mu & = & (\mathds{1}+i\theta^at^a)A_\mu(\mathds{1}-i\theta^at^a)+i(\mathds{1}+i\theta^at^a)\partial_\mu(\mathds{1}-i\theta^at^a)\nonumber\\
& = & A_\mu+i(\theta^at^a) A_\mu-iA_\mu(\theta^at^a)+\partial_\mu(\theta^at^a)\nonumber\\
& = & A_\mu+(\partial_\mu-i[A_\mu,\,\,])(\theta^at^a)=A_\mu+D_\mu[A](\theta^at^a)\,.\nonumber
\eeq
This rewrites
\beq
\delta A^{U_0}_\mu=-iD_\mu[A]\delta U_0\,,\nonumber
\eeq
which implies
\beq
\delta \partial_\mu A^{U_0}_\mu=-i\partial_\mu (D_\mu[A]\delta U_0)\,,\nonumber
\eeq
and from which one reads $\left.\delta \partial_\mu A^{U_0}_\mu /\delta U_0\right|_{U_0=\mathds{1}}$ and deduces that, up to an non-relevant multiplication constant, $J[A]={\rm det}\,\partial_\mu D_\mu[A]$.\\\\

\noindent{{\bf Problem 3.2:} The Faddeev-Popov gauge-fixing part reads now
\beq
J_{\bar A}[A]\,\delta(D_\mu[\bar A](A-\bar A))\,,\nonumber
\eeq
with
\beq
J_{\bar A}[A]=\left.{\rm det}\,\frac{\delta D_\mu[\bar A](A^{U_0}-\bar A)}{\delta U_0}\right|_{U_0=\mathds{1}}\,.\nonumber
\eeq
The functional $\delta$ is again taken into account by introducing a Nakanishi-Lautrup field:
\beq
\delta(D_\mu[\bar A] (A_\mu-\bar A_\mu))\propto\int {\cal D}h\,e^{i\int_x h^a(x)D_\mu[\bar A] (A_\mu^a(x)-\bar A_\mu^a(x))}\,.\nonumber
\eeq
As for the Faddeev-Popov determinant, proceeding as in the previous problem, we find
\beq
\delta D_\mu[\bar A] (A^{U_0}_\mu-\bar A_\mu)=-iD_\mu[\bar A] (D_\mu[A]\delta U_0)\,,\nonumber
\eeq
from which one reads $\left.\delta D_\mu[\bar A](A^{U_0}_\mu-\bar A_\mu)/\delta U_0\right|_{U_0=\mathds{1}}$ and deduces that, up to an non-relevant multiplication constant, $J_{\bar A}[A]={\rm det}\,D_\mu[\bar A] D_\mu[A]$. The, introducing ghost and antighost fields as usual, one arrives at
\beq
J_{\bar A}[A]={\rm det}\,D_\mu[\bar A] D_\mu[A]=\int{\cal D}c{\cal D}\bar c\,e^{\int_x\bar c^a(x)D^{ab}_\mu[\bar A] D_\mu^{bc}[A]c^c(x)}\,.\nonumber
\eeq
\\

\noindent{{\bf Problem 3.3:} We have
\beq
D_\mu[A^U](UXU^\dagger) & = & \partial_\mu(UXU^\dagger)-i[UA_\mu U^\dagger+iU\partial_\mu U^\dagger,UXU^\dagger]\nonumber\\
& = & \partial_\mu(UXU^\dagger)-iU[A_\mu,X]U^\dagger+U(\partial_\mu U^\dagger)UXU^\dagger-UX\partial_\mu U^\dagger\nonumber\\
& = & \partial_\mu(UXU^\dagger)-iU[A_\mu,X]U^\dagger-(\partial_\mu U)XU^\dagger-UX\partial_\mu U^\dagger\nonumber\\
& = & U(\partial_\mu X)U^\dagger-iU[A_\mu,X]U^\dagger\nonumber\\
& = & U(\partial_\mu-i[A_\mu,X])U^\dagger\,.\nonumber
\eeq
In the intermediate steps, we have used that $\smash{U(\partial_\mu U^\dagger)=-(\partial_\mu U) U^\dagger}$ as follows from the unitarity of $U$.\\\\

\noindent{{\bf Problem 3.4:} That $\langle A_\mu\rangle_{\bar A}$ is constant and temporal for the considered backgrounds follows from translation invariance and rotation invariance. Now, apply the background gauge symmetry identity to constant color rotations of the form $e^{i\theta\sigma_3/2}$. Since the considered background is clearly invariant under these transformations, the identity turns into}
\beq
e^{i\theta\sigma_3/2}\langle A_\mu\rangle_{\bar A}e^{-i\theta\sigma_3/2}=\langle A_\mu\rangle_{\bar A}\,,\nonumber
\eeq
which is an actual constraint on $\langle A_\mu\rangle_{\bar A}$. More precisely, since the action of $\smash{U=e^{i\theta\sigma_3/2}}$ on the algebra is a rotation by an angle $\theta$ around direction $3$, the only possibility for $\langle A_\mu\rangle_{\bar A}$ is to be aligned with that direction. In other words, it takes a similar form as the background
\beq
\langle A_\mu(x)\rangle_{\bar A}=T\delta_{\mu0}\, r \frac{\sigma_3}{2}\,.\nonumber
\eeq

\section{Effective Action}

\noindent{{\bf Problem 4.1:} We have}
\beq
\frac{\delta S}{\delta\varphi^a(x)}=-\partial^2\varphi^a(x)+m^2\varphi^a(x)+\frac{\lambda}{6}(\varphi^b(x)\varphi^b(x))\varphi^a(x)\nonumber
\eeq
and thus
\beq
\frac{\delta^2 S}{\delta\varphi^a(x)\delta\varphi^b(y)} & = & (-\partial^2+m^2)\delta^{ab}\delta(x-y)\nonumber\\
& + & \frac{\lambda}{6}(\varphi^b(x)\varphi^b(x))\delta^{ab}\delta(x-y)+\frac{\lambda}{3}\varphi^a(x)\varphi^b(y)\delta(x-y)\,.\nonumber
\eeq

\vglue2mm

\noindent{{\bf Problem 4.2:} We have}
For $\varphi(x)=\phi$, the previous result writes
\beq
\frac{\delta^2 S}{\delta\varphi^a(x)\delta\varphi^b(y)} & = & (-\partial^2+m^2)\delta^{ab}\delta(x-y)\nonumber\\
& + & \frac{\lambda}{6}(\varphi^b\varphi^b)\delta^{ab}\delta(x-y)+\frac{\lambda}{3}\varphi^a\varphi^b\delta(x-y)\,.\nonumber
\eeq
whose Fourier transform is
\beq
\left(Q^2+m^2+\frac{\lambda}{6}\phi^a\phi^a\right)\delta^{ab}+\frac{\lambda}{3}\phi^a\phi^b\,.\nonumber
\eeq

\vglue2mm

\noindent{{\bf Problem 4.3:} We have}
\beq
F^a_{\mu\nu}[A+a] & = & \partial_\mu (A^a_\nu+a^a_\nu)-\partial_\nu (A^a_\mu+a^a_\mu)+gf^{abc}(A^b_\mu+a^b_\mu) (A^c_\nu+a^c_\nu)\nonumber\\
& = & \partial_\mu A^a_\nu-\partial_\nu A^a_\mu+gf^{abc}A^b_\mu A^c_\nu\nonumber\\
& + & \partial_\mu a^a_\nu+gf^{abc}A^b_\mu a^c_\nu\nonumber\\
& - & \partial_\nu a^a_\mu+gf^{abc}a^b_\mu A^c_\nu\nonumber\\
& + & gf^{abc}a_\mu^ba_\nu^c\nonumber\\
& = & F_{\mu\nu}^a[A]+D_\mu^{ab}[A]a_\nu^b-D_\nu^{ab}[A]a_\mu^b+gf^{abc}a_\mu^b a_\nu^c\,.\nonumber
\eeq
\\

\noindent{{\bf Problem 4.4:} We write}
\beq
\partial_\mu (X^a(x)Y^a(x)) & = & \partial_\mu (X^a(x))Y^a(x)+X^a(x)\partial_\mu (Y^a(x))\nonumber\\
& = & (\partial_\mu X^a(x)+gf^{acb}A_\mu^cX^b(x))Y^a(x)\nonumber\\
& + & X^a(x)\partial_\mu (Y^a(x))-gf^{acb}A_\mu^cX^b(x)Y^a(x)\nonumber\\
& = & (\partial_\mu X^a(x)+gf^{acb}A_\mu^cX^b(x))Y^a(x)\nonumber\\
& + & X^a(x)\partial_\mu (Y^a(x))+gf^{acb}A_\mu^cX^a(x)Y^b(x)\nonumber\\
& = & (\partial_\mu X^a(x)+gf^{acb}A_\mu^cX^b(x))Y^a(x)\nonumber\\
& + & X^a(x)(\partial_\mu Y^a(x))+gf^{acb}A_\mu^cY^b(x)).\nonumber
\eeq
\\

\noindent{{\bf Problem 4.5:} The matrix relation is trivially checked and then one uses that the determinant of a square block-diagonal matrix is the product of the determinants of the diagonal blocks.}\\\\

\noindent{{\bf Problem 4.6:} Under the Weyl transformation $\smash{r\to -r}$, we have $Q_\kappa=(\omega_n+r\kappa T,q)\to (\omega_n-r\kappa T,q)$. Since $Q_\kappa$ always appears squared, we can absorb the change of sign of $r$ by the change of variables $n\to -n$ in the Matsubara sums. Under $\smash{r\to r+2\pi}$, we have $Q_\kappa=(\omega_n+r\kappa T,q)\to (\omega_n+2\pi T+r\kappa T,q)$. For the mode $\smash{\kappa=0}$ there is no change. For $\smash{\kappa=\pm 1}$, the change can be absorbed by a change of variables $\smash{n\to n\mp 1}$.}\\

\section{Matsubara Sum-Integrals}

\noindent{{\bf Problem 5.1:} We have}
\beq
\frac{1}{Q^2_\kappa+m^2}=\frac{1}{-(i\omega_n+iT\,r\kappa)^2+\varepsilon_q^2}\,,\nonumber
\eeq
and thus
\beq
f(z)=\frac{1}{-(z+iT\,r\kappa)^2+\varepsilon_q^2}=\frac{1}{2\varepsilon_q}\left[\frac{-1}{z+iT\,r\kappa-\varepsilon_q}+\frac{1}{z+iT\,r\kappa+\varepsilon_q}\right].\nonumber
\eeq
This gives
\beq
T\sum_{n\in\mathds{Z}}\frac{1}{Q^2_\kappa+m^2} & = & -\left[-\frac{1}{2\varepsilon_q}n(\varepsilon_q-iT\,r\kappa)+\frac{1}{2\varepsilon_q}n(-\varepsilon_q-iT\,r\kappa)\right]\nonumber\\
& = & \frac{1+n(\varepsilon_q-iT\,r\kappa)+n(\varepsilon_q+iT\,r\kappa)}{2\varepsilon_q}\nonumber\\
& = & \frac{1}{2\varepsilon_q}+ {\rm Re}\,\frac{n(\varepsilon_q-iT\,r\kappa)}{\varepsilon_q}\,.\nonumber
\eeq
\\

\noindent{{\bf Problem 5.2:} We have}
\beq
(Q_\kappa\cdot\bar Q_\kappa)^2+m^2\bar Q_\kappa^2 & = & (\omega^\kappa_n\bar\omega^\kappa_n+q^2)^2+m^2((\bar\omega^\kappa_n)^2+q^2)\nonumber\\
& = & q^4+(2\omega^\kappa_n\bar\omega^\kappa_n+m^2)q^2+((\omega^\kappa_n)^2+m^2)(\bar\omega^\kappa_n)^2\,.\nonumber
\eeq
The square masses $M^2_\pm$ are minus the roots of this quadratic polynomial in $q^2$.\\

\section{Extension to SU($3$)}

\noindent{{\bf Problem 6.1} The idea is to combine $t^3$ and $t^8$ such that
\beq
at^3+bt^8=\frac{1}{2}\left(\begin{array}{ccc}
1 & 0 & 0\\
0 & 0 & 0\\
0 & 0 & -1
\end{array}\right) \quad {\rm and} \quad ct^3+dt^8=\frac{1}{2\sqrt{3}}\left(\begin{array}{ccc}
1 & 0 & 0\\
0 & -2 & 0\\
0 & 0 & 1
\end{array}\right),\nonumber
\eeq
since the first matrix generates the SU($2$) algebra together with $t^4$ and $t^5$, while the second matrix commutes with these generators. This gives the systems
\beq
a+b/\sqrt{3} & = & 1\,,\nonumber\\
-a+b/\sqrt{3} & = & 0\,,\nonumber
\eeq
as well as
\beq
c+d/\sqrt{3} & = & 1/\sqrt{3}\,,\nonumber\\
-c+d/\sqrt{3} & = & -2/\sqrt{3}\,,\nonumber
\eeq
which are solved as $a=1/2$, $b=\sqrt{3}/2$, $c=\sqrt{3}/2$ and $d=-1/2$. To proceed similarly with $t^6$ and $t^7$, we need to consider combinations such that
\beq
a+b/\sqrt{3} & = & 0\,,\nonumber\\
-a+b/\sqrt{3} & = & 1\,,\nonumber
\eeq
as well as
\beq
c+d/\sqrt{3} & = & -2/\sqrt{3}\,,\nonumber\\
-c+d/\sqrt{3} & = & 1/\sqrt{3}\,,\nonumber
\eeq
which are solved as $a=-1/2$, $b=\sqrt{3}/2$, $c=-\sqrt{3}/2$ and $d=-1/2$.\\\\

\noindent{{\bf Problem 6.2} From Problem 6.1, we know that by introducing the combinations $t^\pm\equiv (t^4\pm it^5)/\sqrt{2}$, we have
\beq
[at^3+bt^8,t^\pm]=\pm t^\pm \quad {\rm and} \quad [ct^3+dt^8,t^\pm]=0\,,\nonumber
\eeq
where $a=1/2$, $b=\sqrt{3}/2$, $c=\sqrt{3}/2$ and $d=-1/2$. It follows that
\beq
[t^3,t^\pm]=\pm\frac{1}{2}t^\pm \quad {\rm and} \quad [t^8,t^\pm]=\pm\frac{\sqrt{3}}{2}t^\pm\,,\nonumber
\eeq
from which we find the two roots $\pm(1/2,\sqrt{3}/2)$. Similarly, introducing the combinations $t^\pm\equiv (t^6\pm it^7)/\sqrt{2}$, we know from Problem 6.2 that
\beq
[at^3+bt^8,t^\pm]=\pm t^\pm \quad {\rm and} \quad [ct^3+dt^8,t^\pm]=0\,,\nonumber
\eeq
where $a=-1/2$, $b=\sqrt{3}/2$, $c=-\sqrt{3}/2$ and $d=-1/2$. It follows that
\beq
[t^3,t^\pm]=\mp\frac{1}{2}t^\pm \quad {\rm and} \quad [t^8,t^\pm]=\pm\frac{\sqrt{3}}{2}t^\pm\,,\nonumber
\eeq
from which we find the two roots $\pm(-1/2,\sqrt{3}/2)$.\\\\

\noindent{{\bf Problem 6.3} By construction, any point on one of the reflection axes of the lattice connects to the origin by a vector $s$ such that $s\cdot\alpha\in\mathds{Z}/2$ where the root $\alpha$ is the one orthogonal to the reflection axis. The vertices of the lattice are the intersections of all these reflection axes and, therefore, obey the same condition but for all the possible roots.}\\\\

\noindent{{\bf Problem 6.4:} We have}
\beq
\Phi[A]=\frac{1+2\cos\left(2\pi/3\right)}{3}=\frac{1+2(-1/2)}{3}=0\,.\nonumber
\eeq
\\

\noindent{{\bf Problem 6.5:} We have}
\beq
\Phi[A^C] & = & \frac{1}{3}\Bigg[e^{-i\frac{-4\pi x_8}{\sqrt{3}}}+2e^{i\frac{-2\pi x_8}{\sqrt{3}}}\cos\left(-2\pi x_3\right)\Bigg]\nonumber\\
& = & \frac{1}{3}\Bigg[e^{-i\frac{-4\pi x_8}{\sqrt{3}}}+2e^{i\frac{-2\pi x_8}{\sqrt{3}}}\cos\left(2\pi x_3\right)\Bigg],\nonumber
\eeq
which differs from $\Phi[A]$ only due to the sign in multiplying $x_8$. It follows that $\smash{\Phi[A^C]=\Phi[A]}$ in the case where $\smash{x_8=0}$.\\\\

\noindent{{\bf Problem 6.6:} We just need to replace $r\kappa$ and $\bar r\kappa$ by $r\cdot\kappa$ and $\bar r\cdot\kappa$ in the derivation. The modes $\kappa$ corresponding to zeros to dot contribute, while those corresponding to a root $\alpha$ and its opposite $-\alpha$ contribute the same, with a factor $\alpha_3^2$. We then arrive at}
\beq
& & \frac{T^2}{g^2}\left.\frac{\partial^2 V_{\bar r}(r)}{\partial r_3^2}\right|_{r=\bar r}=m^2\nonumber\\
& & \hspace{0.5cm}+\,g^2\sum_{\alpha}\int_Q(\alpha_3)^2\left\{\left[6\frac{m^2}{q^2}+12+2\frac{q^2}{m^2}\right]\frac{1}{{\bar Q}_\alpha^2+m^2}-\left[1+2\frac{q^2}{m^2}\right]\frac{1}{\bar Q_\alpha^2}\right\}.\nonumber
\eeq
where $\sum_\alpha$ denotes a sum over alph the roots, for instance $\alpha=(1,0)$, $\alpha=(1/2,\sqrt{3}/2)$ and $\alpha=(-1/2,\sqrt{3}/2)$. Since eventually $\bar r$ is taken equal to $\bar r_c$ which has no component along direction $8$, we fine
\beq
& & \frac{T^2}{g^2}\left.\frac{\partial^2 V_{\bar r_c}(r)}{\partial r_3^2}\right|_{r=\bar r_c}=m^2\nonumber\\
& & \hspace{0.5cm}+\,g^2\int_Q\left\{\left[6\frac{m^2}{q^2}+12+2\frac{q^2}{m^2}\right]\frac{1}{{\bar Q}_+^2+m^2}-\left[1+2\frac{q^2}{m^2}\right]\frac{1}{\bar Q_+^2}\right\}.\nonumber\\
& & \hspace{0.5cm}+\,\frac{g^2}{2}\int_Q\left\{\left[6\frac{m^2}{q^2}+12+2\frac{q^2}{m^2}\right]\frac{1}{{\bar Q}_{+/2}^2+m^2}-\left[1+2\frac{q^2}{m^2}\right]\frac{1}{\bar Q_{+/2}^2}\right\}.\nonumber
\eeq

\end{document}